%% file: main-resub.tex
\documentclass[twocolumn,aps,prb,floatfix,letterpaper]{revtex4}
\usepackage{graphicx,rotating,setspace,wrapfig,lscape,amsmath,
  amsfonts,amssymb,ulem}
 \usepackage{comment}
\usepackage{xr}
\usepackage[singlelinecheck=false ]{caption}
\usepackage[usenames,dvipsnames]{color}
\usepackage{newfloat}
\usepackage[tight]{subfigure}
\usepackage{caption,subcaption}
\usepackage{textcomp}
\usepackage{tikz}
\usetikzlibrary{decorations.pathreplacing}
\usetikzlibrary{colorbrewer}
\usetikzlibrary{arrows}
\usetikzlibrary{arrows.meta}
\usetikzlibrary{calc}
\usetikzlibrary{positioning}
\usepackage{makecell}
\usepackage{physics}
\usepackage{blochsphere}
\usetikzlibrary{quantikz}
\usetikzlibrary{trees}
\usepackage{soul}
\usepackage{tabularx}
\usepackage{mathtools}
\usepackage{float}
\usepackage{soul}
\usepackage {ctable}
\definecolor{LightCyan}{rgb}{0.88,1,1}
\usepackage{multirow}
\usepackage{newfloat}
\usepackage{cellspace}
\DeclareFloatingEnvironment[
    fileext=loa,
    listname=List of Algorithms,
    name=ALGORITHM,
    placement=tbhp,
]{algorithm}
\DeclareFloatingEnvironment[name=Box,within=none,placement=tbp]{CGBox}

\mathchardef\mhyphen="2D

\begin{document}


\normalem
\title{Quantum circuit and mapping algorithms for wavepacket dynamics: case study of anharmonic hydrogen bonds in protonated and hydroxide water clusters}
\author{Debadrita Saha,}
\affiliation{Department of Chemistry, and the Indiana University Quantum Science and Engineering Center (IU-QSEC), Indiana University, 800 E. Kirkwood Ave, Bloomington, IN-47405}
\author{Philip Richerme}   
\affiliation{Department of Physics, and the Indiana University Quantum Science and Engineering Center (IU-QSEC),
Indiana University, Bloomington, IN-47405}
\author{Srinivasan S. Iyengar\email{iyengar@iu.edu},}
\affiliation{Department of Chemistry, Department of Physics, and the Indiana University Quantum Science and Engineering Center (IU-QSEC), Indiana University, 800 E. Kirkwood Ave, Bloomington, IN-47405}
\date{\today}
\begin{abstract}
The accurate computational study of wavepacket nuclear dynamics is considered to be a classically intractable problem, particularly with increasing dimensions. Here we present two algorithms that, in conjunction with other methods developed by us, will form the basis for performing quantum nuclear dynamics in arbitrary dimensions. For one algorithm, we  present a direct map between the Born-Oppenheimer Hamiltonian describing the wavepacket time-evolution and the control parameters of a spin-lattice Hamiltonian that describes the dynamics of qubit states in an ion-trap quantum computer. This map is exact for three qubits, and when implemented, the dynamics of the spin states emulate those of the nuclear wavepacket. However, this map becomes approximate as the number of qubits grow. In a second algorithm we present a general quantum circuit decomposition formalism for such problems using a method called the Quantum Shannon Decomposition. This algorithm is more robust and is exact for any number of qubits, at the cost of increased circuit complexity. The resultant circuit is implemented on IBM's quantum simulator (QASM) for 3-7 qubits. 
In both cases the wavepacket dynamics is found to be in good agreement with the classical result and the corresponding vibrational frequencies obtained from the wavepacket density time-evolution, are in agreement to within a few tenths of a wavenumbers.
\end{abstract}

\maketitle

\twocolumngrid

\section{Introduction}
\label{introduction}
Protonated\cite{johnson-jordan-21mer,admp-21mer,admp-21mer-2}, neutral\cite{Klein-21mer} and hydroxide-rich water 
clusters\cite{AsthagiriOH,TuckermanOH,AgmonOH,Johnson-Zundel-OH-H2O-quantum,JohnsonOH-Science,mundy2009hydroxide} have been widely studied, both experimentally\cite{Miyazaki,Shen-SHG-1989,Shen-SFG-1997,hypo-iodous,protonatedaniline} and theoretically\cite{tuckerman1997quantumnaturehbond,Tuckermanwater,H5O2+XCcomp,Lobaugh:96} due to their broad significance in multiple areas including atmospheric\cite{atmosph-clusters1,atmosph-clusters2,McEwanPhillips:Review}, biological\cite{bio-clusters1,bio-clusters2,bio-clusters3,SSHB-Enzymes,SSHB-3} and condensed phase chemistry\cite{Marx:00,frag-PFOA}. 
There have been substantial efforts devoted to the understanding of the relatively high efficiency
of organic reactions on the surface of water droplets
\cite{Sharpless-ONwater,Gajewski-water-Claisen,Breslow-Diels-Alder-water-1,Breslow-Diels-Alder-water-2,Breslow-Organic-Water,Buckingham-Methane-water-tangential,Organic-OnWater-review,Organic-OnInWater,Aldol-OnWater,Arylation-Thiazoles-OnWater,Hydrogenation-OnWater,Moloney-OnWater,Non-HB-water-Hydrophobic-interface,Marcus-Onwater,Marcus-Freeoh-Onwater,Marcus-Air-Water-Sfg-2011,Kuhne-Water-vapor-interface,Kuhne-OnWater,Kuhne-PI-OnWater}. Protonated water clusters in polymer electrolyte membrane fuel cells\cite{Annurev-matsci-PBI} mediate proton transfer.

Experimental studies on vibrational properties of such hydrogen bonded cluster systems have blossomed due to the development of sophisticated cluster-based measurement techniques such as Argon-tagged single photon action spectral methods\cite{Johnson-2007-Science-Me2OH+} and the Infra-red Multiple Photon Dissociation (IRMPD)\cite{DTMoore-diglyme} approach. Both gas phase single-photon \cite{Johnson-2007-Science-Me2OH+,Johnson-Jordan-Zundel-Science}
and multiple-photon \cite{Meijer-FELIX,DTMoore-diglyme,Asmis-Zundel,Saykally-Lysine,Neumark-BrHBr-vib} vibrational action spectroscopic techniques
have been crucial in deciphering the precise signatures that contribute to dynamics
and spectroscopy in hydrogen bonded systems.
However, the accurate computational modeling of the processes involved in
these experiments requires both the quantum mechanical treatment of electronic as well as nuclear motion.
Due to the light mass of the transferring proton involved in such hydrogen bonded systems, these systems often exhibit quantum effects such as hydrogen tunneling and zero-point energy\cite{Klinman-Chemrevs-2006,SHS-PCET-SLO1,htrans,PCET-Nocera-ARPC-1998, PCET-Mayer-ARPC-1998,qwaimd-SLO-1,SLO-measurement}, and the correlated behavior of multiple nuclear degrees of freedom\cite{HDMeyer-Zundel-1,Johnson-Jordan-Zundel-Science,Johnson-Jordan-Zundel-JCP,Vib1,Manthe1992-kh,mctdh-meyer,Meyer-MCTDH-DFCO,VSCF,McCoy,HDMeyer-Zundel-2,Meyer-MCTDH-Zundel,Bowman-DMD-Zundel,mccoy:064317}. In such scenarios, classical approaches such as approximating nuclear motion to be harmonic about the equilibrium geometry \cite{Pople-2nd-derivs,Pople-2nd-derivs-freqs} fail to accurately predict the static and dynamic properties of such hydrogen-bonded systems \cite{HDMeyer-Zundel-1,qwaimd-wavelet,Johnson-Jordan-Zundel-Science,johnson-jordan-21mer,Me2O2D+Xiaohu}. A full quantum mechanical treatment of the nuclear degrees of freedom and its associated electronic interactions is often needed\cite{HDMeyer-Zundel-1,SADAF,HDMeyer-Zundel-2,HDMeyer-Zundel-3,Bowman-DMD-Zundel,Meyer-MCTDH-Zundel} to account for such quantum nuclear effects, including multi-dimensional effects, in the computation of their molecular properties. This then requires the computation of potential energy surfaces for describing the nuclear degrees of freedom in such systems, and the explicit time evolution of the quantum nuclear wave packet on the potential landscape. However, the study of such multi-dimensional quantum nuclear processes is complicated by: (a) the steep algebraic computational scaling of accurate electron correlation methods\cite{schlegel1991computational}, and (b) the exponential scaling of quantum nuclear dynamics with the number of quantum nuclear degrees of freedom\cite{Wyatt-Zhang,hibbs,MCTDH-Meyer1,Nielsen-Chuang-QuantComp,Feynman-Comp,Berman-Comp-complexity,Ohrn,kuppermann,talezer,monte,RPMD,Nicole-TN, Nicole-Shannon}.

Over the years several classical algorithms have been developed to improve the computational scaling of electronic\cite{Ayala-LS-MP2,Schutz-LOS-MP2,Head-Gordon-LS-MP2,Neese-Pavosevic-SparseMaps,fragAIMD-CC} and nuclear dynamics\cite{MCTDH-Meyer1,TIW,neo-noci,qwaimd,Burghardt2020MLMCTDH,frag-AIMD-multitop} problems. Orthogonally, recent years have also seen the development of quantum algorithms for performing electronic structure calculations on near-term quantum hardware. Promising quantum algorithms \cite{abrams1997simulation,Aspuru-Guzik-Science-2005,wang2008quantum,wecker2015progress,mcclean2016theory,o2016scalable,Martinez-VQE-Solver-Excited-States,frag-QC-Harry,smart2021quantum,SSI-Review1-QC-ES-QN,frag-QC-2} and experiments \cite{Lanyon_Photonic-quantum-comp,lu2011simulation,peruzzo2014variational,kandala2017hardware,hempel2018quantum,nam2020ground,Google-12qubit-HF} that address systems with strongly correlated electrons, and 
quantum simulations of vibronic spectra \cite{sparrow2018simulating,wang2020efficient,huh2015boson,wang2023quantum}, wavepacket evolution through conical intersections \cite{wang2022observation,nature2023geometricphase-1,nature2023geometricphase-2,gambetta2021exploring,kassal-nonadiab-2024,mazziotti2024conical}  and algorithms for reduced dimensional reactive scattering studies \cite{xing2023hybrid,kale2024simulation} do not appropriately describe quantum nuclear effects within hydrogen-bonded systems emanating from anharmonic effects and mode-coupling effects\cite{Me2O2D+Xiaohu,admp-21mer,admp-21mer-2,Johnson-2007-Science-Me2OH+,Johnson-Jordan-Zundel-Science,Scott-proj,johnson-jordan-21mer}. 
Steps in the direction of developing a quantum algorithm for simulating explicit quantum wave packet dynamics on anharmonic electronic structure-based potential energy surfaces were taken in Refs. \onlinecite{Debadrita-Mapping-1D-3Qubits,Sandia-1D-3Qubits,IonQ-Anurag}. In Ref. \onlinecite{Debadrita-Mapping-1D-3Qubits}, a Hamiltonian mapping protocol was introduced to simulate the quantum nuclear dynamics in proton transfer reactions on spin-lattice quantum simulators. This was successfully implemented on Sandia National Labs' Quantum Scientific Computing Open User Testbed (QSCOUT) ion-trap system as reported in Ref. \onlinecite{Sandia-1D-3Qubits} to study the vibrational spectra of the transferring proton in a short-strong hydrogen bonded system. A generalization of this scheme to higher nuclear dimensions using a tensor-network based formalism has recently been introduced in Ref. \onlinecite{IonQ-Anurag,TN-Miguel-Anurag}. A method to perform a parallel quantum computation of quantum wavepacket dynamics on a distributed set of ion-trap systems, along with implementation on IonQ's 11-qubit Harmony ion-trap quantum systems is discussed in Ref. \onlinecite{IonQ-Anurag}.

Quantum algorithms for simulating processes on quantum simulators can be broadly divided into two major classes: Hamiltonian or analog quantum simulation and unitary or quantum circuit decomposition (digital). In this paper, we discuss both—an analog Hamiltonian-based and a quantum circuit decomposition-based algorithm—to study spectroscopic features arising from anharmonic vibrations in small water clusters using quantum wavepacket dynamics. 
Hamiltonian or analog quantum simulation proceeds by mapping a desired Hamiltonian onto the quantum device’s Hamiltonian. This is done by programming the control parameters of the quantum device Hamiltonian. The mapping protocol in Ref. \onlinecite{Debadrita-Mapping-1D-3Qubits} simulates a chemical dynamics Hamiltonian on an ion-trap quantum computer by computing the control parameters of an Ising Hamiltonian, which describes the dynamics of ions in an ion-lattice. As a result, the ion-lattice dynamics directly correspond to the dynamics of the chemical system. 
The mapping protocol, however, is inherently approximate and only works well for systems with specific symmetries for a small number of qubits; errors grow with increasing number of qubits in this algorithm. Quantum circuit decomposition is more commonly used to write the unitary time-evolution operator corresponding to the Hamiltonian of a system in terms of universal quantum gate sets. The circuit model, in theory, can be extended to an arbitrary number of qubits but suffers from increased circuit depth and exponentially increasing number of entangling gates as we scale up to a higher number of qubits. This results in increased measurement errors and loss of contrast in the measured probabilities due to the low fidelities of the entangling gates. For a more accurate implementation of quantum nuclear dynamics problems, we discuss a  quantum circuit model-based approach to studying wavepacket dynamics. This decomposition technique is based on Quantum Shannon Decomposition (QSD)\cite{QSD-Shende}. 

The paper is organized as follows: In Section \ref{IsingH-structure}, we discuss the Hamiltonian mapping protocol that was developed in Ref. \onlinecite{Debadrita-Mapping-1D-3Qubits}. This approach is exact for three qubits, but becomes inherently  approximate as the number of qubits grow. Hence, in Section \ref{QSD_theory}, we outline the Quantum Shannon circuit decomposition technique that is used to decompose the unitary propagator corresponding to the nuclear Hamiltonian. The algorithm presented in Section \ref{QSD_theory} yields a quantum circuit with the number of entanglement gates close to the theoretical lower bound\cite{QSD-Shende} and works for an arbitrary number of qubits. Both techniques use the first quantized form of the nuclear Hamiltonian computed in a grid representation. Numerical results are presented for both algorithms in Section \ref{results_qsim} and \ref{results_qsd} for the anharmonic molecular vibrations of the shared proton in short-strong hydrogen bonds\cite{SSHB-1,SSHB-2,SSHB-3,SSHB-Warshel-1} that are present in protonated and hydroxide-rich water
clusters. These include explicit numerical propagation of both the molecular dynamics problem, the mapped spin-lattice dynamics governed by Ising-type Hamiltonian as obtained from the mapping protocol, and quantum circuit decomposition using QSD. The Quantum Shannon Decomposition approach is implemented on Qiskit, and results from quantum simulation on a classical computer are provided here. 
The systems considered in Section \ref{results} are the H$_5$O$_2^+$ and H$_3$O$_2^-$ ions. These low-barrier, short-strong hydrogen bonded systems\cite{SSHB-Warshel-1,SSHB-NielsonReview,JohnsonOH-Science,AgmonOH,Johnson-Zundel-OH-H2O-quantum,H2O6OH-Xiaohu,AQC} are of fundamental significance in proton-transfer processes, and have a critical role in the enhanced mobility of protons and deuterons in condensed phase, in biological ion-channels and enzymes, and in fuel cells\cite{Bruce-Hudson-perchlorate,Haile-CsHSO4-Entropy-2007,Haile-2007-Faraday,Suzuki-H1-Cs-PRB-2006}.
We inspect the vibrational properties of these systems using quantum algorithms presented here. Conclusions are given in Section \ref{conclusion}. 
\section{Quantum algorithms for wavepacket dynamics: mapping protocols for spin-lattice simulations}\label{IsingH-structure}
The generalized form of the Ising Hamiltonian can be used to describe the interactions of spin-states on a spin lattice. The mapping protocol introduced in Ref. \onlinecite{Debadrita-Mapping-1D-3Qubits} relates a quantum nuclear Hamiltonian, involving the nuclear kinetic energy and the Born-Oppenheimer potential surface obtained from electronic structure theory, to a generalized Ising Hamiltonian realizable on a range of quantum systems including ion trap quantum simulators. 
The inherent symmetries of the Ising Hamiltonian which describes the dynamics of effective spin-states on a lattice and that of the Born-Oppenheimer potential and nuclear kinetic energy are exploited to arrive at the map. In the following sections, we provide the transformations and their geometric interpretations that expose these symmetries in the Ising (Section \ref{basispermutation}) and the nuclear Hamiltonian (Section \ref{Transformations_Hmol}). The parameters of the Ising Hamiltonian that are programmable on a lattice of ions are computed from the quantum nuclear Hamiltonian and the algorithm is summarized in Section \ref{hamiltonian-map-summary}. For details, the reader is directed to Ref \onlinecite{Debadrita-Mapping-1D-3Qubits}.
\subsection{The role of geometric structure in generalized Ising Hamiltonians towards achieving a map to quantum nuclear dynamics}\label{basispermutation}
Any two-level quantum system can be mapped to a spin-1/2 particle in an effective magnetic field. In the ion trap used in Ref. \onlinecite{Sandia-1D-3Qubits}, the qubit levels encoded onto the $^2$S$_{1/2}$ state, namely, $\ket{F=0,m_F=0}$ and $\ket{F=1,m_F=0}$ hyperfine `clock' states of $^{171}$Yb$^+$ ions are mapped onto the two levels of a spin-1/2 particle and denoted as $\ket{0}$ and $\ket{1}$, respectively \cite{olmschenk2007manipulation}. 
In its most generalized form, the Ising Hamiltonian that can be implemented on the ion-trap is given by, 
\begin{eqnarray}
{\cal H}_{IT} &=& 
\sum_{i =1}^{N-1}
\sum_{j>i}^N
\left\{ J_{ij}^{x}\sigma_{i}^{x}\sigma_{j}^{x} +
J_{ij}^{y}\sigma_{i}^{y}\sigma_{j}^{y} +
J_{ij}^{z}\sigma_{i}^{z}\sigma_{j}^{z} 
\right\}
+ \nonumber \\ &&
\sum_{i=1}^{N}
\left\{ 
B_{i}^{x} \sigma_{i}^{x} +
B_{i}^{y} \sigma_{i}^{y} +
B_{i}^{z} \sigma_{i}^{z}
\right\}
\label{HIT-gen}  
\end{eqnarray}
where $\left\{ \sigma_i^{x}, \sigma_i^{y}, \sigma_i^{z} \right\}$ are the Pauli spin operators on the $i$-th lattice site along the respective spatial direction. The energy gap between the states at each ion, $i$, and their relative orientations are controlled by local effective magnetic fields, $\left\{ B_{i}^{x}, B_{i}^{y}, B_{i}^{z}\right\}$, and the spin-spin coupling between different lattice sites, $i$ and $j$, is controlled using laser pulses, also spatially non-isotropic, and represented as $\left\{ J_{ij}^{x}, J_{ij}^{y}, J_{ij}^{z}\right\}$.
It is critical to note that the expression above is more general than the form of the Ising Hamiltonian commonly used in condensed matter physics \cite{onsager1944Ising,mccoy1973Ising,wilson1971rgIsing}, NMR, other zero-field splitting studies \cite{anderson1964field} where only nearest-neighbor interactions or spin-lattice sites within a certain spatial distance may interact and magnetic fields only across certain directions are considered, and transverse Ising models\cite{cohen2014d,D-Wave-AQC-orig,D-Wave-AQC} implemented for adiabatic quantum computing for electronic structure studies\cite{Kais-Ising}.
The magnetic fields and inter-site coupling parameters, $\left\{ J_{ij}^{\gamma}, B_{i}^{\gamma}\right\}$ with $\gamma \in \left\{x, y, z\right\}$, form a set of programmable parameters that can be manipulated to simulate a general form of the Ising-type Hamiltonian, as in Eq. (\ref{HIT-gen}), on a lattice of ions.

The Ising Hamiltonian structure exhibits inherent symmetries in terms of these control parameters, that can be exploited to map a class of Hamiltonians to the quantum simulator. In Ref. \onlinecite{Debadrita-Mapping-1D-3Qubits}, we show how a certain permutation of the computational basis in which such Hamiltonians are conventionally represented, reveals a block structure of the Hamiltonian.  Here, we provide a geometric visualization of the computational basis set ordering scheme through a generalized representation of the Bloch sphere. Through this geometric representation, we illustrate the classification of the computational basis that exposes the inherent symmetries in the Ising Hamiltonian. Consequently, we discuss and illustrate the structure of the Ising Hamiltonian in this ordered basis set.

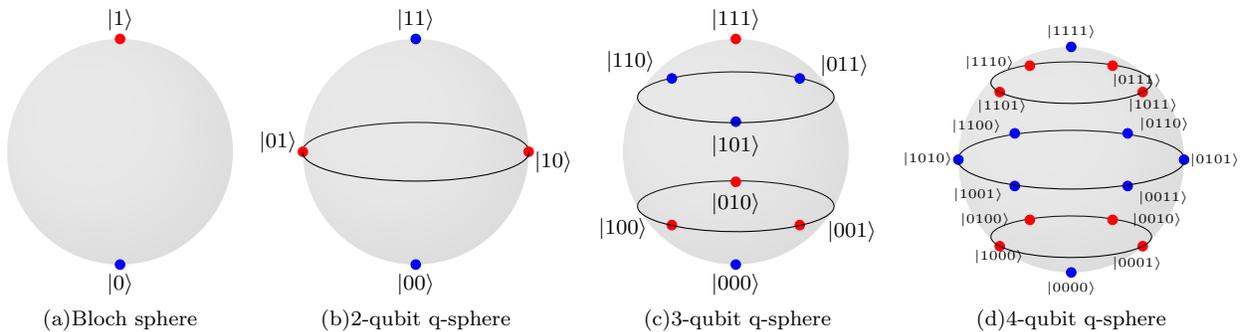
\begin{figure*}[tbp]
    \begin{minipage}{\textwidth}
    \subfigure[Bloch sphere]{\input{qsphere_1qubit}}
    \subfigure[2-qubit q-sphere]{\input{qsphere_2qubits}}
    \subfigure[3-qubit q-sphere]{\input{qsphere_3qubits}}
    \subfigure[4-qubit q-sphere]{\input{qsphere_4qubits}}
    \caption{ \label{bloch1_4} A generalized Bloch sphere for an arbitrary number of qubits, along with the classification of basis states shown using red and blue colors. }
    \end{minipage}
\end{figure*}
Conventionally, the spin-lattice Hamiltonian is represented in the $2^N$-dimensional space of spin-$\frac{1}{2}$ states where each spin state corresponds to a two-level system that can either be up or down and equivalently $\ket{0}$ or $\ket{1}$. A Bloch sphere (Figure \ref{bloch1_4}(a)) provides a geometrical representation of all pure states of a single spin system. To generalize the Bloch sphere representation for a single spin to a higher number of spins, we borrow the idea of q-sphere from IBM Qiskit \cite{qiskit2024}, to geometrically represent all $2^N$ spin states of an $N-$qubit system. Figure \ref{bloch1_4} complements our discussion. 

The q-sphere for an $N-$qubit system ($N \ge 2$) is divided into rungs as follows. We begin by placing the all-down state $\ket{00 \cdots 0}$ at the bottom pole or the $0^{th}$ rung of the q-sphere. We then apply the total spin raising operator on this state, one spin at a time to obtain all other states in the higher rungs of the sphere. The action of the total spin raising operator once on the $0^{th}$ rung yields all $N$-states that correspond to a single spin up and $N-1$ spins down. These are represented on a single plane close to the  $\ket{00 \cdots 0}$. See Figure \ref{bloch1_4}. The action of the total spin-raising operator on all states of this rung yields all possible $N(N-1)/2$ states of the $2^{nd}$ rung, and these states are represented on the next plane. As we go to higher qubits, all possible $^N$C$_{n}$ states for the $n^{th}$ rung are obtained. This way, all basis states in the spin-lattice system are represented on the q-sphere and the number of states in each rung of the q-sphere are identical to those encountered in Pascal's triangle. All combinations of spin states with the same total $S_{z}$-value occupy the same rung.

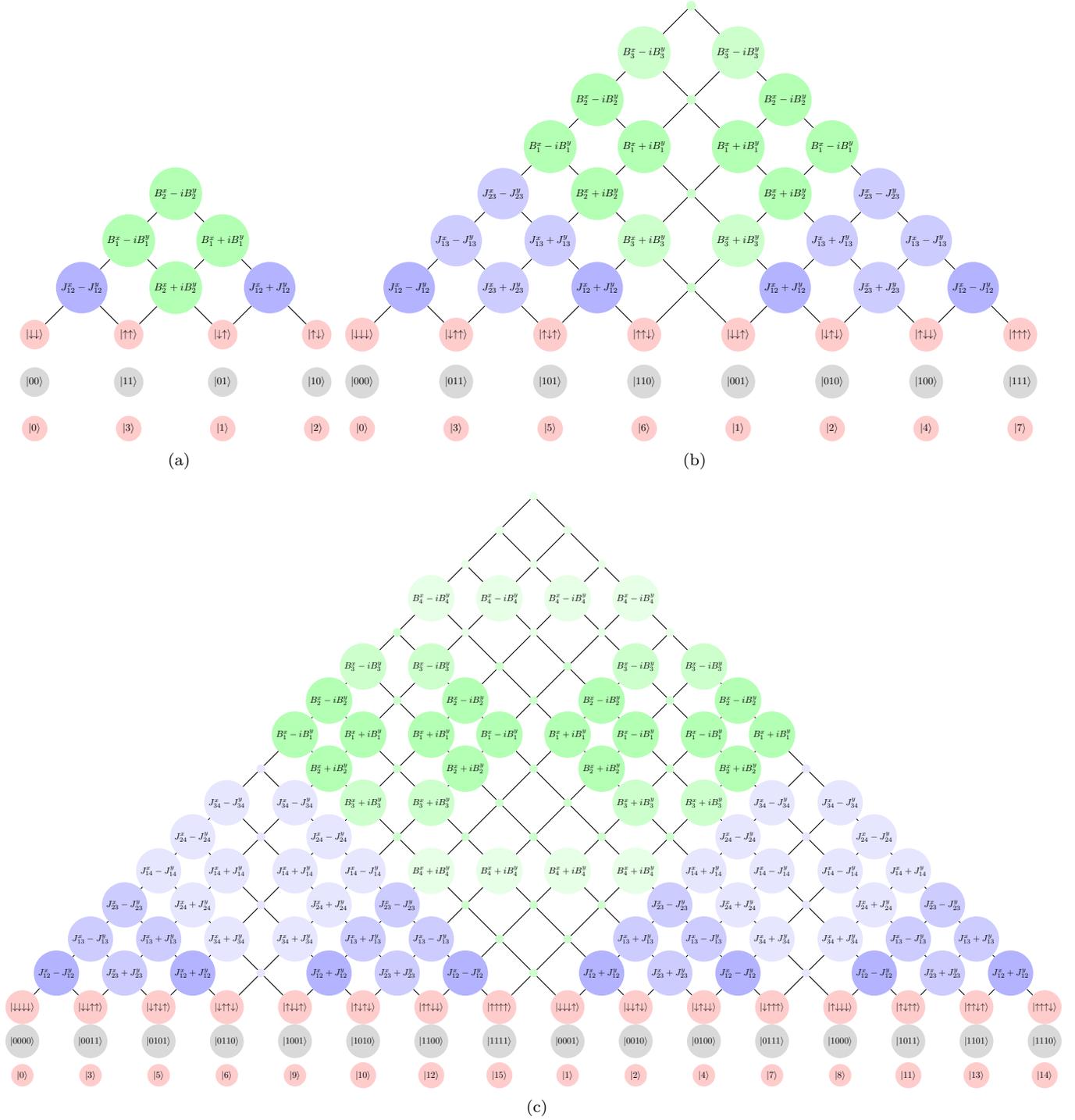
\begin{figure*}[tbp]
\subfigure[]{\input{2qubit-Ising}}
\subfigure[]{\input{3qubit-Ising}}
\subfigure[]{\input{4qubit-Ising}}
\caption{\label{Ising-block-graph-main} Recursive block structure of the Ising Hamiltonian in Eq. (\ref{HIT-gen}) for three (a) and four (b) qubits. The upper triangular portion of the Hamiltonian matrix is shown (excluding the diagonal). The computational basis  is partitioned into odd, $\left\{ {\bf S^+}^{2n-1} \ket{00\cdots}\right\}$, and even, $\left\{ {\bf S^+}^{2n} \ket{00\cdots}\right\}$, spans of the total spin raising operators. The interaction between states,  $\ket{i}$ and $\ket{j}$ is the $ij^{th}$ matrix element of the ion trap Hamiltonian. For example in (b), $\bra{0101} {\cal H}_{IT}$$\ket{1111} =  J_{13}^{x} - J_{13}^{y}$. The off-diagonal block that couples the odd and even spans of the total spin-raising operators are marked in green. Zero coupling is represented with a ``dot''.}
\end{figure*}

The geometrical representation of the $N-$qubit basis states using the q-sphere representation lends itself naturally to the classification of basis states we discuss for representing the Ising Hamiltonian. In Ref.\onlinecite{Debadrita-Mapping-1D-3Qubits}, we note that the basis vectors created from acting an even number of lattice-site spin raising operators, $\left\{ S_i^+ \right\}$ on the full downspin state, $\ket{00\cdots}$, yield the set, $\left\{ \ket{00 \cdots};  S_i^+ S_j^+\ket{00 \cdots}; S_i^+ S_j^+ S_k^+ S_l^+\ket{00 \cdots}; \cdots \right\}$, that are grouped as part of one block of the ion-trap Hamiltonian and those that are obtained using an odd number of raising operators:  $\left\{S_i^+\ket{00 \cdots}; S_i^+ S_j^+ S_k^+\ket{00 \cdots}; \cdots \right\}$ are grouped into a second block. The states belonging to the same block are represented using the same color (blue or red) in Figure \ref{bloch1_4}.
As can be seen in the q-sphere representation in Figure \ref{bloch1_4}, the action of the odd and even number of site-specific spin-raising operators on the all down spin state $\ket{00..00}$ correspond to its alternating rungs. The alternating rungs can now be grouped to form the two basis set blocks in which the Ising Hamiltonian has a block structure as discussed below and shown in Figure \ref{Ising-block-graph-main}.
The diagonal elements of the Ising Hamiltonian, when represented in the computational basis, are linear combinations of all $\left\{J_{ij}^{z}, B_{i}^{z}\right\}$ parameters. Figure \ref{Ising-block-graph-main} does not show these to maintain clarity. Since, inside each block of the permuted computational basis, the bases differ by at least two spin flips, the bases inside each block are connected by spin-spin coupling parameters $\left\{J_{ij}^{x}, J_{ij}^{y}\right\}$ (shown in shades of purple in Figures \ref{Ising-block-graph-main}a-c). 
Between the two blocks, the pairs of bases that differ by a single spin flip or an odd number of flips are coupled by the $\left\{B_{i}^{x}, B_{i}^{y}\right\}$ parameters and hence those form the elements of the off-diagonal blocks of the Ising Hamiltonian in this permuted basis (as shown in green shades in \ref{Ising-block-graph-main}a-c). 
The structure derived here is completely general for N qubits as can be seen in Figures \ref{Ising-block-graph-main} (a), (b), and (c) for two, three, and four qubits, respectively. 

Furthermore, the diagonal and off-diagonal blocks for the $N$-qubit Ising Hamiltonian can be recursively obtained from those of the $N-1$ qubit Hamiltonian. This recursive structure within each block of the Ising Hamiltonian is illustrated in Figures \ref{Ising-block-graph-main} (a), (b), and (c) using a gradient in colors, purple for the diagonal blocks with $J_{ij}^{x}\pm J_{ij}^{y}$ elements and green for the off-diagonal blocks with $B_{i}^{x}\pm iB_{i}^{y}$ matrix elements. For example, the recursion can be seen as follows for the diagonal block containing elements with spin-spin interaction terms. We begin with the elements of the diagonal block for a 2-qubit Ising Hamiltonian as in Figure \ref{Ising-block-graph-main} (a), $J_{ij}^{x}\pm J_{ij}^{y}$ for $i=1$, and $j=2$ corresponding to the only spin-spin interaction in a two-qubit system shaded the darkest in all Figures \ref{Ising-block-graph-main} (a), (b), and (c). The introduction of a third qubit introduces all possible spin-spin interactions $J_{i3}^{x}\pm J_{i3}^{y}$ for all $i \le 2$ as shown with the next shade of purple in Figure \ref{Ising-block-graph-main} (b). Similarly, all possible spin-spin interactions  $J_{i4}^{x}\pm J_{i4}^{y}$ for $i \le 3$ also appear in the diagonal blocks of the 4-qubit Ising Hamiltonian as shown using the lightest shade of purple in Figure \ref{Ising-block-graph-main} (c). The 2- and 3-qubit diagonal block elements are nested in the 4-qubit Ising Hamiltonian. Therefore, with each additional qubit, the structure of the $N-1$ qubit Ising Hamiltonian is preserved, and blocks containing $J_{iN}^{x}\pm J_{iN}^{y}$ for all $i \le N-1$, the interaction of the $N^{th}$ spin with the $N-1$ spins are added. The off-diagonal blocks also have a similar recursive structure, wherein with the addition of a qubit, a block with elements $B_{N}^{x}\pm iB_{N}^{y}$ is added to the Ising Hamiltonian. This is made clear in Figures \ref{Ising-block-graph-main} (a)-(c) with the gradation in green used for $B_{1}^{x}\pm iB_{1}^{y}$, $B_{2}^{x}\pm iB_{2}^{y}$, and $B_{N}^{x}\pm iB_{N}^{y}$ elements in the Hamiltonians. This block-form of the Ising-type Hamiltonian and the associated structure in Figure \ref{Ising-block-graph-main}, is a significant general result\cite{Debadrita-Mapping-1D-3Qubits}. To the best of our knowledge, such a structure of the general Ising model was first discussed in Ref \onlinecite{Debadrita-Mapping-1D-3Qubits}, and we see that this analysis is critical for mapping arbitrary problems.


\subsection{Nuclear Hamiltonian: Transformations that yield a block diagonal structure}
\label{Transformations_Hmol}
\begin{figure}
    \centering
    \includegraphics[width=\columnwidth]{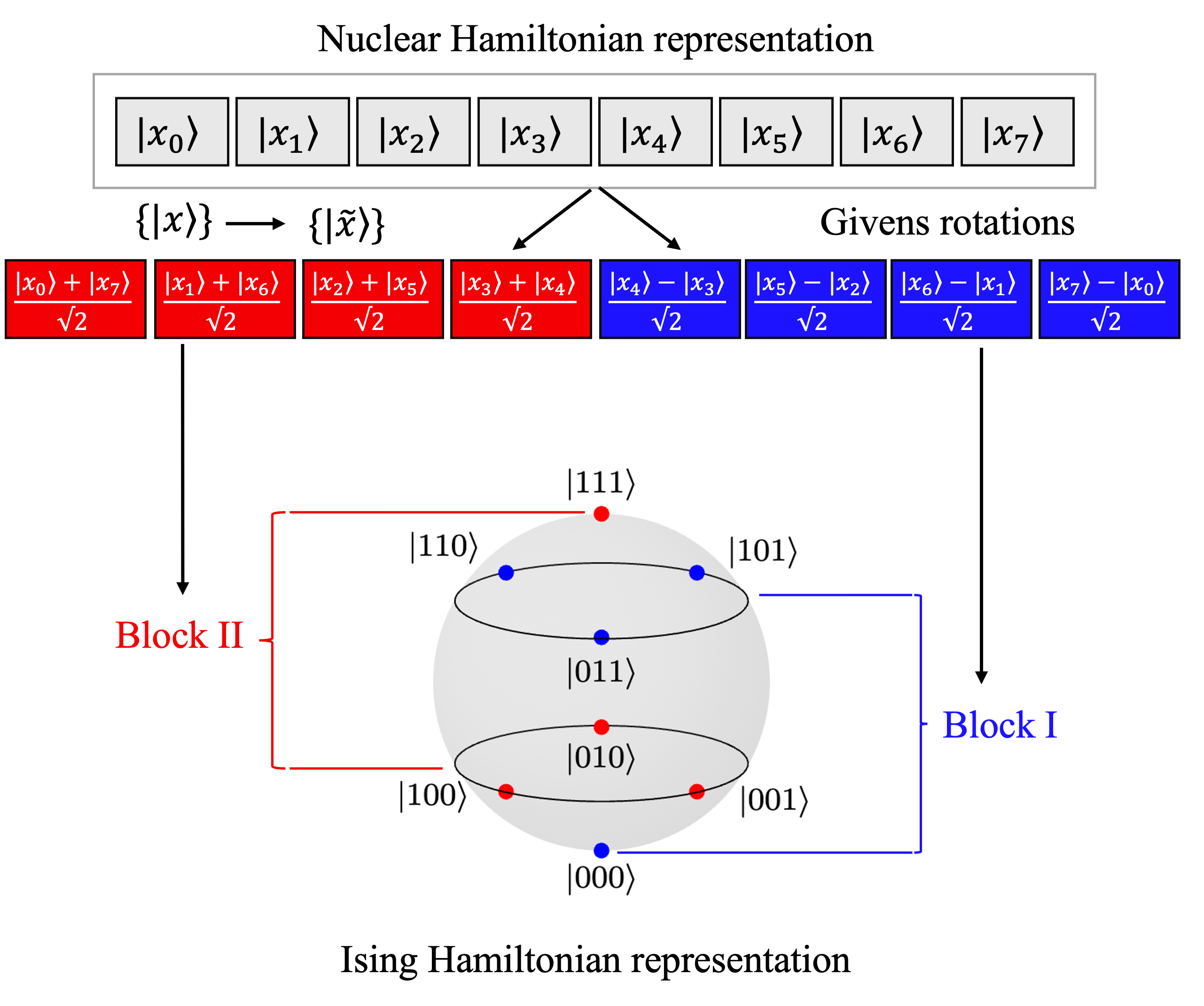}
    \caption{The classification of the Givens-transformed grid bases, and the permuted computational bases in the q-sphere representation. Note that the Givens rotations result in symmetric and anti-symmetric combinations of pairs of symmetrically located grid basis states. The map between the transformed Hamiltonians results from a map between the corresponding blocks of the basis of representation.}
    \label{basis-transform-map}
\end{figure}
The nuclear Hamiltonian is computed using a coordinate representation, where the dimension along the donor-acceptor axis is discretized into $2^{N}$ points. The matrix elements of the nuclear Hamiltonian, $\mathcal{H}_{Mol}$ in this coordinate representation, $\{\ket{x_{i}}\}$ is therefore given by
\begin{align}
    \bra{x} {{\cal H}}_{Mol} \ket{x^{\prime}} 
    = 
    K(x,x^{\prime}) + V(x)\delta \left( {x-x^{\prime}} \right)
    \label{HDAF-molQD}
\end{align}
The potential energy for the quantum dimensions is computed using standard electronic structure methods for which further details for each system are provided in the results section. The nuclear kinetic energy in this grid representation is approximated using the analytic banded Toeplitz Distributed Approximating Functionals (DAFs) \cite{DAFprop-PRL,discreteDAF}. 
\begin{align}
K(x,x^{\prime}) =& K(\left\vert x-x^{\prime}\right\vert) =
\frac{-\hbar^2}{4m\sigma^3\sqrt{2\pi}}
\exp \left\{ -\frac{ {\left( x - x^\prime \right)}^2}
{2 {\sigma}^2} \right\} \nonumber \\ & \sum_{n=0}^{M_{DAF}/2} {\left( \frac{-1}{4} \right)}^n \frac{1}{n!} 
H_{2n+2} \left( \frac{ x - x^\prime }{ \sqrt{2} \sigma} \right).
\label{DAFfreeprop+derivative}
\end{align} 
where, $H_{2n+2}\left( \frac{ x - x^\prime }{ \sqrt{2} \sigma} \right)$ are the even order Hermite polynomials that only depend on the spread separating the grid basis vectors, $\ket{x}$ and $\ket{x^{\prime}}$, and $M_{DAF}$ and $\sigma$ are parameters that together determine the accuracy and efficiency of the resultant approximate kinetic energy operator. The nuclear Hamiltonian, $\mathcal{H}_{Mol}$ in this coordinate representation, $\{\ket{x_{i}}\}$, therefore has a banded Toeplitz structure due to the structure of the kinetic energy when expressed in terms of DAFs. This banded-Toeplitz representation of the DAF approximation for the kinetic energy operator in eq. \ref{DAFfreeprop+derivative}, where the property of its matrix elements, $K_{ij} \equiv K{(\vert} i-j {\vert )}$, has a critical role in reducing the nuclear Hamiltonian to the form of ${\cal H}_{IT}$ as discussed in Ref. \onlinecite{Debadrita-Mapping-1D-3Qubits} and summarized in Section \ref{hamiltonian-map-summary}.

The Householder and Givens transformations are commonly used matrix transformations that allow arbitrary matrices to be reduced to canonical forms\cite{golub2013matrix}. 
Here, we use a sequence of Givens transformations to transform the banded Toeplitz form of the nuclear kinetic energy operator into a block diagonal form, $\mathcal{\tilde{H}}_{Mol}$ that is commensurate with that of the Ising Hamiltonian in the permuted spin basis, as discussed in Section \ref{IsingH-structure}. The details of the exact transformation are presented in section A of the SI. 
Furthermore,  this block diagonal form is maintained, when the one-dimensional potential energy surface belongs to the $C_{s}$ point group, whose only symmetry operations are identity and reflection about a mirror plane. For cases where such symmetry does not exist, the more general set of transformations shown in Section \ref{QSD_theory} may be employed. The Givens matrix elements are the characters of this point group and thus with each Givens transformation, a rotation is effected in the two-dimensional plane of the pair of symmetrically located grid basis states. 
The resulting Givens transformed basis, $\{\ket{\tilde{x_{i}}}\}$ from the corresponding grid basis, $\{\ket{x_{i}}\}$ is illustrated in Figure \ref{basis-transform-map} for the case of eight grid points. We, therefore, exploit the banded Toeplitz symmetry of the DAF kinetic energy operator and symmetry of the potential energy surface. The details of this transformation can be found in SI section A and Ref. \onlinecite{Debadrita-Mapping-1D-3Qubits}, where the explicit transformations of each matrix element of the transformed Hamiltonian are given in terms of the elements of the banded Toeplitz Hamiltonian and the elements of the Givens matrices.

\subsection{Obtaining ion-trap parameters, $\{B^{z}_{i},J^{z}_{ij}\}$ from transformed nuclear Hamiltonian}\label{hamiltonian-map-summary}
As seen from the discussions in Sections \ref{basispermutation} and \ref{Transformations_Hmol}, the Ising Hamiltonian and the nuclear Hamiltonian both have block structures resulting from the permutation of the computational basis and the Givens transformation of the grid basis, respectively. Owing to their commensurate structures, a direct map between each element of the nuclear Hamiltonian, $\mathcal{H}_{Mol}$ in the $\{\ket{\tilde{x_{i}}}\}$ basis to the corresponding elements of the Ising Hamiltonian, $\mathcal{H}_{IT}$ can be generated to compute the Ising Hamiltonian parameters. Our goal is to use the diagonal and off-diagonal elements of $\tilde{\mathcal{H}}_{Mol}$
to obtain the sets of ion-trap parameters $\{B^{z}_{i},J^{z}_{ij}\}$ and $\{J^{x}_{ij},J^{y}_{ij}\}$, respectively. This establishes a map between the givens transformed grid basis and the permuted computational basis, as is shown in Figure \ref{basis-transform-map}.
The mapping expression between the elements of the molecular Hamiltonian and the corresponding elements of the ion-trap Hamiltonian may be written as
    \begin{align}
       {\bra{\tilde{x}}} \mathcal{H}_{Mol} {\ket{\tilde{x}'}} \equiv  {\bra{\tilde{\lambda}}} {\cal H}_{\text {IT}} {\ket{\tilde{\lambda}'}} 
        \label{Diag-B}
    \end{align} 
for ${\ket{\tilde{\lambda}}}$ corresponding to the computational bases in blocks I or II for the Ising Hamiltonian (as discussed in Section \ref{basispermutation} and shown in Figures \ref{bloch1_4} and \ref{basis-transform-map}).
Using Eq. (\ref{HDAF-molQD}) and the transformations discussed in Section \ref{Transformations_Hmol} (detailed in SI A) to write the elements of $\mathcal{\tilde{H}}_{Mol}$ and the corresponding Ising Hamiltonian matrix elements for Eq. (\ref{HIT-gen}), for the left and right sides of Eq. (\ref{Diag-B}) corresponding to the diagonal elements only, we obtain
\begin{align}
        \left[ {K}(x_{i},x_{i}) - {K}(x_{i},x_{n-i}) \right] + \frac{1}{{2}} \left[ V(x_{i}) + V(x_{n-i}) \right] \nonumber \\ = \sum_{j=1}^{N} (-1)^{\tilde{\lambda}_{j}} B_{j}^{z} + \sum_{j=1}^{N-1} \sum_{k>j}^{N} (-1)^{\tilde{\lambda}_{j}\oplus\tilde{\lambda}_{k}} J_{jk}^{z} \nonumber \\ {\text for } \; i \leq n/2 \label{Diag-B1-1} \\
        \left[ {K}(x_{i},x_{i}) + {K}(x_{i},x_{n-i}) \right] + \frac{1}{{2}} \left[ V(x_{i}) + V(x_{n-i}) \right] \nonumber \\ = \sum_{j=1}^{N} (-1)^{\tilde{\lambda}_{j}} B_{j}^{z} + \sum_{j=1}^{N-1} \sum_{k>j}^{N} (-1)^{\tilde{\lambda}_{j}\oplus\tilde{\lambda}_{k}} J_{jk}^{z} \nonumber \\ {\text for } \; i > n/2 
        \label{Diag-B1}
\end{align} 
where $\oplus$ on the right side denotes the addition modulo 2, 
$\tilde{\lambda}_{j}$ is the $j^{th}$ bit of the bit representation of $\ket{\tilde{\lambda}}$ with values $0$ or $1$ as shown in Figures \ref{bloch1_4} and \ref{Ising-block-graph-main}. 

\begin{figure*}[tbp]
    \centering
    \includegraphics[width=\textwidth]{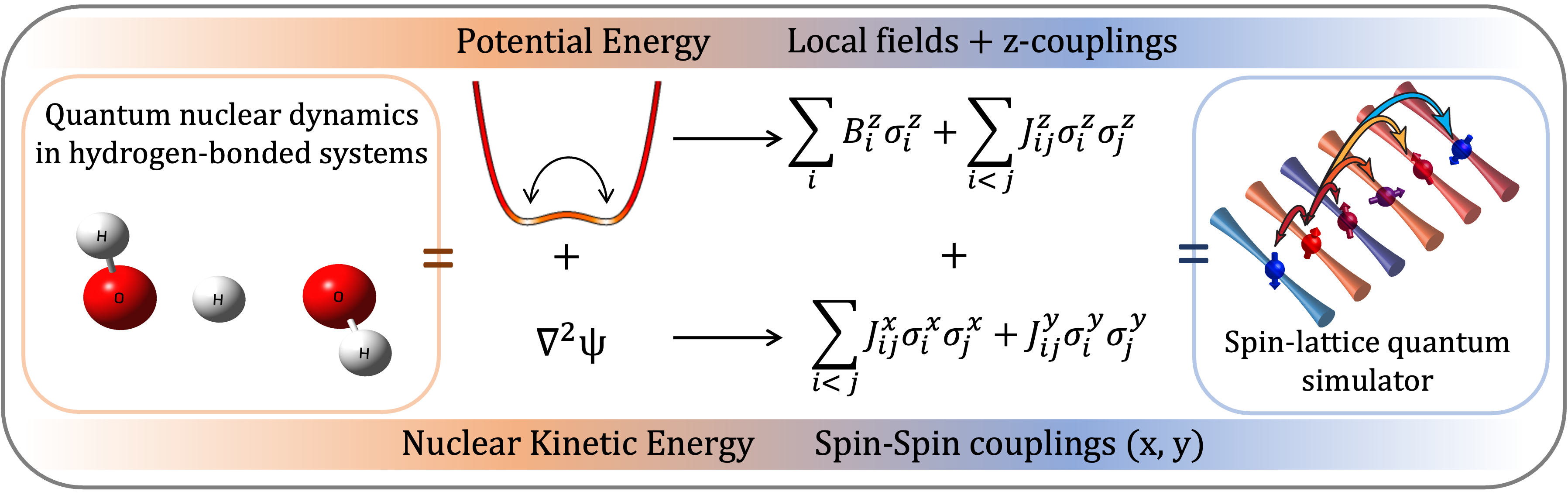}
    \caption{An outline of the mapping algorithm: The algorithm converts the Born-Oppenheimer potential surface and kinetic energy terms in a quantum-nuclear problem to a set of controllable ion-trap parameters, $\left\{ \left\{ B_i^z \right\}; \left\{ J_{ij}^x, J_{ij}^y, J_{ij}^z \right\} \right\}$, and facilitates the dynamical evolution of quantum states in an ion trap.}
    \label{hamiltonianmap_summary}
\end{figure*}
A detailed discussion on this map is provided in Ref. \onlinecite{Debadrita-Mapping-1D-3Qubits} where it is shown that the ion-trap control parameters, $\left\{ B_i^z;  J_{ij}^z \right\}$ are specific Hadamard transforms of ${\bra{\tilde{x}}} \mathcal{H}_{Mol} {\ket{\tilde{x}}}$. In a similar manner, the off-diagonal elements of $\mathcal{\tilde{H}}_{Mol}$ are mapped to the corresponding $\mathcal{H}_{IT}$ elements to obtain the $\left\{ J_{ij}^x;  J_{ij}^y \right\}$ parameters. The map is approximate for a higher number of qubits, for which the error estimates are provided in Ref. \onlinecite{Debadrita-Mapping-1D-3Qubits}. In Figure \ref{hamiltonianmap_summary}, the map is illustrated for the H$_3$O$_2^-$ system to be studied later in the publication. 

The reasons behind the intrinsically approximate nature of the mapping algrithm are discussed in detail in Ref. \cite{Debadrita-Mapping-1D-3Qubits}. We summarize the main features here. 
In essence, as the number of qubits $N$ increases, the Ising Hamiltonian parameters ($B$ and $J$ handles in Eq. (1)) scale as,
\begin{align}
    \left\{ N + N(N-1)/2 \right\} + \left\{ N(N-1) \right\} \rightarrow {\cal O} \left( N^2 \right),
    \label{IT-handles1}
\end{align}
Here (a) the first quantity, $\left\{ N + N(N-1)/2 \right\}$, refers to the parameters, $\left\{ B_{i}^z;  J_{ij}^z \right\}$, that form the diagonal elements of the Ising matrix, (b) the second quantity on the left, $\left\{ N(N-1) \right\}$, refers to the parameters, $\left\{ J_{ij}^x \pm J_{ij}^y \right\}$, that control the off-diagonal elements. 
This scaling and structure of the spin-lattice Hamiltonian\cite{Debadrita-Mapping-1D-3Qubits}, may restrict  the mapping of a general unitary operator, since a general $2^N \times 2^N$ unitary  matrix may have ${\cal O} \left( 2^N \right)$ independent elements. Thus, as the number of qubits increases, the mapping algorithm becomes more and more approximate as the number of equations given by Eqs. (\ref{Diag-B1-1}) and (\ref{Diag-B1}) exceeds the number of spin-lattice parameters $\left\{ B_{i}^\gamma;  J_{ij}^\gamma \right\}$. However, it may also be possible to reduce the number of actual paramenters within the $2^N \times 2^N$ unitary  matrix  and these aspects will be considered in future publications\cite{IonQ-Anurag}. 

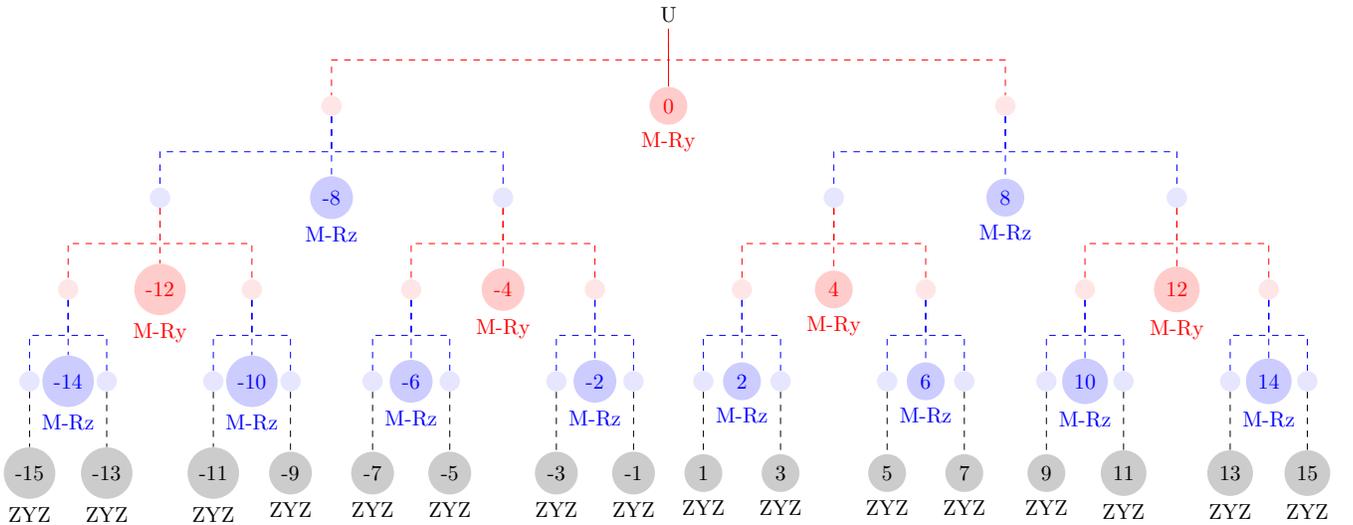
\begin{figure*}[tbp]
    \centering
    \resizebox{\textwidth}{!}{
    \input{qsdfull_tree_3qubits}}
    \caption{Decomposition of a 3-qubit unitary, U, into one and two-qubit gates. The decomposition involves alternate layers of CSD (red) and VDW (blue). M-Rz(Ry) are $N$ and $N-1$ qubit multiplexed Rz(Ry) gates which can be further decomposed into a set of CNOT and Rz(Ry) gates. The ultimate layer (gray) involves the decomposition into single qubit gates. 
    }
    \label{3qubits_decompositiontree}
\end{figure*}

\section{Quantum Shannon Decomposition: Reduction of arbitrary unitary operations into quantum circuits}\label{QSD_theory}
The mapping protocol necessitates that the number of  $\left\{ B_{i}^{x}, B_{i}^{y}, B_{i}^{z}\right\}$, and $\left\{ J_{ij}^{x}, J_{ij}^{y}, J_{ij}^{z}\right\}$ parameters in the Ising Hamiltonian match the number of parameters in the molecular Hamiltonian. Since we also consider a fine grid for our potential energy surface with an equivalent increase in the number of qubits, it is critical to note that the molecular Hamiltonian matrix size grows exponentially while the number of Ising Hamiltonian parameters only grows quadratically (See Ref. \onlinecite{Debadrita-Mapping-1D-3Qubits}). Consequently, the map between the molecular Born-Oppenheimer Hamiltonian and the Ising Hamiltonian becomes more and more approximate as the number of qubits grows. To address this and achieve accurate treatment of the chemical dynamics process for a larger number of qubits, we introduce a quantum circuit decomposition method here. The unitary propagator corresponding to the molecular Hamiltonian is written as a quantum circuit which is then used to simulate the temporal evolution of the molecular system. It has been shown that an arbitrary unitary matrix can be decomposed into a universal quantum gate set consisting of a few single qubit gates along with a two-qubit entangling gate \cite{barenco1995elementary}. Several matrix decomposition techniques such as the QR, Givens, Householder, and cosine-sine decomposition have been used to obtain quantum circuits for arbitrary unitary operators\cite{barenco1995elementary,mottonen12006decompositions} resulting in universal quantum gate sets. In this section, we adapt the Quantum-Shannon decomposition \cite{QSD-Shende}(QSD) for the quantum circuit decomposition of unitary matrices obtained in quantum chemical dynamics processes. 
The decomposed unitaries are the equivalent to unitary gate operations that can be implemented using standard quantum gates from a universal gate set on quantum hardware. This allows executing an arbitrary unitary operation as a concatenated sequence of universal gates on a given quantum hardware. 
The decomposition scheme involves two well-known matrix decomposition schemes - the cosine-sine decomposition (CSD)\cite{golub,sutton2009csd,paige1994historyCSD}and the joint eigenvalue-based decomposition of block diagonal unitaries referred to here as VDW \cite{mottonen12006decompositions} scheme. A brief description of the algorithm is provided below. One key outcome from the algorithm below is that the number of entangling gates in this algorithm can be estimated at $\frac{3}{4}4^{n}-\frac{3}{2}2^{n}$ which is close to the theoretical lower bounds to the number of CNOT gates, $\frac{1}{4}(4^{n}-3n-1)$, as discussed in Ref. \onlinecite{QSD-Shende}. 

\begin{figure*}[tbp]
    \centering
    \input{3qubits_circuit_blackbox}
    \caption{Schematic of the circuit resulting from the Quantum Shannon decomposition for an arbitrary three-qubit unitary matrix. All single qubit operations (on q[2]) are decomposed using the ZYZ decomposition, while all multi-qubit operations are either multi-controlled $R_{y}$ or $R_{z}$ operations as discussed in Section \ref{QSD_theory} and Figure \ref{3qubits_decompositiontree}.}
    \label{3qubits_cicuit_blackbox}
\end{figure*}
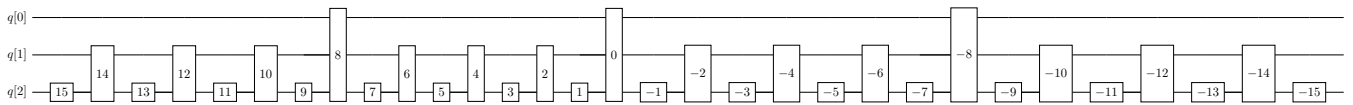
We begin by considering a $2^{N}$ by $2^{N}$ unitary matrix that is to be implemented on a quantum computer using universal gates. We consider the universal gate set $\{R_{y}(\theta), R_{z}(\theta), \text{CNOT}\}$, consisting of single qubit rotation operations, $R_{y}(\theta)$ and $R_{z}(\theta)$ that affect rotations about the y and z axis of the Bloch sphere by arbitrary angles $\theta$, and the two-qubit operation, CNOT. A flow of the decomposition is provided as a tree diagram in Figure \ref{3qubits_decompositiontree}. The CSD (shown in red in Figure \ref{3qubits_decompositiontree}) and VDW  (shown in blue in Figure \ref{3qubits_decompositiontree}) steps iteratively break down the $2^{N}$ by $2^{N}$ unitary to a sequence of these single and two-qubit operations. 
The CSD decomposes the unitary matrix, $U$, into a product of three unitaries with distinct structures,
\begin{align}
    U = \begin{pmatrix}
    \mathbf{L_{0}} & \mathbf{0}\\
    \mathbf{0} & \mathbf{L_{1}}
    \end{pmatrix}\begin{pmatrix}
    \mathbf{C} & -\mathbf{S}\\
    \mathbf{S} & \mathbf{C}
    \end{pmatrix}\begin{pmatrix}
    \mathbf{R_{0}} & \mathbf{0}\\
    \mathbf{0} & \mathbf{R_{1}}
    \end{pmatrix}
    \label{CSD}
\end{align}
where the left($L$) and right($R$) matrices are block diagonal unitaries with blocks $\mathbf{L_{0}}$, $\mathbf{L_{1}}$ and $\mathbf{R_{0}}$, $\mathbf{R_{1}}$, respectively. 
$\mathbf{C}$ and $\mathbf{S}$ are diagonal matrices with entries $cos(\alpha_{i})$ and $sin(\alpha_{i})$, respectively for $i = 0 - 2^{N-1}$. This cosine-sine ($CS$) matrix (node $0$ in Figure \ref{3qubits_decompositiontree}) is a multiple-control $R_{y}$ gate. A multi-control \cite{QSD-Shende} gate can be cast as an N-qubit generalization of a conditional gate in a quantum circuit, wherein each of the $2^{N-1}$ conditions implemented using $N-1$ control qubits results in a different unitary operation on the $N^{th}$ target qubit. These may be thought of as $CCC \cdots R_{y}$ operations. A multi-control $R_{y}$ can, therefore, be thought of as a conditional operation of a $R_{y}(\alpha_{i})$ rotation on the target qubit depending on the $i^{th}$ condition enforced by $N-1$ qubits. 
This simultaneous operation of the $2^{N-1}$ $R_{y}(\alpha_{i})$'s for all values in $\left\{ \alpha_{i} \right\}$ can be decomposed further and implemented using $2^{N-1}$ CNOT's and $2^{N-1}$ $R_{y}(\theta_{i})$ rotation gates. The rotation angles $\theta_{i}$ can be obtained from linear combinations of $\alpha_{i})$'s using a Gray code method as outlined in Ref. \onlinecite{mottonen2004quantum}.
Classically, one can think about this as $2^{N-1}$ conditionality statements, one condition representing a specific $R_{y}$ rotation, only, on a quantum computer these are meant to be executed in parallel. While such operations may in principle represent the true power of quantum devices in the future, in the NISQ era\cite{Nai-Hui-Hybrid-QC,Preskill2018-NISQ}, these operations are deeply limited by the number of CNOT gates required to implement them.
 
The block diagonal unitaries on the left and right in Eq. (\ref{CSD}) (red empty nodes in Figure \ref{3qubits_decompositiontree}), are $N$-qubit conditional gates, known as quantum multiplexors, with a single qubit control and corresponding conditional operation on $N-1$ target qubits. At this point, it is important to reiterate and note the difference in the structures of the block diagonal $L$, $R$ matrices and the $CS$ matrix and the resulting multi-qubit gates. While the N-qubit operation corresponding to the $CS$ unitary matrix, as discussed earlier, has multiple controls (precisely $2^{N-1}$ for an $N$ qubit unitary), and a single target qubit, that for the $L$, $R$ matrices have a single control 
with $N-1$ target qubits. This decomposition technique thus exploits these two generalizations to multi-qubit gates to build the entire quantum circuit corresponding to $U$. The VDW transformation, as given by eq. \ref{VDW}, is a technique for decomposing such quantum multiplexor gates (obtained from the $L$ and $R$ matrices above), with a single control and multiple target qubits, into basic quantum circuit unitaries. The L and R matrices from Eq. (\ref{CSD}) are thus further decomposed into, 
\begin{align}
    L = \begin{pmatrix}
    \mathbf{L_{0}} & \mathbf{0}\\
    \mathbf{0} & \mathbf{L_{1}}
    \end{pmatrix} = \begin{pmatrix}
    \mathbf{V} & \mathbf{0}\\
    \mathbf{0} & \mathbf{V}
    \end{pmatrix}\begin{pmatrix}
    \mathbf{D} & \mathbf{0}\\
    \mathbf{0} & \mathbf{D^{\dagger}}
    \end{pmatrix}\begin{pmatrix}
    \mathbf{W} & \mathbf{0}\\
    \mathbf{0} & \mathbf{W}
    \end{pmatrix}
    \label{VDW}
\end{align}
The VDW is critical in that left and right block diagonal matrices containing $\mathbf{V}$ and $\mathbf{W}$, respectively correspond to $N-1$ qubit operations in a quantum circuit. Therefore, each VDW step in the decomposition halves the dimension of the unitary matrix. This step is shown in blue in Figure \ref{3qubits_decompositiontree} and the $\mathbf{V}$ and $\mathbf{W}$ matrices are blue nodes in Figure \ref{3qubits_decompositiontree}). The $\mathbf{V}$ and $\mathbf{W}$ matrices are further decomposed using CSD in the next step. The diagonal matrix, $\mathbf{D}$, (node -8 for $L$ and 8 for $R$ in Figure \ref{3qubits_decompositiontree}) is a multi-control $R_{z}$ gate (similar to the multi-control $R_{y}$ operation for the $CS$ matrix) and can be decomposed further into a sequence of $2^{N-1}$ CNOT's and $2^{N-1}$ $R_{z}(\theta)$ gates, similar to the decomposition of the cosine-sine matrix as a multiplexed $R_{y}(\theta)$. As noted above, these multiplexed $R_{z}(\theta)$ operations are also akin, now, to $2^{N-2}$ conditionality statements. 
Following $N-1$ steps of alternating CSD and VDW steps, the resultant matrices are single-qubit unitaries, which are further decomposed using the ZYZ scheme for arbitrary single-qubit unitaries. The multiplexors are decomposed into single qubit rotation gates and CNOT gates using a gray code implementation as outlined in Ref. \onlinecite{mottonen2004quantum}. Thus, the QSD yields a quantum circuit of multi-controlled $R_{y}(\theta)$ and $R_{z}(\theta)$ gates, and single qubit unitaries, which have standard procedures to be further decomposed into the gates in the universal gate set $\{R_{y}, R_{z}, CNOT\}$. The numbered nodes of Figure \ref{3qubits_decompositiontree} correspond to these multi-controlled and single qubit unitaries and for clarity, the circuit schematic corresponding to the decomposition is provided in Figure \ref{3qubits_cicuit_blackbox}. Note the order is reversed in the quantum circuit for showing the action of $U$ on an initial qubit state.

The number of gates in any standard implementation of this decomposition is estimated from the implementation of the multi-controlled $R_{y}(\theta)$ and $R_{z}(\theta)$ gates. This scheme has the advantage that when a fixed gate set is used to construct the circuit, the resulting circuits have a constant circuit depth 
for an arbitrary $2^N$ by $2^N$ dimensional unitary. 
This implementation assumes the use of the universal gate set $\{R_{y}, R_{z}, CNOT\}$ as is apparent from the summary above, but is not limited by this gate set alone. 
Each $2^N$ by $2^N$ multiplexed-Ry(Rz) requires $2^{N-1}$ CNOTs and $2^{N-1}$ Ry(Rz) gates using the gray code implementation as discussed in \onlinecite{QSD-Shende}. There is an inherent structure in the decomposition scheme that is evident from the decomposition of general multiplexed $R_{y}$ and $R_{z}$ gates as shown in reference \onlinecite{QSD-Shende}. This structure of the decomposition is irrespective of the inherent symmetries present in the unitary itself. 

\section{Proton stretch dynamics in H$_5$O$_2^+$ and H$_3$O$_2^-$ using the mapping protocol and using Quantum Shannon Decomposition}
\label{results}
We examine the map in Section \ref{IsingH-structure} and the quantum circuit decomposition method in Section \ref{QSD_theory} by simulating and comparing the quantum dynamics of the shared proton in the protonated 
 Zundel (H$_5$O$_2^+$) and the corresponding hydroxide water clusters (H$_3$O$_2^-$). In examining the mapping protocol, we simulate and compare the dynamics of both the molecular systems and their corresponding ion-trap dynamics, on classical hardware, independently. The Ising model Hamiltonian that we consider for validating the mapping protocol is for a trapped ion system with three qubits. In simulating the proton-transfer dynamics for each of the systems, we study the time evolution of initial nuclear wavepacket states prepared in the respective permuted basis representations for the molecular and Ising model Hamiltonians. As stated, the parameters in the Ising Hamiltonian are determined, and thus controlled, by the pre-computed matrix elements of the molecular Hamiltonian. In implementing the circuit decomposition technique, we simulate the unitary time evolution of the transferring proton in the water clusters by decomposing the unitary propagator, $e^{-i \hat{H} t/\hbar}$, for each value of $t$ into a sequence of quantum gates using the Quantum Shannon decomposition method detailed in Section \ref{QSD_theory}. We implement the resulting quantum circuits on IBM's QASM simulator using their software development kit, Qiskit. In the following sections, we first introduce the molecular systems we consider to validate our mapping protocol and circuit decomposition scheme. In Section \ref{compdetails_pes}, we outline the computational details of the potential energy surfaces and the quantum nuclear Hamiltonian, followed by the time evolution of the quantum nuclear wavepacket, for both the systems under consideration in Sections \ref{results_qsim} and \ref{results_qsd}. 

Water clusters are an important class of molecular systems found in many constrained environments such as biological membranes, enzyme active sites,\cite{baciou1995interruption,Guo} and ion channels\cite{domene2003potassium}. Water-mediated “proton wires,” for example, are routinely invoked to explain charge transport across cell membranes and the primary charge-separation step in photosynthesis. They may also be confined within carbon nanotubes leading close to ballistic transport,\cite{WW3,ww11,hummer_rasaiah_noworyta_2001} and are a critical aspect of polymer electrolyte fuel cells.\cite{ye2012water} The lighter mass of the hydrogen nucleus makes quantum nuclear effects critical in all such cases;\cite{Cptuckerman3,H+OH-solv, schmittVothProtonTransport1999, schmitt1998multistate} additionally the multidimensional quantum nuclear effects in such systems arising from the vibrational coupling between the proton-transfer dimensions and other orthogonal modes is also known to be critical in such problems.\cite{johnson-jordan-21mer,Johnson-Jordan-Zundel-Science,Johnson-Jordan-Zundel-JCP,HDMeyer-Zundel-1} 

The chemical systems considered in this publication are specific small protonated and hydroxide-rich water clusters. The isolated H$_5$O$_2^+$ and H$_3$O$_2^-$ ions are two of the most fundamental structures involved in the proton-transfer process. The anionic H$_3$O$_2^-$ complex is especially interesting because it involves a strong low-barrier hydrogen bond (LBHB),\cite{JohnsonOH-Science,AgmonOH,Johnson-Zundel-OH-H2O-quantum,H2O6OH-Xiaohu,AQC} a phenomenon often introduced to explain the surprisingly high rates of some enzyme-catalyzed reactions \cite{kemp2021lbhbprotein,Sshb-Serine-Protease,SSHB-2,SSHB-Warshel-1}. Based on the studies of small molecules, such short-strong HBs are often formed between functional groups with comparable pKa's, and often result in a zero point energy for the shared hydrogen higher than the barrier height energy for proton transfer. The Zundel cation is a small prototypical system with a proton shared between two water molecules forming a short, strong hydrogen bond\cite{Johnson-Jordan-Zundel-JCP,Johnson-Jordan-Zundel-Science,HDMeyer-Zundel-1,Scott-proj,Bowman-DMD-Zundel,HDMeyer-Zundel-2,HDMeyer-Zundel-3}. This system plays a fundamental role in the understanding of processes such as the enhanced mobility of protons and deuterons in condensed phase aqueous environments, in biological systems, and several problems of interest in materials chemistry, such as protonic conductorsand fuel cells\cite{Bruce-Hudson-perchlorate,Haile-CsHSO4-Entropy-2007,Haile-2007-Faraday,Suzuki-H1-Cs-PRB-2006} .
Due to their central role in aqueous charge transport, the H$_5$O$_2^+$ and H$_3$O$_2^-$ ions have been extensively investigated with electronic structure theory and quantum nuclear dynamics\cite{HDMeyer-Zundel-1,HDMeyer-Zundel-2,HDMeyer-Zundel-3}, and both display stable configurations where one hydrogen atom resides between the two oxygen atoms (e.g., [H$_2$O···H···OH$_2$]$^+$ and [HO···H···OH]$^-$). 


\subsection{Computation of the nuclear Hamiltonians describing proton transfers in H$_3$O$_{2}^{-}$ and Zundel clusters}\label{compdetails_pes}
We compute one-dimensional potential energy surfaces for the intra-molecular proton transfer mode in both H$_3$O$_2^-$ and H$_5$O$_2^+$,  by first locating a stationary point for both systems where the proton is symmetrically shared between the donor and acceptor groups. For the case of H$_3$O$_2^-$, this corresponds to the transition state, with one imaginary frequency of the Hessian matrix corresponding to the vibrational mode along the intra-molecular proton transfer direction. For the case of the protonated Zundel, however, the geometry where the donor and acceptor atoms symmetrically share the proton corresponds to a minimum.
Standard electronic structure methods are employed to perform these computations. Born-Oppenheimer potential energy surfaces for one-dimensional proton motion along the donor-acceptor axis are computed at these stationary point geometries. For that purpose, we choose a one-dimensional grid along the donor-acceptor axis with $2^{N}$ number of equally spaced grid points, symmetrically located about the stationary point (grid center). We perform electronic structure calculations at these points, on a classical computing platform, at the level of theory mentioned in Table \ref{PES_details} for the range of $N = 3$ to $N = 7$. 
Details of the electronic structure methods and total grid lengths for these potential energy surface calculations are provided in Table \ref{PES_details}. 
\begin{table*}[]
    \centering
    \begin{tabular*}{\textwidth}{@{\extracolsep{\fill}}lllll}
    \hline \hline
    System & DA-distance & Level of theory & Grid Spread & No. of gridpoints\\
    \hline
    H$_3$O$_{2}^{-}$& 2.42\AA & CCSD/6-311++G(d,p) & 0.66\AA & 8,16,32,64,128\\
    H$_5$O$_{2}^{+}$& 2.39\AA & B3LYP/6-311++G(d,p) & 0.66\AA & 8,16,32,64,128\\
    \hline \hline
    \end{tabular*}
    \caption{Computational details for computing the potential energy surfaces for the transferring proton in both water clusters. In all cases the grid spacing ranges from 0.083\AA\; (for 8 grid points, that is three qubits) to 0.005\AA\; (for 128 grid points, that is 7 qubits). The finer grid spacings essentially approach the continuous limit given the mass of the proton and associated de Broglie wavelength. Also see Figure \ref{PES_h3o2_ZPE} for associated smoothness of potential.}
    \label{PES_details}
\end{table*}

\subsection{Quantum simulation of proton-transfer dynamics in H$_3$O$_{2}^{-}$ and Zundel using the mapping protocol}
\label{results_qsim}
\begin{table*}[]
    \centering
    \begin{tabular*}{\textwidth}{@{\extracolsep{\fill}}ll|lll|lll}
    \hline \hline
    \multirow{2}{*}{Initial Wavepacket} & \multirow{2}{*}{Parameters} & \multicolumn{3}{c|}{H$_3$O$_{2}^{-}$}& \multicolumn{3}{c}{H$_5$O$_{2}^{+}$} \\
    & & $\left|\braket{\chi_0}{\psi}\right|^2$ & E\footnote[1]{in units of kcal/mol} & $\epsilon$\footnote[2]{as in Eq.(\ref{proberror_txavg})}  & $\left|\braket{\chi_0}{\psi}\right|^2$ & E \footnotemark[1]& $\epsilon$\footnotemark[2] \\
    \hline
     & & & & & & & \\
    $\psi_L(x;0) = \delta (x-x_0)$ & $x_0 = $ donor site & $0.27\%$ & 28.1 & $10^{-15}$ & $0.08\%$ & 35.1& $10^{-7}$\\
    $\psi_G(x;0) = \exp\qty[-\frac{(x-\mu)^2}{2\sigma^2}]$ & $\sigma = 0.1$\AA, $\mu = 0.0$\AA & $87.7\%$& 2.41 & $10^{-15}$ & $93.71\%$& 2.6 & $10^{-7}$\\
    $\psi_T(x;0) = \sum\limits_{j} \exp\qty[-E_j/kT]\chi_j(x)$ & $T=300K$ & $99\%$ & 1.34 & $10^{-15}$ &  $99.92\%$ & 1.92 & $10^{-7}$\\
    \hline \hline
    \end{tabular*}
    \caption{Parameters and characteristics of the initial wavepacket states considered for the transferring proton in the water cluster systems. The ground state, $\ket{\chi_{0}}$, overlap of for each $\ket{\psi(x;0)}$ and corresponding energy are reported for the case of $N=3$. The mean absolute errors in probability, $\epsilon$, computed between the classical propagated probability density, $\rho_{C}(x)$, and the probability density, $\rho_{Q}(x)$, computed using the mapping protocol in Section \ref{IsingH-structure}, are provided.}
    \label{InitWPs}
\end{table*}
\begin{figure}
    \centering
    \includegraphics[width=\columnwidth]{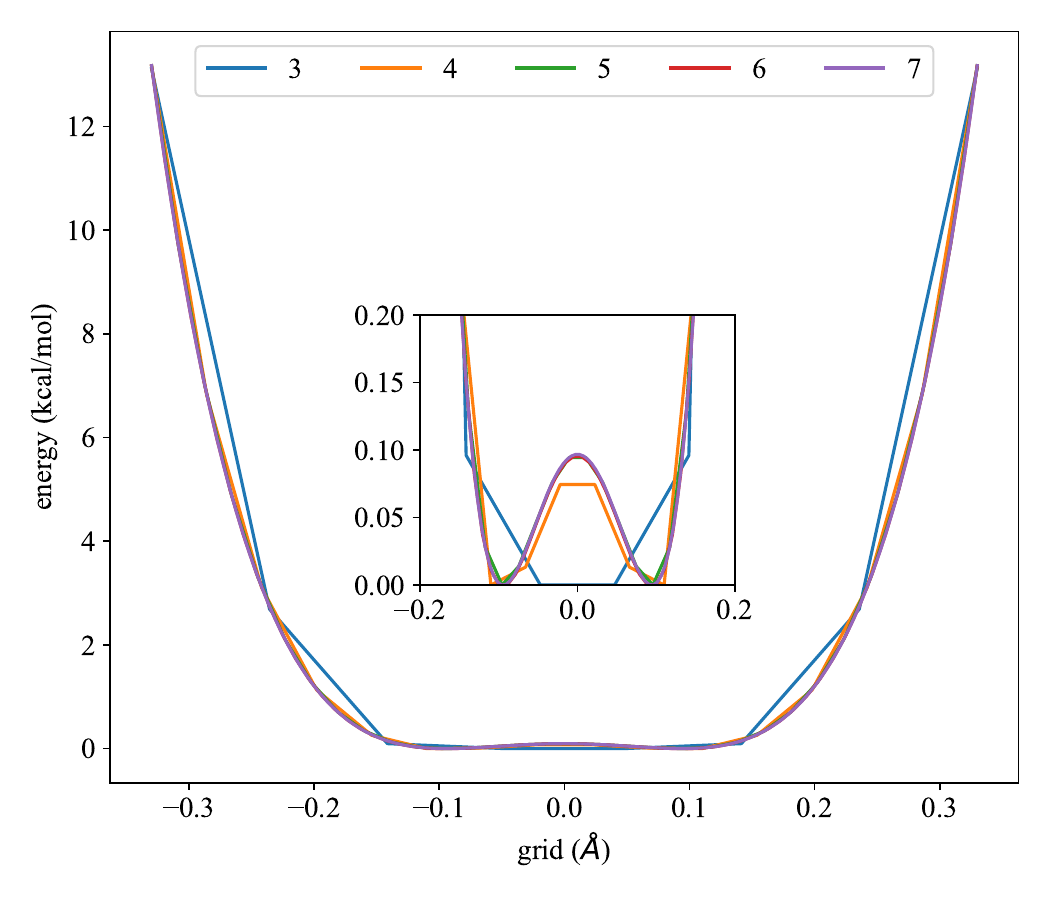}
    \caption{The one-dimensional potential energy surfaces for the proton along the donor-acceptor axis in H$_3$O$_{2}^{-}$ computed using a range of grid separations (refer Table \ref{PES_details}). The surfaces converge with increasing gridpoint densities (64 and 128 for $N=6$ and $N=7$ respectively). The low barrier height is shown in the inset for all grid point densities.} 
    \label{PES_h3o2_ZPE}
\end{figure}

We examine the map by simulating the quantum dynamics of the water clusters and the ion-trap dynamics, on classical hardware, independently. For that, we choose the initial wavepacket state for the transferring proton in both systems in three different ways in the grid representation ($x$), the details of which are provided in Table \ref{InitWPs}. The three initial wavepackets considered here are designed to probe a broad range of energy. The wavepacket, $\psi_L(x;0)$ is particularly harsh given that it samples almost the full eigenspectrum, whereas the other two choices populate the lower regions of the energy spectrum. 
The corresponding Givens transformed wavepacket initial state is considered for time evolution using the block diagonal molecular Hamiltonian in the mapping protocol. Given the direct map described in Section \ref{hamiltonian-map-summary} between the permuted computational basis and the Givens transformed molecular grid basis, the initial wavepackets for the Ising Hamiltonian are chosen analogously to the initial wavepacket of the molecular system. 
The wavepacket state, $\psi_L(x;0)$, initialized on the end of the grid close to the donor site (as in Table \ref{InitWPs}), corresponds to $\left\{\frac{\ket{\tilde{x}_{0}} - \ket{\tilde{x}_{7}}}{\sqrt{2}} \right\}$ in the Givens transformed basis. The corresponding initial state for the Ising Hamiltonian is chosen analogously to be $\left\{\frac{\ket{\tilde{\lambda}_{0}} - \ket{\tilde{\lambda}_{7}}}{\sqrt{2}} \right\}$ in the permuted computational basis, $\{\ket{\tilde{\lambda}}\}$. $\psi_G(x;0)$ and $\psi_T(x;0)$ are symmetric about the center of the grid and are mostly concentrated on the first block of the Givens transformed basis. The spin-lattice and molecular wavepackets are then independently propagated according to the transformed Ising and molecular Hamiltonians, and compared to gauge the accuracy of the quantum simulation. 

Given the map between the matrix representation of the Ising Hamiltonian and the Givens transformed molecular Hamiltonian (in Eq. \ref{Diag-B}), as discussed in Ref. \onlinecite{Debadrita-Mapping-1D-3Qubits}, the ion-trap hardware initial wavepacket state is directly propagated by the choice of $\left\{ B_{i}^\gamma; J_{ij}^\gamma \right\}$ for arbitrary time-segments. 
The time-dependent probabilities resulting from the projection of the resultant time-dependent wavepacket on the computational basis, at each interval of time, are used to compute the difference between the classical and quantum algorithms:
\begin{align}
    \epsilon = \frac{1}{T}\int dt \frac{1}{2^N}\int dx \left\vert \rho_Q (x) - \rho_C (x) \right\vert
    \label{proberror_txavg}
\end{align}
where $\rho_Q (x)$ and $\rho_C (x)$ are the quantum and classical values of the wavepacket density, $N$ is the number of qubits (ions), and $T$ is the total simulation time. The resultant errors are provided in Table \ref{InitWPs}.

Given the exact match between the spin-lattice dynamics and the quantum chemical dynamics, the features present in ion-trap dynamics must also exist in the chemical dynamics problem. Thus through the isomorphism constructed above, our algorithm allows the ability to extract properties of the chemical systems from the corresponding Ising Hamiltonian dynamics on the ion trap. 

\subsection{Quantum simulation of proton-transfer dynamics in H$_3$O$_{2}^{-}$ and Zundel using Quantum Shannon Decomposition}\label{results_qsd}
\begin{table*}[]
    \begin{tabular*}{\textwidth}{@{\extracolsep{\fill}}c|ccccc|ccccc}
    \hline \hline
     & \multicolumn{5}{c|} {$\epsilon$ (Eq. (\ref{proberror_txavg}))  for H$_5$O$_{2}^{+}$} & \multicolumn{5}{c}{$\epsilon$ (Eq. (\ref{proberror_txavg})) for H$_3$O$_{2}^{-}$ }  \\
    Initial Wavepacket & $N=3$ & $N=4$ & $N=5$ & $N=6$ & $N=7$& $N=3$ & $N=4$ & $N=5$ & $N=6$ & $N=7$ \\
    \hline
    $\psi_L(x;0)$ & 6$\times 10^{-4}$ & 5$\times 10^{-4}$ & 4$\times 10^{-4}$ & 3$\times 10^{-4}$ & 2$\times 10^{-4}$& 7$\times 10^{-4}$ & 5$\times 10^{-4}$ & 4$\times 10^{-4}$ & 3$\times 10^{-4}$ & 2$\times 10^{-4}$ \\
    $\psi_{G}(x;0)$ & 6$\times 10^{-4}$ & 5$\times 10^{-4}$ & 4$\times 10^{-4}$ & 3$\times 10^{-4}$ & 2$\times 10^{-4}$ & 7$\times 10^{-4}$ & 5$\times 10^{-4}$ & 4$\times 10^{-4}$ & 3$\times 10^{-4}$ & 2$\times 10^{-4}$\\
    $\psi_T(x;0)$& 6$\times 10^{-4}$ & 5$\times 10^{-4}$ & 4$\times 10^{-4}$ & 3$\times 10^{-4}$ & 2$\times 10^{-4}$ & 7$\times 10^{-4}$ & 5$\times 10^{-4}$ & 4$\times 10^{-4}$ & 3$\times 10^{-4}$ & 2$\times 10^{-4}$\\
    \hline \hline
    \end{tabular*}
    \caption{The mean absolute errors in probability (Eq. (\ref{proberror_txavg})) computed between the classical propagated probability density $\rho_{C}(x)$ and the QASM simulated probability densities, $\rho_{Q}(x)$, using the circuit decomposition method summarized in Section \ref{QSD_theory}. The number of shots is $1000$. Errors are reported for different initial nuclear wavepacket states of the transferring proton in both water clusters for all cases of $N = 3-7$. As the number of shots is increased, the error decreases as noted in Figure \ref{shots-errors}.}
    \label{QSD_errors_H3O2_H5O2}
\end{table*}

\begin{figure*}[htbp]
    \subfigure[Initial state $\psi_{T}(x,0)$ for H$_5$O$_{2}^{+}$]{\includegraphics[width=0.32\linewidth]{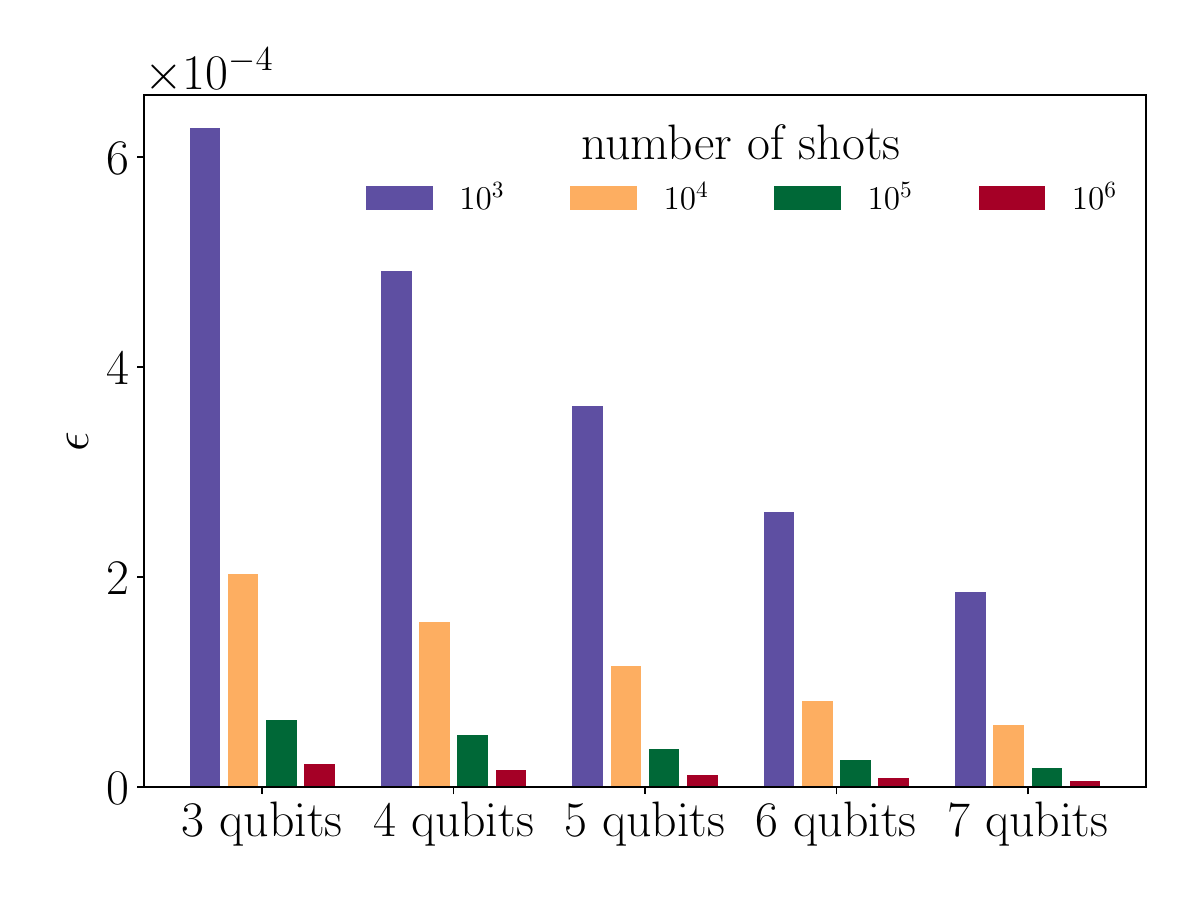}}
    \subfigure[Initial state $\psi_{G}(x,0)$ for H$_5$O$_{2}^{+}$]{\includegraphics[width=0.32\linewidth]{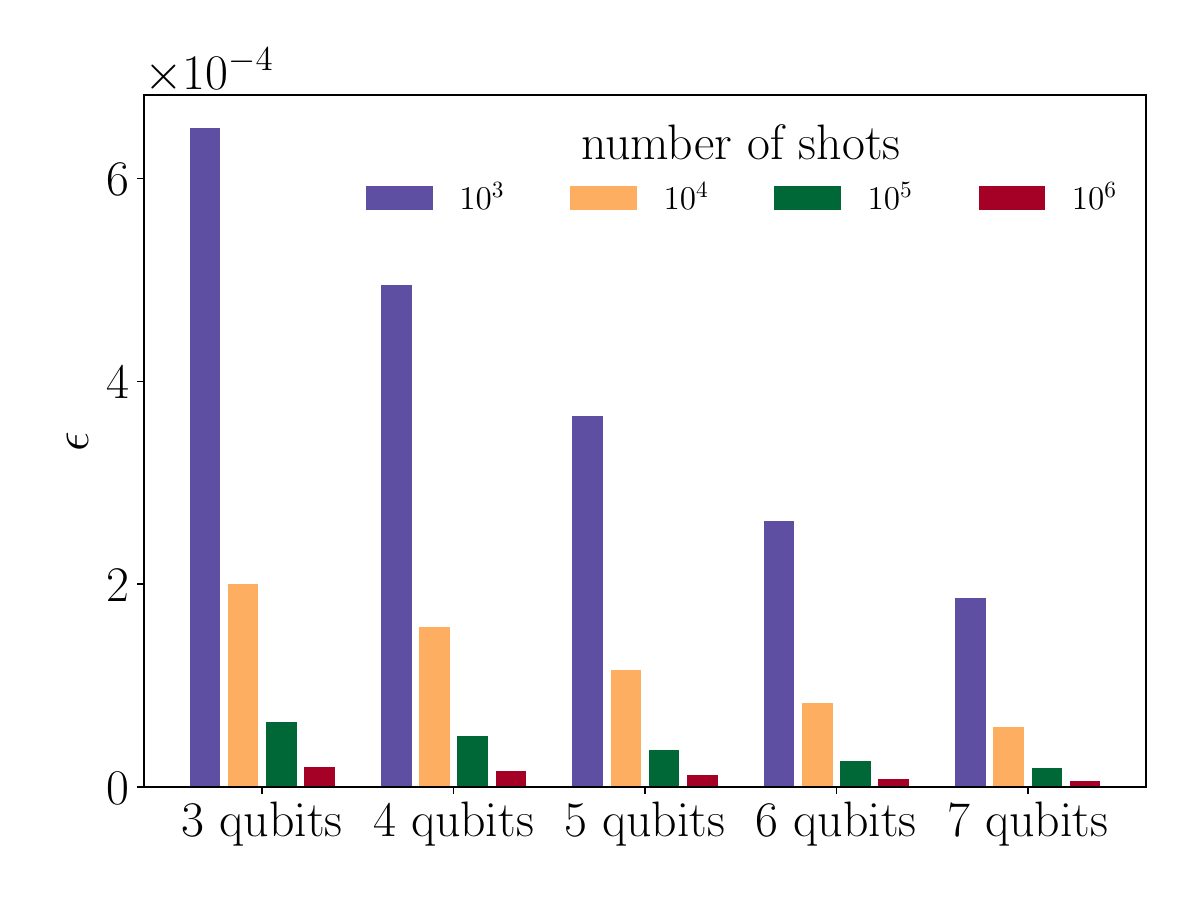}}
    \subfigure[Initial state $\psi_{L}(x,0)$ for H$_5$O$_{2}^{+}$]{\includegraphics[width=0.32\linewidth]{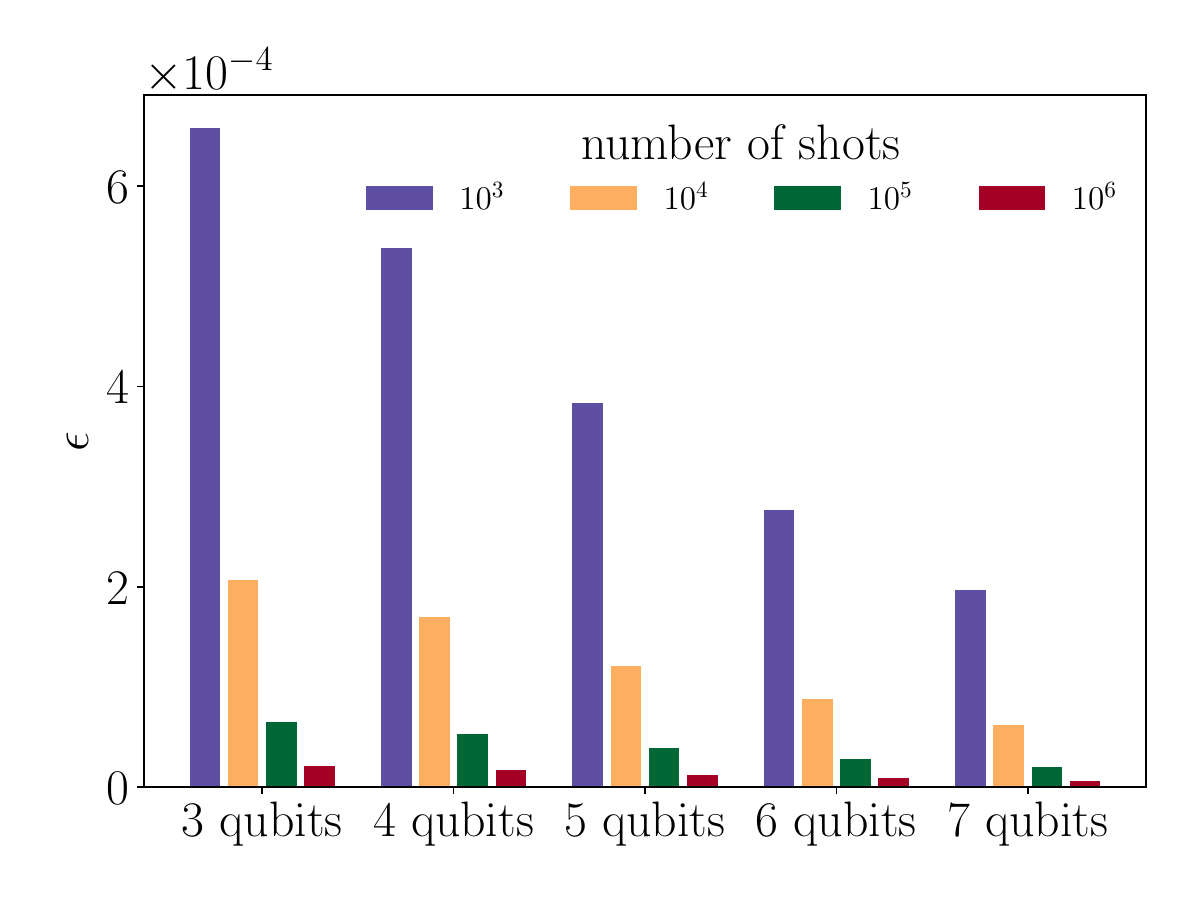}}
    \subfigure[Initial state $\psi_{T}(x,0)$ for H$_3$O$_{2}^{-}$]{\includegraphics[width=0.32\linewidth]{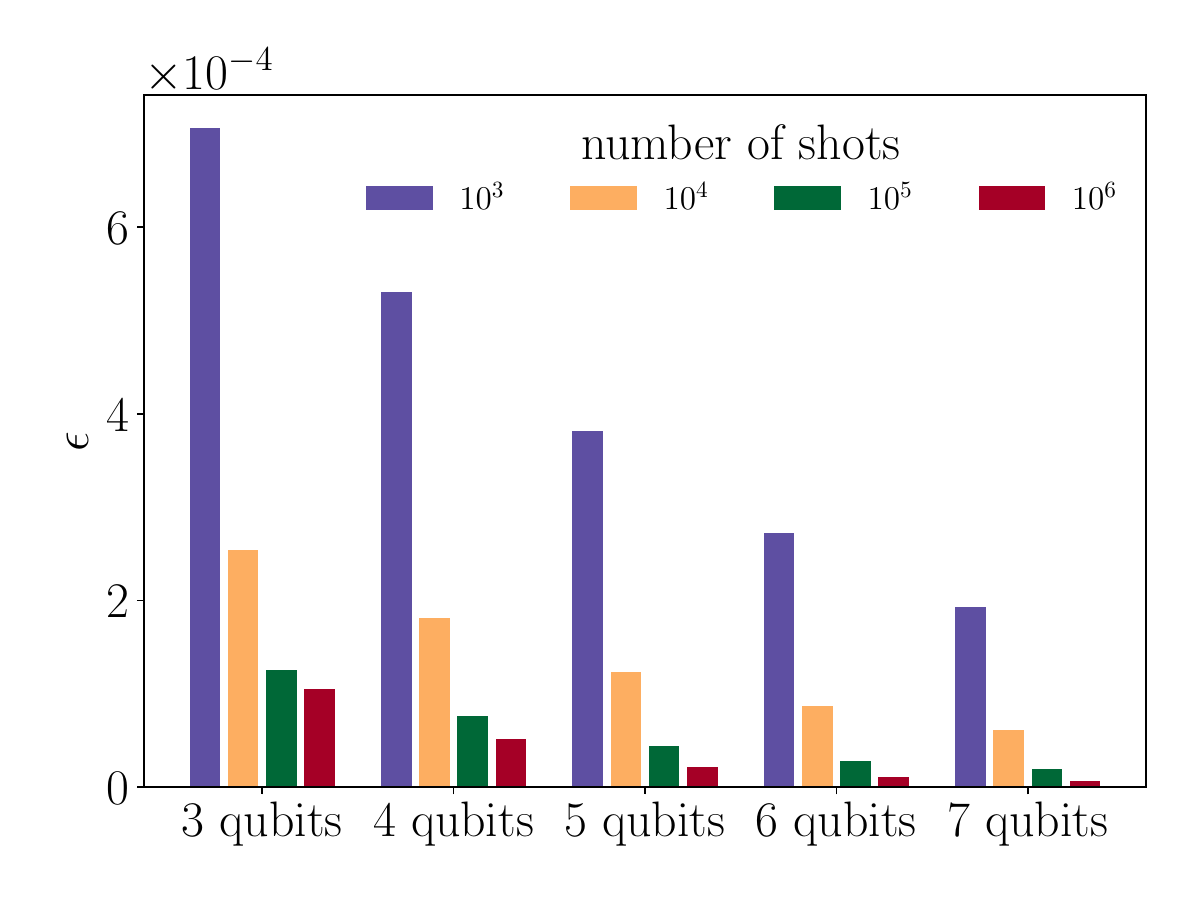}}
    \subfigure[Initial state $\psi_{G}(x,0)$ for H$_3$O$_{2}^{-}$]{\includegraphics[width=0.32\linewidth]{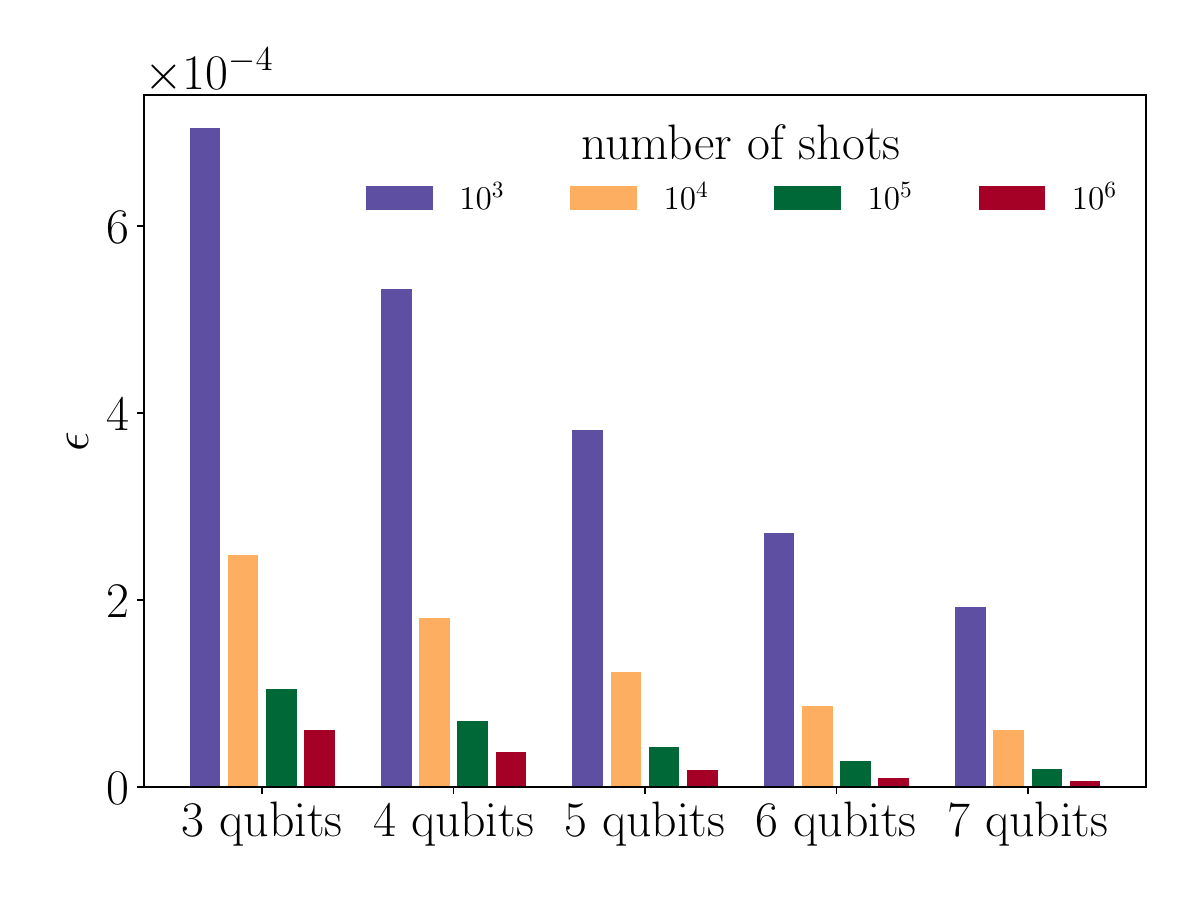}}
    \subfigure[Initial state $\psi_{L}(x,0)$ for H$_3$O$_{2}^{-}$]{\includegraphics[width=0.32\linewidth]{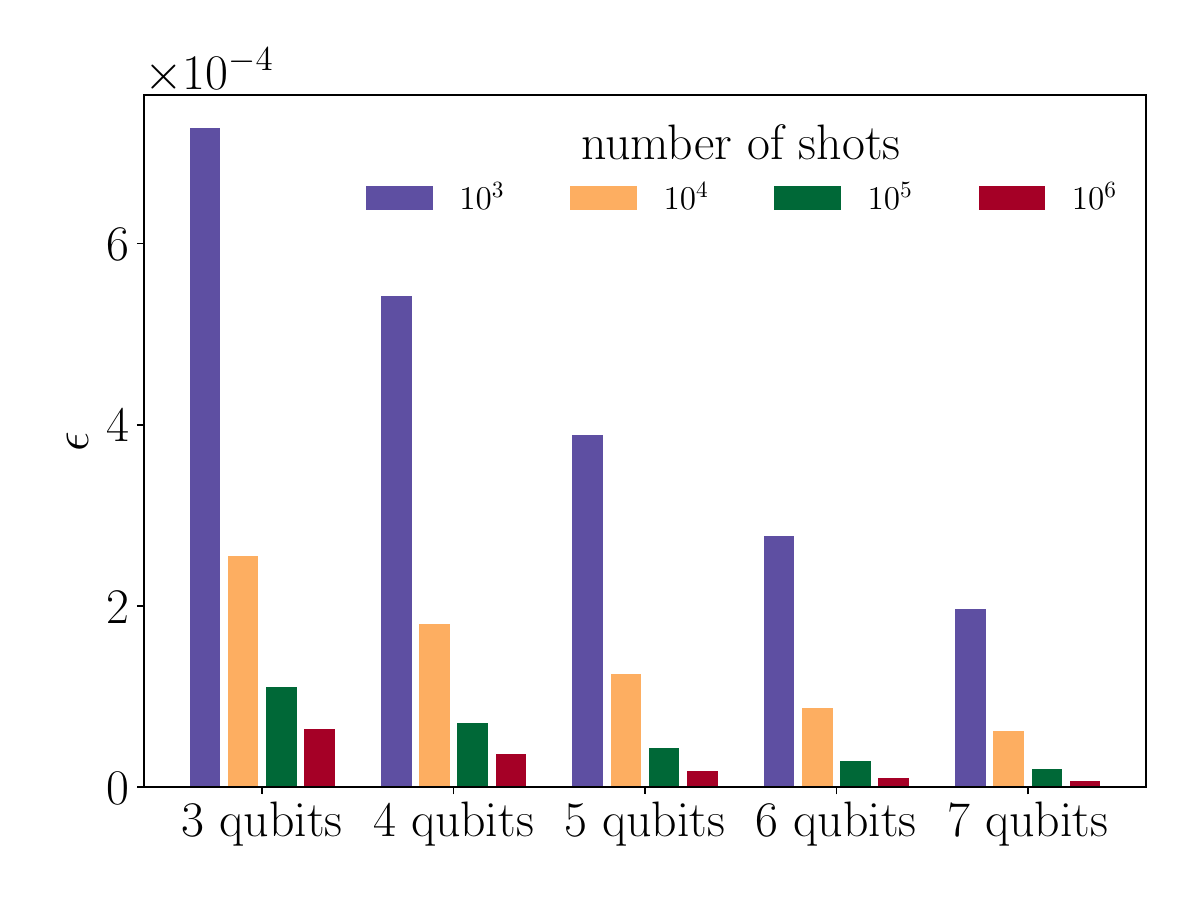}}
    \caption{Vertical axes show probability errors, ($\epsilon$ in Eq. (\ref{proberror_txavg})) for $N=3-7$ qubits and for the different initial nuclear wavepacket states as a function of the number of measurement shots on IBM's QASM simulator for both chemical systems. Clearly the error reduces drastically with increasing shots.}
    \label{shots-errors}
\end{figure*}
\begin{figure}[htbp]
    \centering
    \includegraphics[width=0.48\columnwidth]{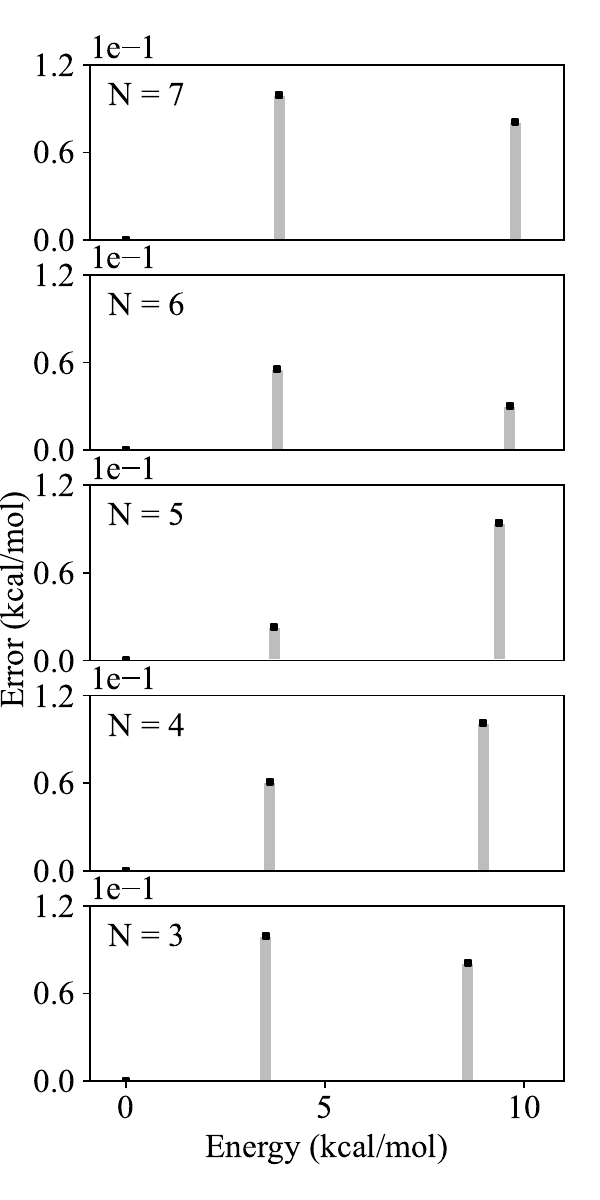}
    \includegraphics[width=0.48\columnwidth]{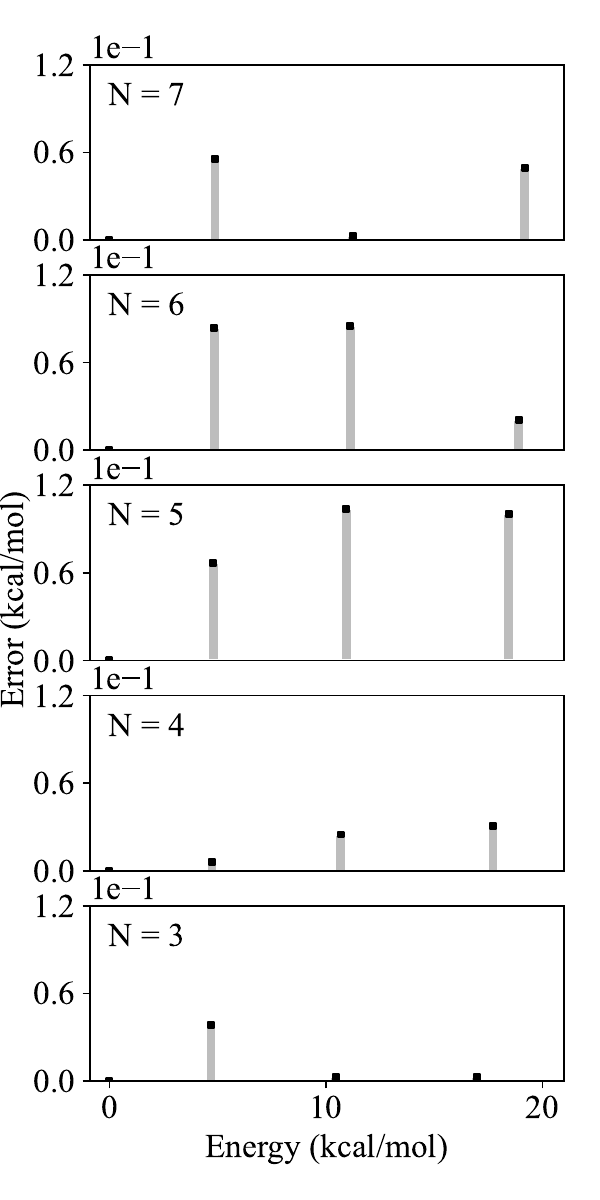}
    \caption{\label{QSD_freqdiff_errors} Eigenenergy differences between the classically computed (exact diagonalization) and QASM computed eigenenergies for the first three and four eigenvalues for H$_5$O$_{2}^{+}$(left) and  H$_3$O$_{2}^{-}$(right).}
\end{figure}

The proton dynamics in the water clusters as simulated using the mapping protocol is exact for the case of $N=3$. However, the number of gridpoints in that case does not capture the important characteristics of the potential energy surface as is clear from Figure \ref{PES_h3o2_ZPE}. 
We therefore simulate the circuit decomposed unitary evolution of the proton wavepacket for more accurate representations of the chemical problem for $N = 3$ to $7$ qubits. 
The unitary propagator, $e^{-i \hat{H} t/\hbar}$, for each value of $t$ are decomposed into a sequence of quantum gates using the Quantum Shannon decomposition method detailed in Section \ref{QSD_theory}. 
This decomposition will result in a significantly lower number of gates since the number of gates remains constant than in a Trotter-based decomposition where the circuit depth and gate counts double with subsequent Trotter steps. 
We implement the resulting quantum circuits on IBM's QASM simulator using their software development kit, Qiskit. 
Probability densities are measured after unitary evolution at each time step. 
The errors in probability densities for the circuit implementation and classical propagation are computed using Eq. (\ref{proberror_txavg}) and reported in Table \ref{QSD_errors_H3O2_H5O2} for the proton transfer dynamics in both the water clusters. Since the QASM simulator emulates the behavior of an actual quantum device, the precision of the estimated probabilities depends on the number of measurement shots used for each time step. For about 1000 shots per time step, the error in the probabilities is of the order of $10^{-4}$, across all qubit cases, as reported in Table \ref{QSD_errors_H3O2_H5O2}. We also report probability errors for an increasing number of measurement shots and see agreement up to $10^{-6}$. Figure \ref{shots-errors} shows how the error improves as we increase the number of shots from $10^{3}$ to $10^{6}$ for all qubit cases. 
Furthermore, we compute the Fourier transform of the measured time-evolved probability densities, $\rho_{Q}(x,t)$, which results in a spectrum with Fourier peaks corresponding to the frequency differences between the energy eigenvalues. The relevant details are provided in SI Section D. We can extract the oscillation frequencies of the shared proton in both chemical systems from these Fourier spectra of the frequency differences. 
In Figure \ref{QSD_freqdiff_errors}, we compare the lower level eigenenergies obtained from the Fourier transform of the time-evolved probabilities on QASM to those obtained from the corresponding exact diagonalization for all bound eigenstates for both the chemical systems for qubit cases $N=3-7$. The absolute errors between the QASM simulated eigen energies and the exact diagonalization results are well below $1$ kcal/mol 
for the first four and three energy levels for H$_5$O$_{2}^{+}$ and  H$_3$O$_{2}^{-}$, respectively.
The errors in the frequency spectrum are not affected by the increase in the number of shots since the frequency resolution depends inversely on the total time of evolution, which is consistent for all cases considered here. An increase to 6–7 qubits, with a total time interval of several hundred femtoseconds and approximately $10^{3}$ measurement shots, provides sufficient contrast in the measured probabilities to capture the relevant dynamics with accuracy beyond typical chemical or spectroscopic precision for hydrogen-bonded systems like those considered here.

\section{Conclusions}
\label{conclusion}
The promise of solving exponentially complex problems efficiently using quantum computing hardware and associated quantum computing algorithms software is a rapidly evolving research frontier \cite{Preskill2018-NISQ}.   
While we are in the  early stages of this  quantum revolution, there are a wide set of scientific and technological areas that can benefit from such developments. However, true progress in such areas can only be achieved by a rigorous study and understanding of the electronic structure and dynamics of complex materials, thus requiring accurate treatment of electron correlation effects in conjunction with a rigorous treatment of quantum nuclear effects\cite{Klinman-Chemrevs-2006,qwaimd-SLO-1,SLO-measurement,raghavachari1989fifth,SHS-PCET-ACR-2009,Hoffman-N2-H2-redelim,MHG-DFT-Rungs-2017}. 

In this paper, we discuss (a) the Hamiltonian mapping protocol detailed in 
Ref. \onlinecite{Debadrita-Mapping-1D-3Qubits} and (b) a quantum circuit method based on the Quantum Shannon decomposition for simulating quantum nuclear dynamics. Using the two methods discussed here we simulate the time evolution of the quantum nuclear wavepacket corresponding to the shared-proton degree of freedom in a short-strong hydrogen bond in small water clusters. The Hamiltonian mapping is a general but approximate mapping procedure between a
quantum chemical dynamics problem, constructed on a single Born-Oppenheimer surface, and an ion-trap quantum simulator where the dynamics are dictated by a generalized form of a spin-lattice or Ising model Hamiltonian. This is exact for a small number of qubits while it becomes approximate for a larger number of qubits with quantitative error measures. The quantum circuit decomposition technique, on the other hand, is in principle exact for a higher number of qubits but practical implementation of the circuits remains a challenge on near-term quantum architectures due to the exponential increase in the number of entangling gates in the circuit decomposition.

The key step involved in facilitating our Hamiltonian map is the partitioning of the coupled qubit space into two zones that we illustrate using the qsphere representation of the computational basis. Once the coupled qubit computational basis set is partitioned in such a way, the Ising model Hamiltonian reduces into a block form thus allowing the possibility to map all problems that may be written in a similar block form. The Quantum Shannon decomposition method, on the other hand, reduces any arbitrary unitary into a compact sequence of quantum gates from a universal gate set. The decomposition is also formally exact for an arbitrary dimensional unitary. This can also treat chemical problems with arbitrary potential energy surfaces and yields a number of entangling gates close to the expected theoretical lower bound.

We consider intra-molecular proton-transfers in hydroxide and protonated water clusters and show how such problems can be mapped to an ion-trap system, and also show that the dynamics can be simulated using a quantum circuit decomposition method on a quantum simulator. General quantum nuclear dynamics problems, however, have unsymmetric potential energy surfaces and are generally performed in higher dimensions. Critical extensions to higher quantum nuclear dimensions have been implemented using tensor networks in Refs. \onlinecite{TN-Miguel-Anurag} and \onlinecite{IonQ-Anurag}. The methods discussed here will become critical in extending our mapping protocol to general potentials in higher dimensions, as will be considered in future publications. 

\section{Acknowledgments}
This research was supported by the National Science
Foundation grants CHE-2102610 (SSI), OMA-1936353 (SSI and PR) and CHE-2311165 (SSI and PR). 
 

\include{refs}

\end{document}


\title{SUPPORTING INFORMATION \\
Quantum circuit and mapping algorithms for wavepacket dynamics: case study of anharmonic hydrogen bonds in protonated and hydroxide water clusters}

\author{Debadrita Saha,}
\affiliation{Department of Chemistry, and the Indiana University Quantum Science and Engineering Center (IU-QSEC), Indiana University, 800 E. Kirkwood Ave, Bloomington, IN-47405}
\author{Philip Richerme}   
\affiliation{Department of Physics, and the Indiana University Quantum Science and Engineering Center (IU-QSEC),
Indiana University, Bloomington, IN-47405}
\author{Srinivasan S. Iyengar\email{iyengar@iu.edu},}
\affiliation{Department of Chemistry, Department of Physics, and the Indiana University Quantum Science and Engineering Center (IU-QSEC), Indiana University, 800 E. Kirkwood Ave, Bloomington, IN-47405}
\date{\today}

\maketitle

Here we discuss details regarding the unitary transforms that yield a block structure for the nuclear Hamiltonian and thus make the map between the ion-trap Hamiltonian and quantum nuclear Hamiltonian possible, electronic structure based potentials surfaces for H$_3$O$_{2}^{-}$ and H$_5$O$_{2}^{+}$, initial wavepacket details, a new correlation function based idea to compute vibrational eigenenergies, and more details on the vibrational spectra obtained from quantum simulations. 

\section{Unitary transformations that yield the Block structure of the nuclear Hamiltonian, for symmetric potentials, to make these commensurate with and mappable to the spin-lattice Hamiltonian, ${\cal H}_{IT}$}
\label{Transformations_Hmol-SI}
The nuclear Hamiltonian, ${\cal H}^{Mol}$ from Eq. (2) in the paper, 
has a banded Toeplitz structure due to the kinetic energy being expressed in terms of DAFs. 
\begin{figure}
    \includegraphics[width=0.5\columnwidth]{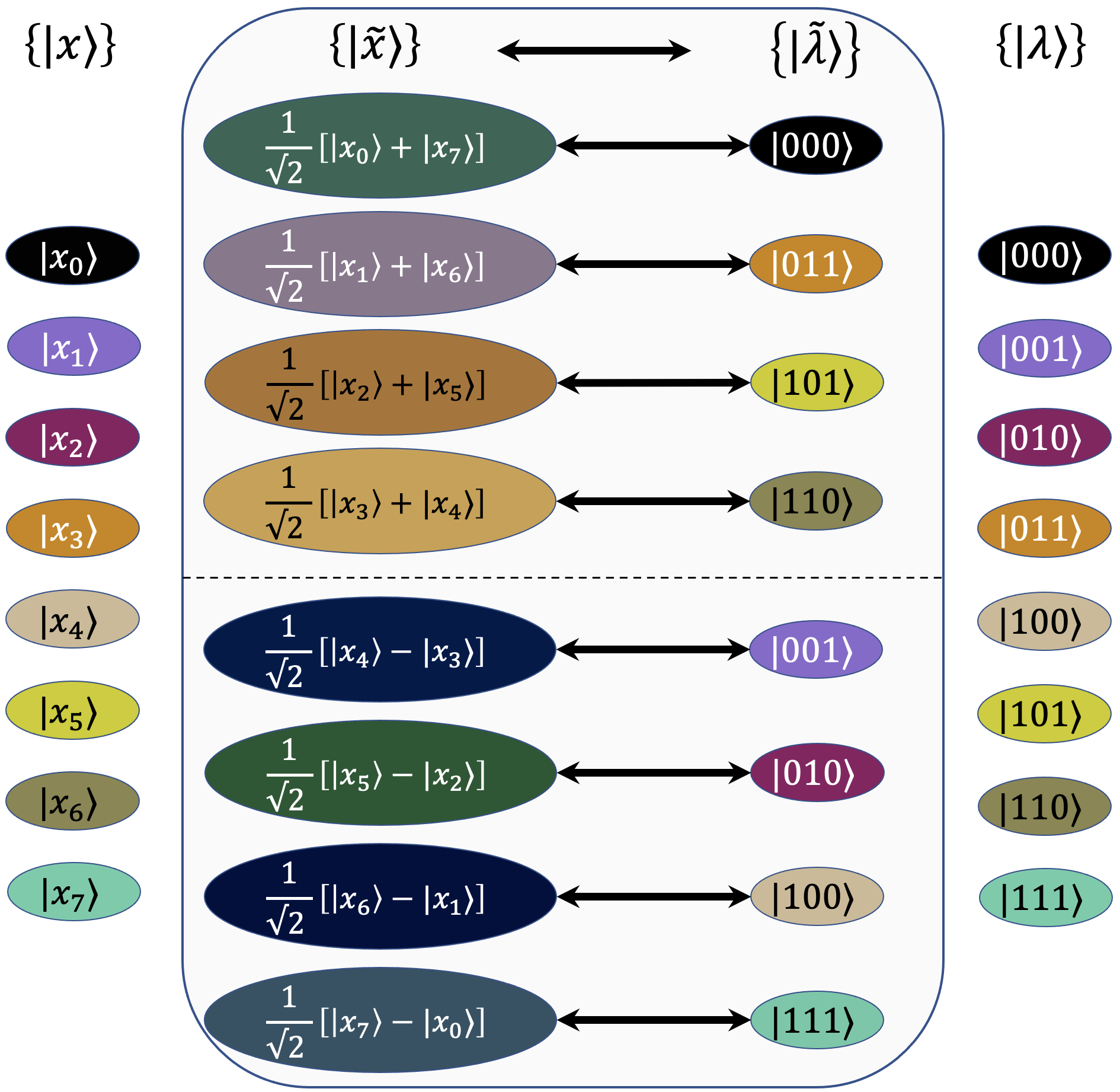}
    \caption{An illustration of the mapping of the Givens transformed grid basis state representation, $\ket{\tilde{x}}$ (Eq. (\ref{Givens-transformed-basis})), for the discrete quantum nuclear Hamiltonian to the permuted computational basis state representation, $\ket{\tilde{\lambda}}$ (Section IIA in paper), for the Ising model Hamiltonian. The respective basis states map shown here for the case of 3 qubits holds true and can be generalized to an arbitrary number of qubits. The dashed line in the middle separates the two blocks of each Hamiltonian.
    } 
    \label{fig:x-lambda-3qubits}
\end{figure}
\begin{figure*}[htb!]
    \centering
    \includegraphics[width=0.8\textwidth]
    {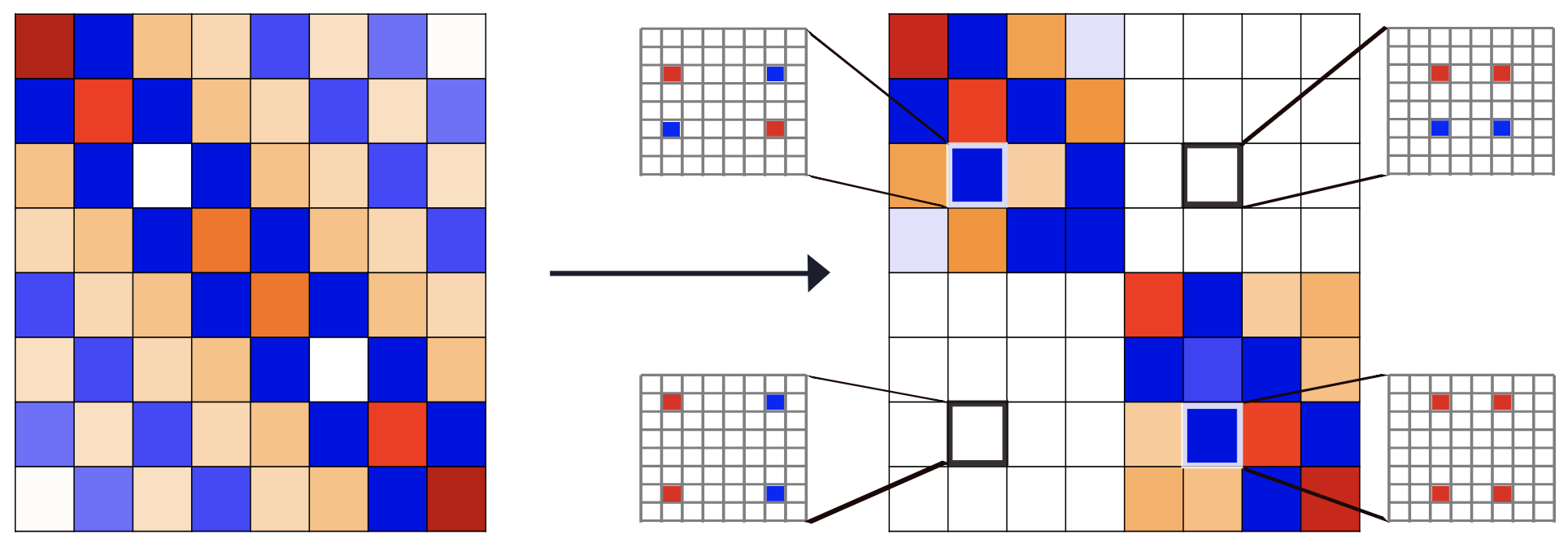}
    \caption{An illustration of the block-diagonalization  of the nuclear Hamiltonian, as captured by Eq. (\ref{Htilde-il}).  The original Hamiltonian, ${\cal H}^{Mol}$ is on the left, whereas the transformed $\Tilde{\cal{H}}^{Mol}$ is shown on the right. 
    On the right side, specific matrix elements from each block of  $\tilde{{\cal H}}_{il}^{Mol}$ are highlighted to illustrate Eqs. (\ref{Htilde-ii}) and (\ref{Htilde-il-od}). These highlighted elements of $\tilde{{\cal H}}_{il}^{Mol}$ are obtained by combining  elements of ${\cal H}^{Mol}$, as per Eq. (\ref{Htilde-il}), and these are marked using red and blue squares in zoomed in representations matrix elements in ${\cal H}^{Mol}$. The blue (negative) and red (positive) indicate the phase of the corresponding elements of ${\cal H}^{Mol}$, as obtained from $\alpha_{i}$ in Eqs. (\ref{Htilde-il}), (\ref{Htilde-ii}) and (\ref{Htilde-il-od}). 
    }
    \label{Unitary_transformH}
\end{figure*}
The unitary transform that leads to the block structure of the nuclear Hamiltonian, similar to the structure of the Ising Hamiltonian, can be expressed as a product of Givens rotations. 
The effect of the Givens rotations on the grid basis states is to create symmetric and anti-symmetric superpositions of pairs of grid basis states that are reflections about the grid center. 
To explain this, we introduce a uniform one-dimensional set of grid points, $\left\{ \ket{x_i} \right\}$ where $i \in (0,n)$, such that the Givens transformed grid basis, $\left\{ \ket{\tilde{x}_i} \right\}$, may be represented as 
\begin{eqnarray}
        \ket{\tilde{x}_{i}} &\equiv& \frac{1}{\sqrt{2}}\left[{\ket{x_{i}} + \ket{x_{n-i}}}\right], \quad 0 \leq i < (n+1)/2 \label{Givens-transformed-basis-1} \\
     &\equiv& \frac{1}{\sqrt{2}}\left[{\ket{x_{i}} - \ket{x_{n-i}}}\right], \quad (n+1)/2 \leq i \leq n
     \label{Givens-transformed-basis}
\end{eqnarray}
where $n= 2^N-1$. The grid basis and the Givens transformed grid basis, for a three-qubit system, are represented on the left columns of Figure \ref{fig:x-lambda-3qubits}.
Equations \ref{Givens-transformed-basis-1} and \ref{Givens-transformed-basis} form two mutually orthogonal subspaces and are represented in the top and bottom portions of Figure \ref{fig:x-lambda-3qubits}, separated by the dashed line. These subspaces block diagonalize the nuclear Hamiltonian for symmetric potentials. This process is illustrated for a three-qubit system ($2^3$-grid points) in Figure \ref{Unitary_transformH}.

The $il^{th}$ matrix element of the resultant molecular Hamiltonian in the Givens transformed grid basis 
is explicitly written as
  \begin{align}
      \tilde{\mathcal{H}}^{Mol}_{il} = \frac{1}{2}&\left(\mathcal{H}^{Mol}_{i,l} + \alpha_{l}\mathcal{H}^{Mol}_{i,n-l} + \alpha_{i}\mathcal{H}^{Mol}_{n-i,l} + \right. \nonumber \\ & \left. \hphantom{(}\alpha_{i}\alpha_{l}\mathcal{H}^{Mol}_{n-i,n-l}\right),
      \label{Htilde-il}
  \end{align}
where $\alpha_{i}=\text{sgn}\left[i-(n/2)\right]$. 
The elements of the diagonal blocks of $\tilde{\mathcal{H}}^{Mol}$ (matrix on the right in Figure (\ref{Unitary_transformH})) are obtained from Eq. (\ref{Htilde-il}) as 
   \begin{align}
      \tilde{\mathcal{H}}^{Mol}_{il} &= \frac{1}{2}\left(\mathcal{H}^{Mol}_{i,l} + \alpha_{i}\mathcal{H}^{Mol}_{i,n-l} + \alpha_{i}\mathcal{H}^{Mol}_{n-i,l} + \right. \nonumber \\ & \left. \hphantom{= \frac{1}{2}(} \mathcal{H}^{Mol}_{n-i,n-l}\right) \nonumber \\ &= \left[ {K}(x_{i},x_{l}) + \alpha_{i} {K}(x_{i},x_{n-l}) \right] + \nonumber \\ &  \hphantom{= \frac{1}{2}(}\frac{1}{2} \left[ V(x_{i}) + V(x_{n-l}) \right] \delta_{i,l}
      \label{Htilde-ii}
  \end{align}
The elements of the unitary transform, $\alpha_{i}$ are, in fact, the characters of the $C_s$ point group. The right hand side of the above equation, therefore, represents a symmetry adapted transformation of the nuclear Hamiltonian, and the term $\frac{1}{2} \left[ V(x_{i}) + V(x_{n-i}) \right]$, symmeterizes the potential energy surface in one-dimension. 
By extension, for the elements of the  off-diagonal blocks of $\tilde{\mathcal{H}}^{Mol}$ in Figure \ref{Unitary_transformH}, $\alpha_{l} = -\alpha_{i}$ and   \begin{align}
      \tilde{\mathcal{H}}^{Mol}_{il} =& \frac{1}{2}\left(\mathcal{H}^{Mol}_{i,l} - \alpha_{i}\mathcal{H}^{Mol}_{i,n-l} + \alpha_{i}\mathcal{H}^{Mol}_{n-i,l} - \right. \nonumber \\ & \left. \hphantom{\frac{1}{2}(}\mathcal{H}^{Mol}_{n-i,n-l}\right) \nonumber \\ =& 
      \frac{1}{2} \left[ V(x_{i}) - V(x_{n-l}) \right]  
      \delta_{i,n-l}
      \label{Htilde-il-od}
  \end{align}
where the kinetic energy contribution is identically zero due to the Toeplitz nature of the kinetic energy operator (Eq. (2) in the paper), and only the anti-symmetric portion of the potential, $\frac{1}{2} \left[ V(x_{i}) - V(x_{n-l}) \right]$, contributes to the anti-diagonal part of $\tilde{\mathcal{H}}^{Mol}$. Thus for symmetric potentials such as those considered here, Eq. (\ref{Htilde-il-od}) is identically zero. This observation becomes useful for our approach here.
\clearpage
\section{Role of grid spacing and electronic structure theory in the computation of potential energy surfaces for the proton transfer in H$_3$O$_{2}^{-}$ and H$_5$O$_{2}^{+}$}
The level of theory used for the computation of potential energy surfaces for the two systems is CCSD/6-311++G(d,p) and B3LYP/6-311++G(d,p) respectively, as given in Table I of the paper. Here, we present the potential energy surfaces as computed using a hierarchy of electronic structure theories to justify the use of a higher level of theory (CCSD) for H$_3$O$_{2}^{-}$. The level of theory helped to capture the anharmonicities in the potential for H$_3$O$_{2}^{-}$. As can be observed from the figures  below the MP2 surface develops a slight barrier height for 
$\Delta x=0.021$\AA, while the barrier height is significant for the CCSD surface with  $\Delta x=0.044$\AA. The anharmonic nature of the CCSD surface is not captured for the case of $\Delta x=0.094$\AA \ implying that the width of the barrier is smaller than the grid spacing $\Delta x$.
\begin{figure*}[htbp]
    \subfigure[$\Delta x=0.094$\AA]{\includegraphics[width=0.19\linewidth]{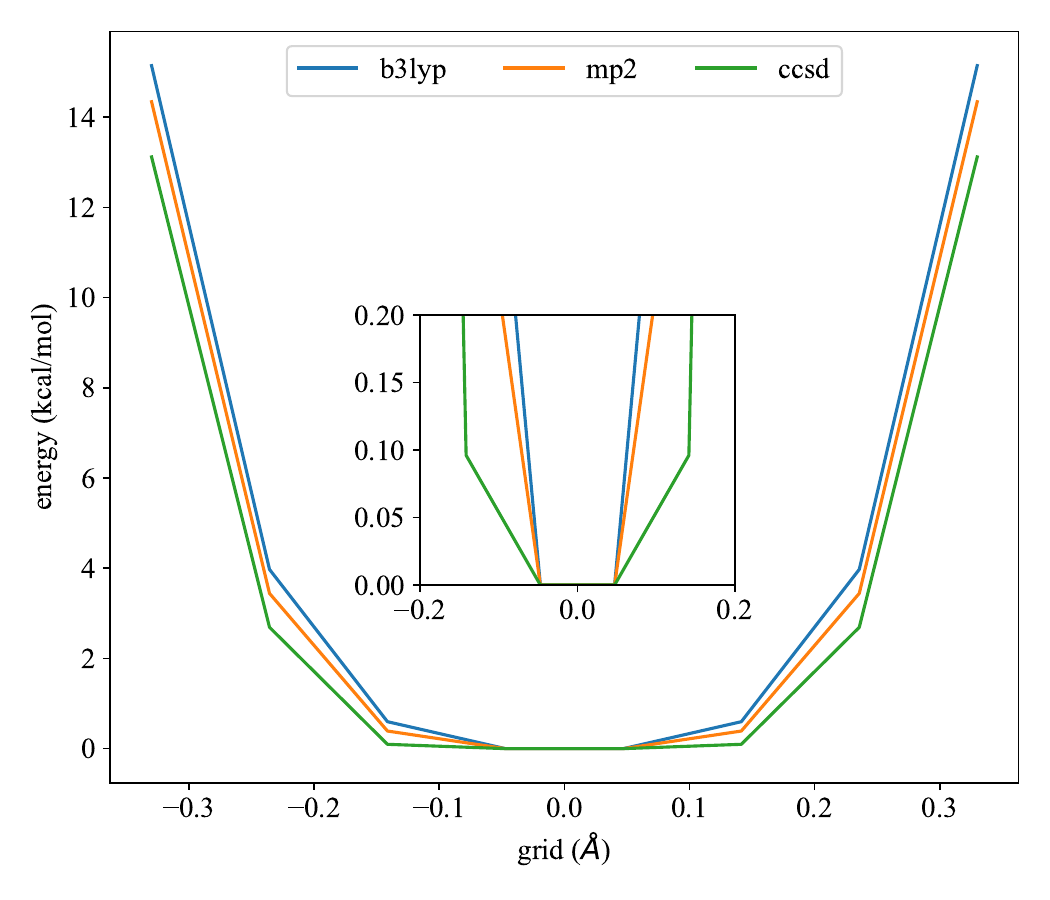}}
    \subfigure[$\Delta x=0.044$\AA]{\includegraphics[width=0.19\linewidth]{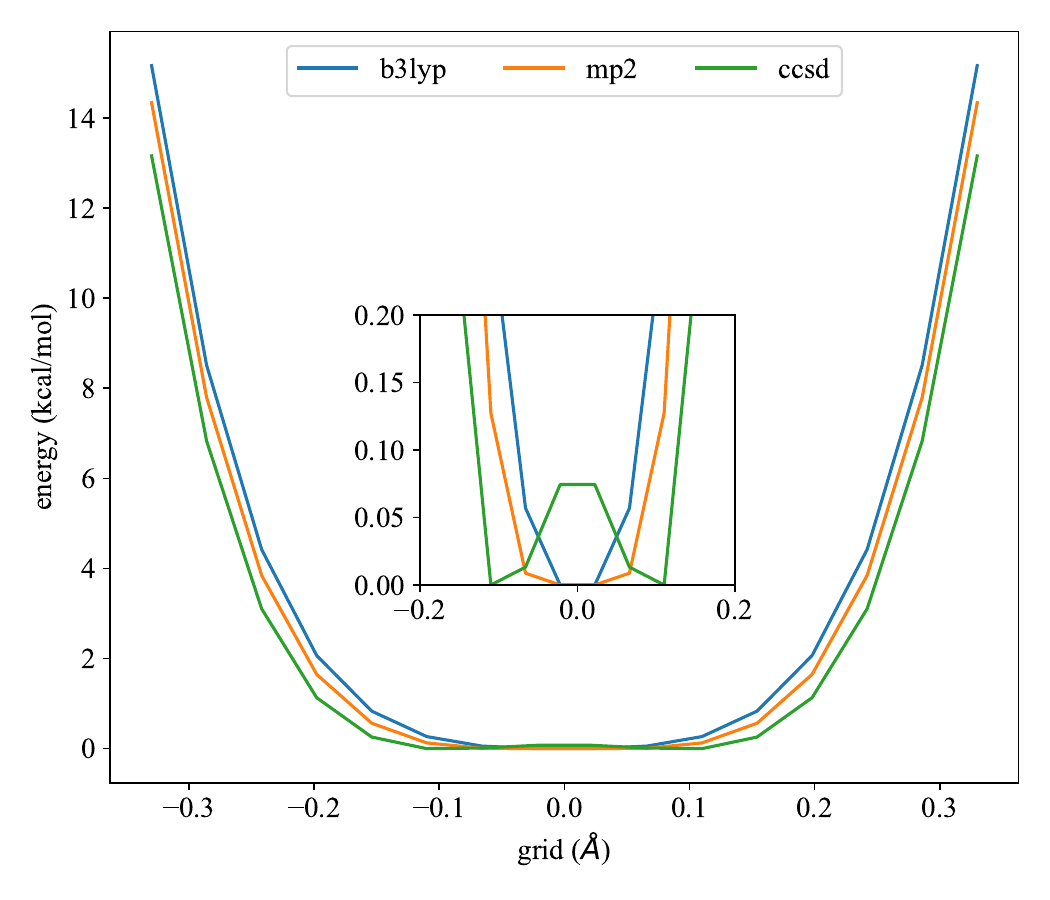}}
    \subfigure[$\Delta x=0.021$\AA]{\includegraphics[width=0.19\linewidth]{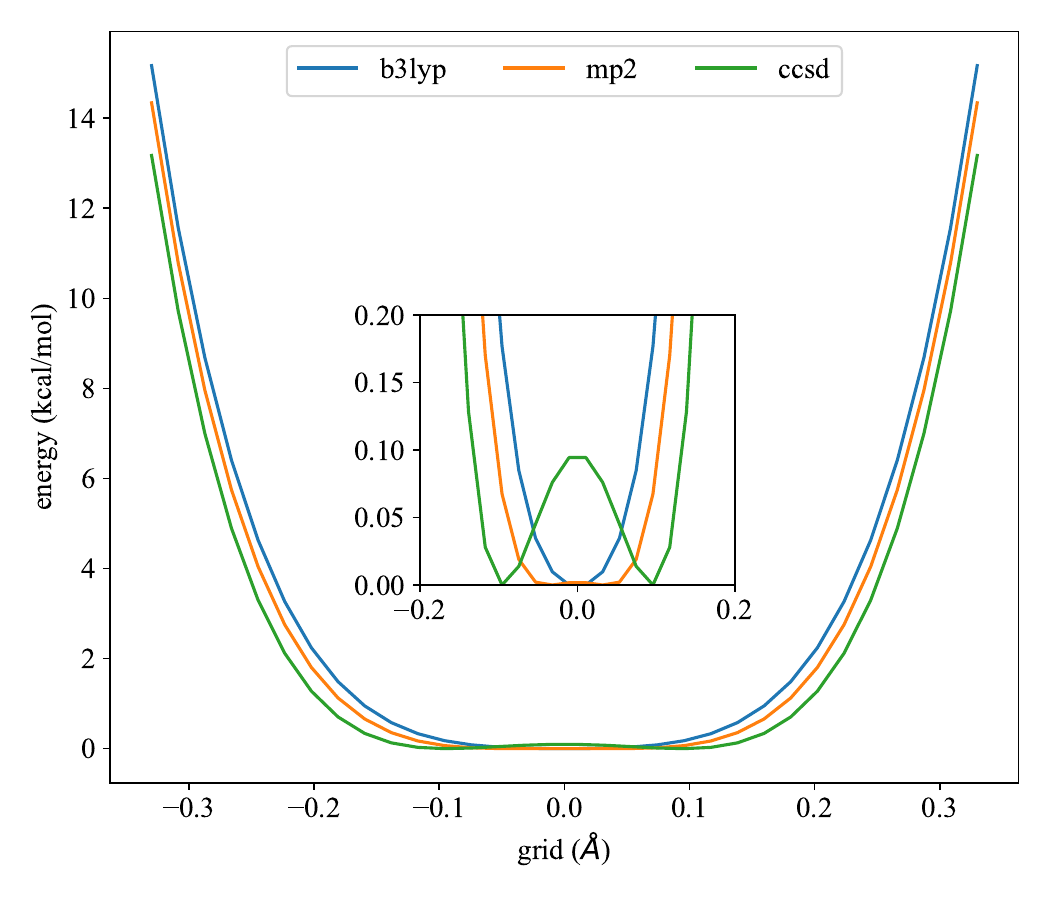}}
    \subfigure[$\Delta x=0.01$\AA]{\includegraphics[width=0.19\linewidth]{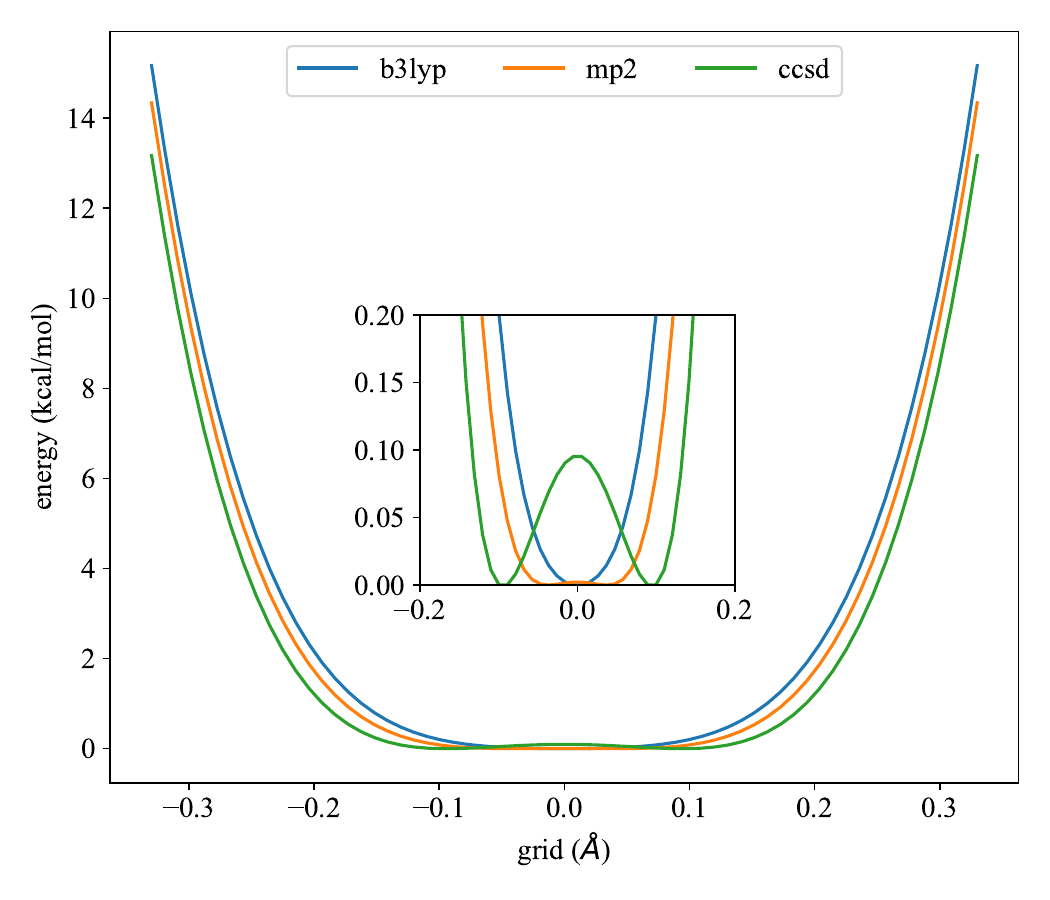}}
    \subfigure[$\Delta x=0.005$\AA]{\includegraphics[width=0.19\linewidth]{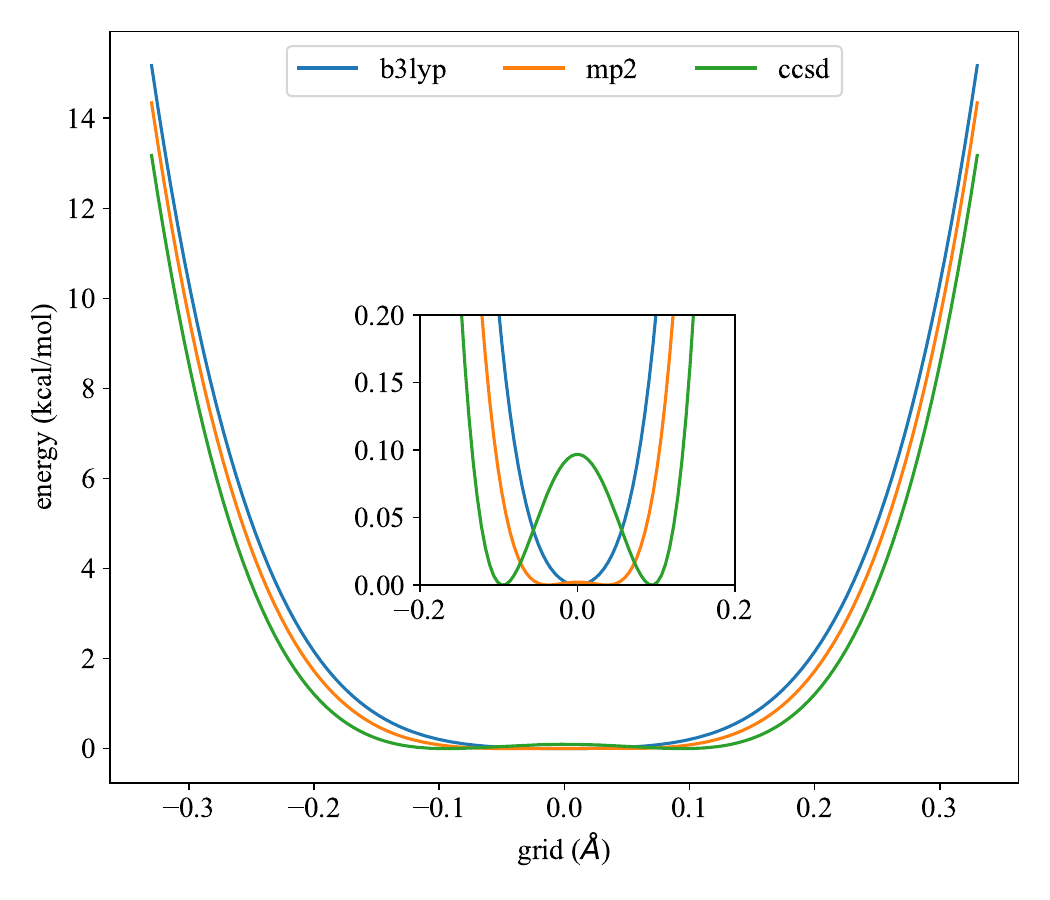}}
    \caption{The potential energy surface for the proton in H$_3$O$_{2}^{-}$ as computed with a hierarchy of electronic structure methods (along with basis set 6-311++G(d,p)) for a range of grid spacing ($\Delta x$). The grid length is chosen to be $0.66$ \AA (recorded in Table I of the paper) and $2^N$ grid points for an increasing number of qubits $N = 3$ to $7$ are chosen for (a)-(e).}
    \label{PES_h3o2-theory}
\end{figure*}

\begin{figure*}[htbp]
    \subfigure[$\Delta x=0.094$\AA]{\includegraphics[width=0.19\linewidth]{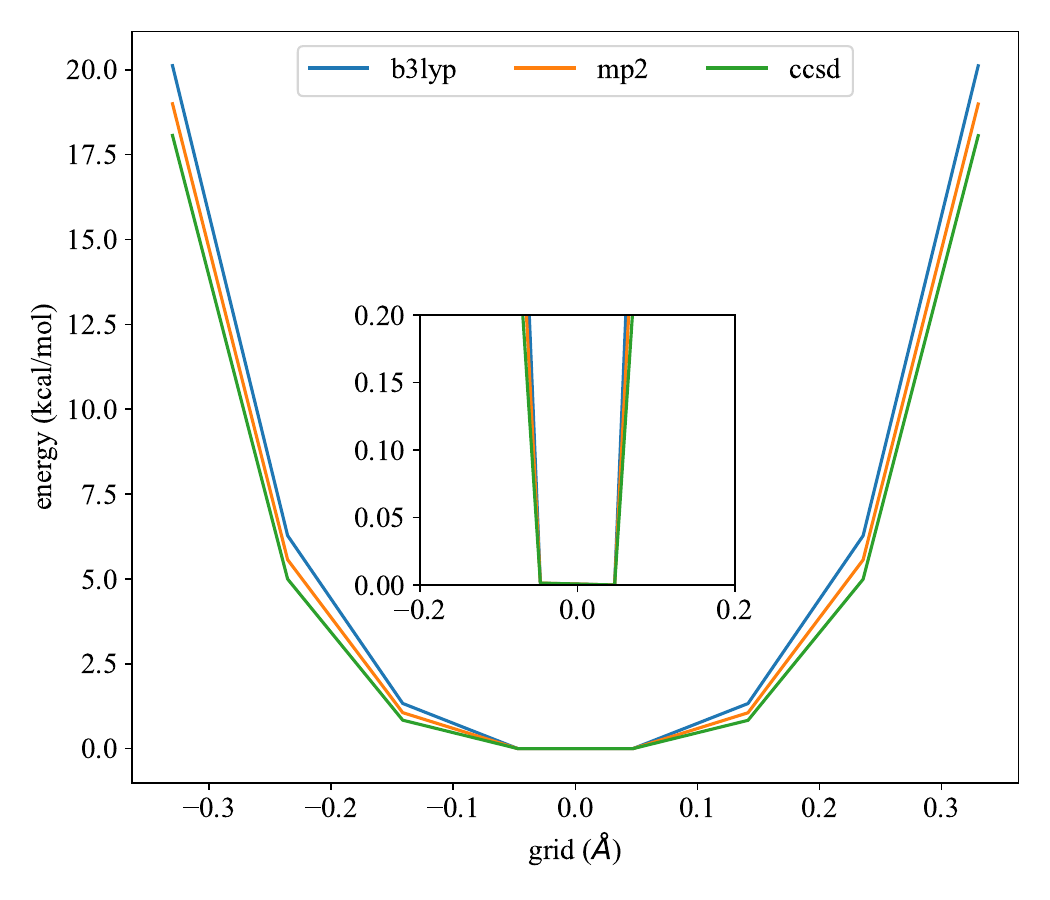}}
    \subfigure[$\Delta x=0.044$\AA]{\includegraphics[width=0.19\linewidth]{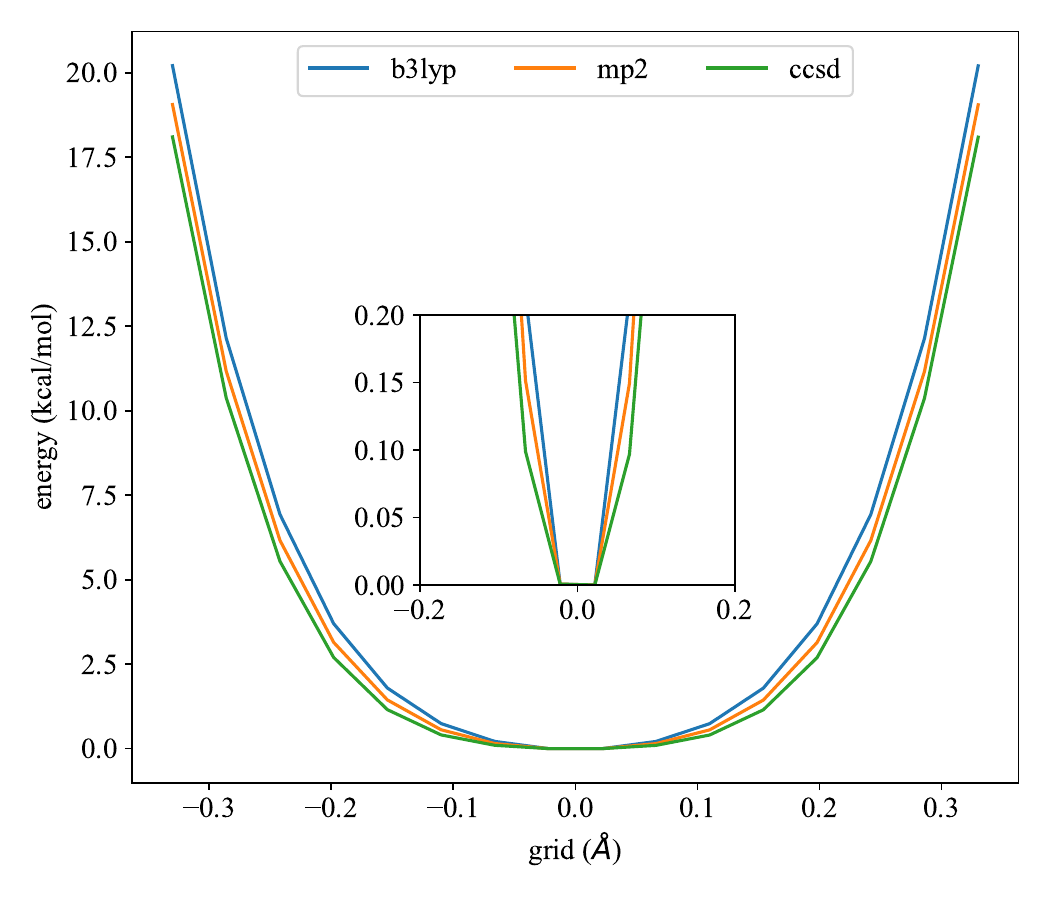}}
    \subfigure[$\Delta x=0.021$\AA]{\includegraphics[width=0.19\linewidth]{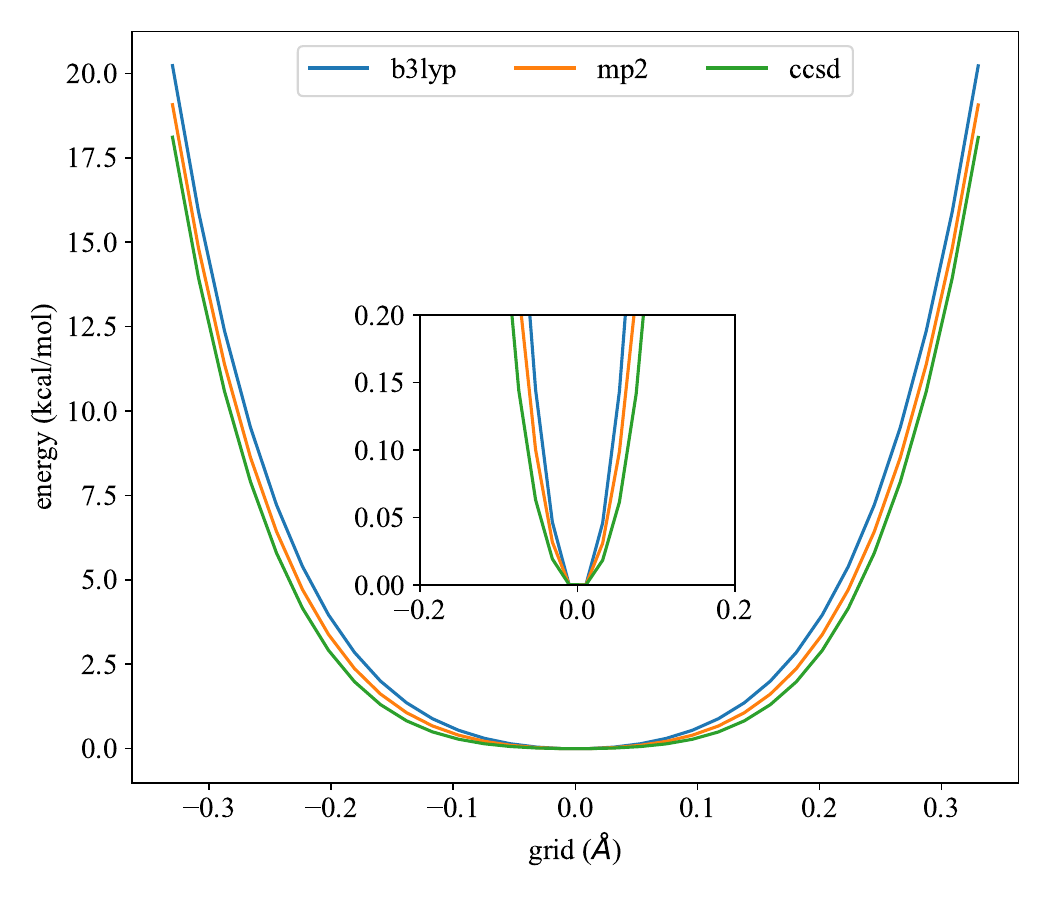}}
    \subfigure[$\Delta x=0.01$\AA]{\includegraphics[width=0.19\linewidth]{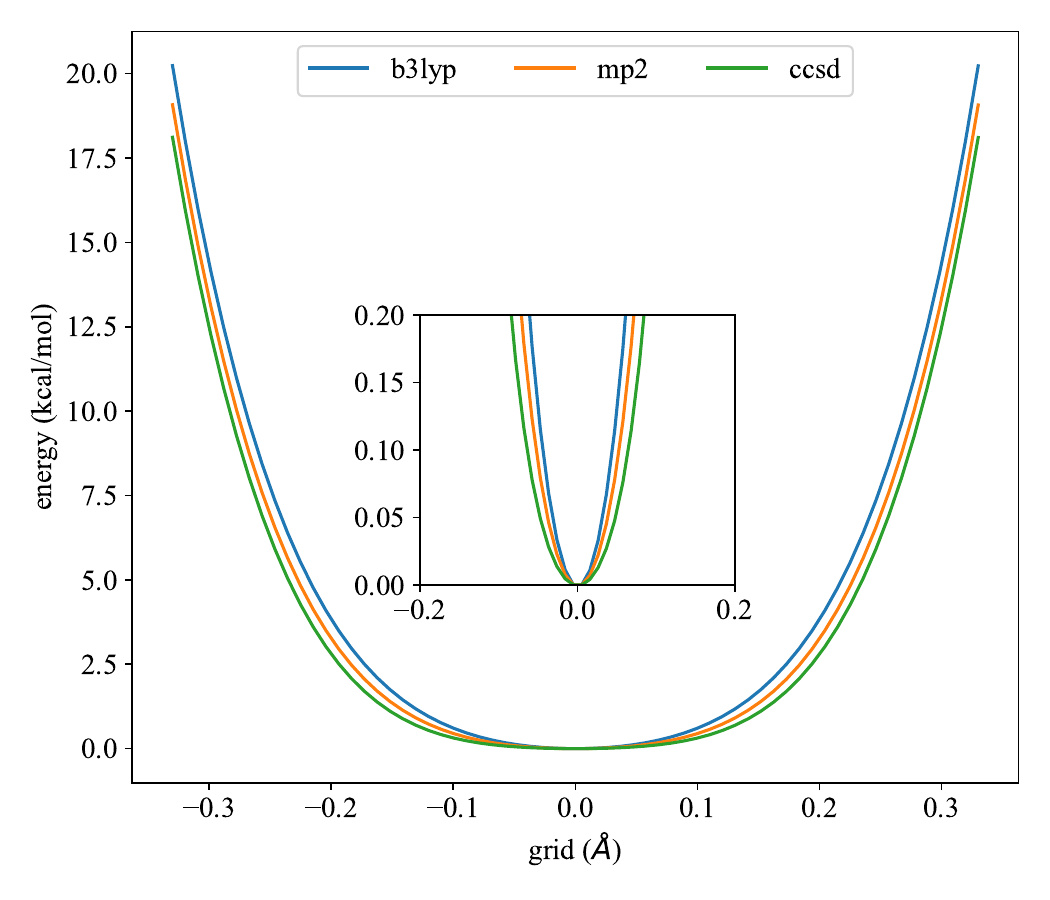}}
    \subfigure[$\Delta x=0.005$\AA]{\includegraphics[width=0.19\linewidth]{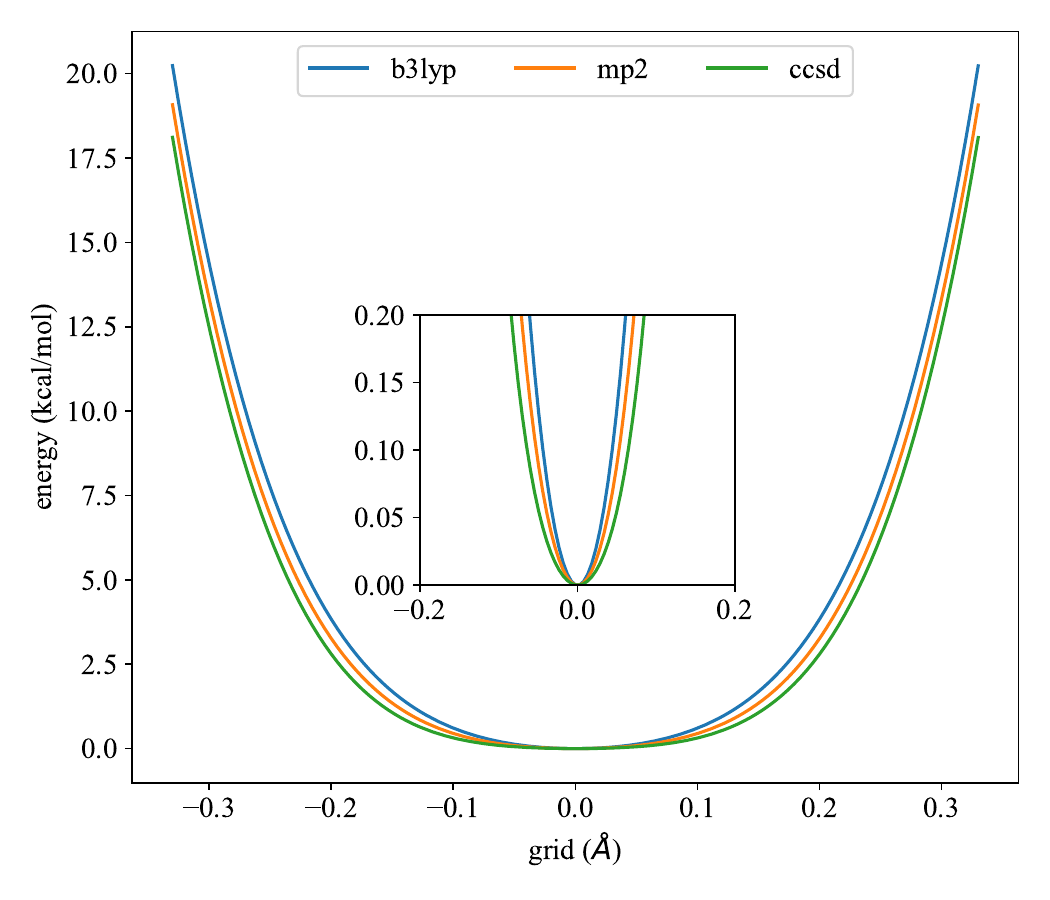}}
    \caption{The potential energy surface for the proton in H$_5$O$_{2}^{+}$ as computed with a hierarchy of electronic structure methods (along with basis set 6-311++G(d,p)) for a range of grid spacing ($\Delta x$). The grid length is chosen to be $0.66$ \AA (recorded in Table I of the paper) and $2^N$ grid points for an increasing number of qubits $N = 3$ to $7$ are chosen for (a)-(e).}
    \label{PES_h5o2+theory}
\end{figure*}

\clearpage

\section{Characteristics of initial wavepacket states for the quantum nuclei for quantum dynamics using circuit decomposition}
\begin{table*}[hbt!]
    \centering
    \begin{tabular*}{\textwidth}{@{\extracolsep{\fill}}ll|llll|llll}
    \hline \hline
    \multicolumn{2}{c|}{H$_3$O$_{2}^{-}$} &\multicolumn{4}{c|}{ $\left|\braket{\chi_0}{\psi}\right|^2$} & \multicolumn{4}{c}{E(kcal/mol)}\\
    \cline{3-10}
     $\psi(x;0)$ & Parameters & $N=4$ & $N=5$& $N=5$ & $N=7$& $N=4$ & $N=5$& $N=6$ & $N=7$ \\
    \hline
    $\psi_L(x;0)$ & $x_0 = $ donor site & $0.07\%$ & $0.01\%$ & $0.002\%$ & $0.0003\%$ & 82.1 & 307.6 & 1229.4 & 4955.4\\
    $\psi_G(x;0)$ & $\sigma = 0.1$\AA, $\mu = 0.0$\AA & $88.3\%$ & $88.9\%$ & $89.3\%$ & $89.6\%$ & 2.46 & 2.47 & 2.52 & 2.42 \\
    $\psi_T(x;0)$ & $T=300K$ & $99.5\%$ & $99.6\%$ & $99.7\%$ & $99.7\%$ & 1.65 & 2.50 & 5.55 & 16.85\\
    \hline \hline
    \end{tabular*}
    \vspace{\baselineskip}
    \begin{tabular*}{\textwidth}{@{\extracolsep{\fill}}ll|llll|llll}
    \multicolumn{2}{c|}{H$_5$O$_{2}^{+}$}&\multicolumn{4}{c|}{ $\left|\braket{\chi_0}{\psi}\right|^2$} & \multicolumn{4}{c}{E(kcal/mol)}\\
    \cline{3-10}
     $\psi(x;0)$ & Parameters & $N=4$ & $N=5$& $N=5$ & $N=7$& $N=4$ & $N=5$& $N=6$ & $N=7$ \\
    \hline
    $\psi_L(x;0)$ & $x_0 = $ donor site & $0.03\%$ & $0.006\%$ & $0.0009\%$ & $0.0001\%$ & 89.16 & 314.70 & 1236.43 & 4962.46\\
    $\psi_G(x;0)$ & $\sigma = 0.1$\AA, $\mu = 0.0$\AA & $93.84\%$ & $94.03\%$ & 94.17$\%$ & $94.26\%$ & 2.7 & 2.72 & 2.78 & 2.68 \\
    $\psi_T(x;0)$ & $T=300K$ & $99.93\%$ & $99.94\%$ & $99.94\%$ & $99.94\%$ & 2.06 & 2.22 & 2.82 & 4.77\\
    \hline \hline
    \end{tabular*}
    \caption{Parameters and characteristics of the initial wavepacket states considered for the transferring proton in the water cluster systems. The ground state overlap and corresponding energy correspond to the cases of $N=4-7$.}
    \label{InitWPs}
\end{table*}

\begin{table*}[]
    \begin{tabular*}{\textwidth}{@{\extracolsep{\fill}}c|ccccc|ccccc}
    \hline \hline
     & \multicolumn{5}{c|} {$\epsilon$ (Eq.(9))  for H$_5$O$_{2}^{+}$ (shots$=10^{4}$)} & \multicolumn{5}{c}{$\epsilon$ (Eq.(9)) for H$_3$O$_{2}^{-}$ (shots$=10^{4}$)}  \\
    $\psi(x;0)$ & $N=3$ & $N=4$ & $N=5$ & $N=6$ & $N=7$& $N=3$ & $N=4$ & $N=5$ & $N=6$ & $N=7$ \\
    \hline
    $\psi_L(x;0)$ & 2$\times 10^{-4}$ & 2$\times 10^{-4}$ & 1$\times 10^{-4}$ & 9$\times 10^{-5}$ & 6$\times 10^{-5}$& 2$\times 10^{-4}$ & 2$\times 10^{-4}$ & 1$\times 10^{-4}$ & 9$\times 10^{-5}$ & 6$\times 10^{-5}$ \\
    $\psi_{G}(x;0)$ & 2$\times 10^{-4}$ & 2$\times 10^{-4}$ & 1$\times 10^{-4}$ & 8$\times 10^{-5}$ & 6$\times 10^{-5}$ & 2$\times 10^{-4}$ & 2$\times 10^{-4}$ & 1$\times 10^{-4}$ & 9$\times 10^{-4}$ & 6$\times 10^{-4}$\\
    $\psi_T(x;0)$& 2$\times 10^{-4}$ & 2$\times 10^{-4}$ & 1$\times 10^{-4}$ & 8$\times 10^{-5}$ & 6$\times 10^{-5}$ & 3$\times 10^{-4}$ & 2$\times 10^{-4}$ & 1$\times 10^{-4}$ & 9$\times 10^{-5}$ & 6$\times 10^{-5}$\\
    \hline
    \end{tabular*}
    \begin{tabular*}{\textwidth}{@{\extracolsep{\fill}}c|ccccc|ccccc}
    \hline
     & \multicolumn{5}{c|} {$\epsilon$ (Eq.(9))  for H$_5$O$_{2}^{+}$ (shots$=10^{5}$)} & \multicolumn{5}{c}{$\epsilon$ (Eq.(9)) for H$_3$O$_{2}^{-}$ (shots$=10^{5}$)}  \\
    $\psi(x;0)$& $N=3$ & $N=4$ & $N=5$ & $N=6$ & $N=7$& $N=3$ & $N=4$ & $N=5$ & $N=6$ & $N=7$ \\
    \hline
    $\psi_L(x;0)$ & 6$\times 10^{-5}$ & 5$\times 10^{-5}$ & 4$\times 10^{-5}$ & 3$\times 10^{-5}$ & 2$\times 10^{-5}$& 1$\times 10^{-4}$ & 7$\times 10^{-5}$ & 4$\times 10^{-5}$ & 3$\times 10^{-5}$ & 2$\times 10^{-5}$ \\
    $\psi_{G}(x;0)$ & 6$\times 10^{-5}$ & 5$\times 10^{-5}$ & 4$\times 10^{-5}$ & 3$\times 10^{-5}$ & 2$\times 10^{-5}$ & 1$\times 10^{-4}$ & 7$\times 10^{-5}$ & 4$\times 10^{-5}$ & 3$\times 10^{-5}$ & 2$\times 10^{-5}$\\
    $\psi_T(x;0)$& 6$\times 10^{-5}$&5$\times 10^{-5}$& 4$\times 10^{-5}$& 3$\times 10^{-5}$& 2$\times 10^{-5}$ & 1$\times 10^{-4}$& 8$\times 10^{-5}$ & 4$\times10^{-5}$ & 3$\times10^{-5}$ & 2$\times 10^{-5}$\\
    \hline
    \end{tabular*}
    \begin{tabular*}{\textwidth}{@{\extracolsep{\fill}}c|ccccc|ccccc}
    \hline
     & \multicolumn{5}{c|} {$\epsilon$ (Eq.(9)) for H$_5$O$_{2}^{+}$ (shots$=10^{6}$)}  & \multicolumn{5}{c}{$\epsilon$ (Eq.(9)) for H$_3$O$_{2}^{-}$(shots$=10^{6}$)}  \\
    $\psi(x;0)$ & $N=3$ & $N=4$ & $N=5$ & $N=6$ & $N=7$& $N=3$ & $N=4$ & $N=5$ & $N=6$ & $N=7$ \\
    \hline
    $\psi_L(x;0)$ & 2$\times 10^{-5}$ & 2$\times 10^{-5}$ & 1$\times 10^{-5}$ & 9$\times 10^{-6}$ & 6$\times 10^{-6}$ & 6$\times 10^{-5}$ & 4$\times 10^{-5}$ & 2$\times 10^{-5}$ & 1$\times 10^{-5}$ & 6$\times 10^{-6}$ \\
    $\psi_{G}(x;0)$ & 2$\times 10^{-5}$ & 2$\times 10^{-5}$ & 1$\times 10^{-5}$ & 8$\times 10^{-6}$ & 6$\times 10^{-6}$ & 6$\times 10^{-5}$ & 4$\times 10^{-5}$ & 2$\times 10^{-5}$ & 1$\times 10^{-5}$ & 6$\times 10^{-6}$\\
    $\psi_T(x;0)$& 2$\times 10^{-5}$ & 2$\times 10^{-5}$ & 1$\times 10^{-5}$ & 8$\times 10^{-6}$ & 6$\times 10^{-6}$ & 2$\times 10^{-4}$ & 5$\times 10^{-5}$ & 2$\times 10^{-5}$ & 1$\times 10^{-6}$ & 7$\times 10^{-6}$\\
    \hline \hline
    \end{tabular*}
    \caption{The mean absolute errors in probability (Eq. (9)) computed between the classical propagated probability density $\rho_{C}(x)$ and the QASM simulated probability densities, $\rho_{Q}(x)$, using the circuit decomposition method summarized in 
    the main article. The number of shots used are $10^{4}$, $10^{5}$, and $10^{6}$ respectively. Errors are reported for different initial nuclear wavepacket states of the transferring proton in both water clusters for all cases of $N = 3-7$. As the number of shots is increased, the error decreases as noted in Figure 
    8 in the paper.}
    \label{QSD_errors_H3O2_H5O2_10000}
\end{table*}

\section{Calculation of vibrational properties from Fourier transforms of the time dynamics of the shared nuclei}\label{results_FT}

To extract the oscillation frequencies of the shared-proton wavepacket in our hydrogen-bonded system, we perform a Fourier transform of the measured time-evolution data for the Zundel and hydroxide water cluster. Mathematically, this operation is equivalent to computing
\begin{align}
\nonumber
    \int e^{i\omega t} |\chi(x,t)|^2 dt & =
    \sum_{i,j} \left[ \int e^{i \omega t} e^{i (E_j-E_i)t/\hbar} dt \right] c_j^* c_i  \phi_j^*(x) \phi_i(x) \\
    & = \sum_{i,j} \delta(\omega-(E_j-E_i)/\hbar) c_j^* c_i  \phi_j^*(x) \phi_i(x),
    \label{FTequation}
\end{align}
where the wavepacket $\chi(x,t)$ is expressed as a linear combination of eigenstates. This produces peaks in the Fourier spectrum corresponding to frequency differences between energy eigenstates, $(E_j-E_i)/\hbar$. This expression is also related to the Fourier transform of the density matrix auto-correlation function, ${\text{Tr}}[\rho(0)\rho(t)]$ as discussed below. The time-dependent density matrix in an eigenstate representation is given by,
\begin{eqnarray}
    \rho(t) &=& \ket{\chi(t)}\bra{\chi(t)} = 
    e^{-\imath {\cal H}t /\hbar}
    \ket{\chi(0)}\bra{\chi(0)}
    e^{\imath {\cal H}t/{\hbar}}
    \nonumber \\
    &=& \sum_{i,j}c_{i}(0)c_{j}^{*}(0) 
    e^{-\imath{\cal H}t/{\hbar}}
    \ket{\phi_{i}}\bra{\phi_{j}}
    e^{\imath {\cal H}t/{\hbar}}
    \nonumber \\
    &=& \sum_{i,j}c_{i}(0)c_{j}^{*}(0) 
    e^{\imath (E_{i}-E_{j}) t/{\hbar}}
    \ket{\phi_{i}}
    \bra{\phi_{j}},
    \label{rhodefn}
\end{eqnarray}
and is used to construct the Fourier transform of the density-density time auto-correlation function, ${\text{Tr}}[\rho(0)\rho(t)]$ as
\begin{widetext}
\begin{eqnarray}
    \int_{-\infty}^{+\infty} dt\, 
    e^{{\imath \omega t}} \; {\text{Tr}}[\rho(0)\rho(t)] &=& \int_{-\infty}^{+\infty} dt\, e^{{\imath \omega t}}
    \; {\text{Tr}}\left[\ket{\chi(0)}\bra{\chi(0)}\sum_{i,j}c_{i}(0)c_{j}^{*}(0) 
    e^{\imath (E_{i}-E_{j}) t/{\hbar}}
    \ket{\phi_{i}}\bra{\phi_{j}}\right] \nonumber \\ 
    &=& \sum_{i,j}|c_{i}(0)|^{2}|c_{j}(0)|^{2} \; \delta \left(\omega-(E_{i}-E_{j})/\hbar \right),
    \label{Density-timecorrelation-FT}
\end{eqnarray}
\end{widetext}
which is further reduced the following using the convolution theorem, 
\begin{align}
    & \int dx  {
    \left\vert 
    \int_{-\infty}^{+\infty} dt  e^{{\imath \omega t}} \;
    \sum_{i,j}c_{i}(0)c_{j}^{*}(0) 
    e^{\imath (E_{i}-E_{j}) t/{\hbar}}
    {\phi_{i}(x)}
    {\phi_{j}(x)} 
    \right\vert}^2 +
    \nonumber \\ & 
    \int dx \,dx^\prime \left\vert \int_{-\infty}^{+\infty} dt  e^{{\imath \omega t}} 
    \sum_{i,j}c_{i}(0)c_{j}^{*}(0) 
    e^{\imath (E_{i}-E_{j}) t/{\hbar}}
    {\phi_{i}(x)}
    {\phi_{j}(x^\prime)} \right\vert^2
    \label{Density-timecorrelation-FT-2} 
\end{align}

\begin{figure*}
    \includegraphics[width=0.48\textwidth]{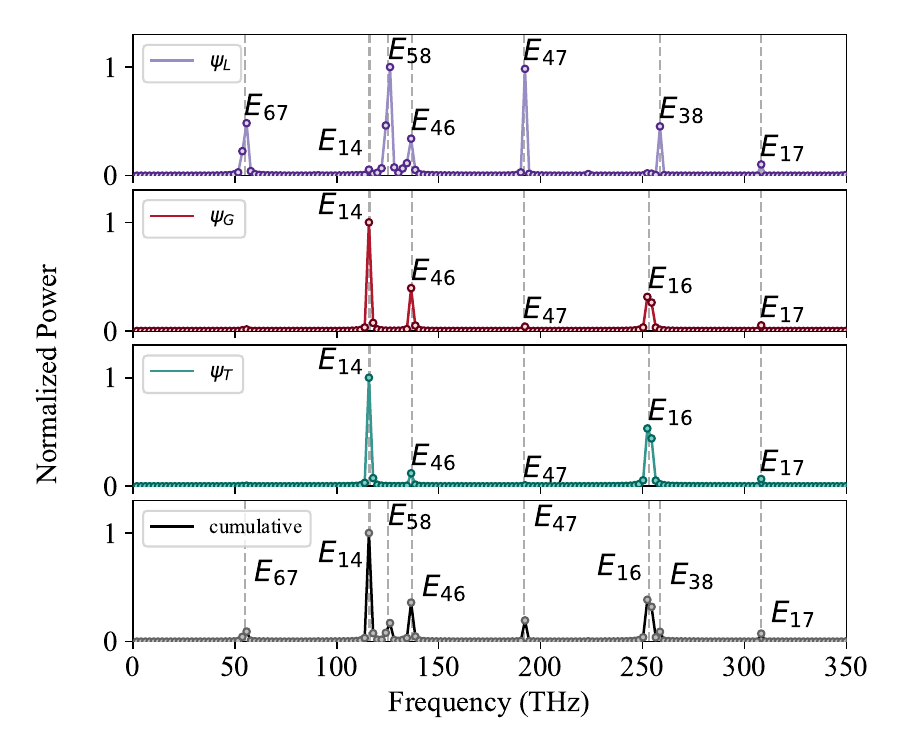}
    \includegraphics[width=0.48\textwidth]{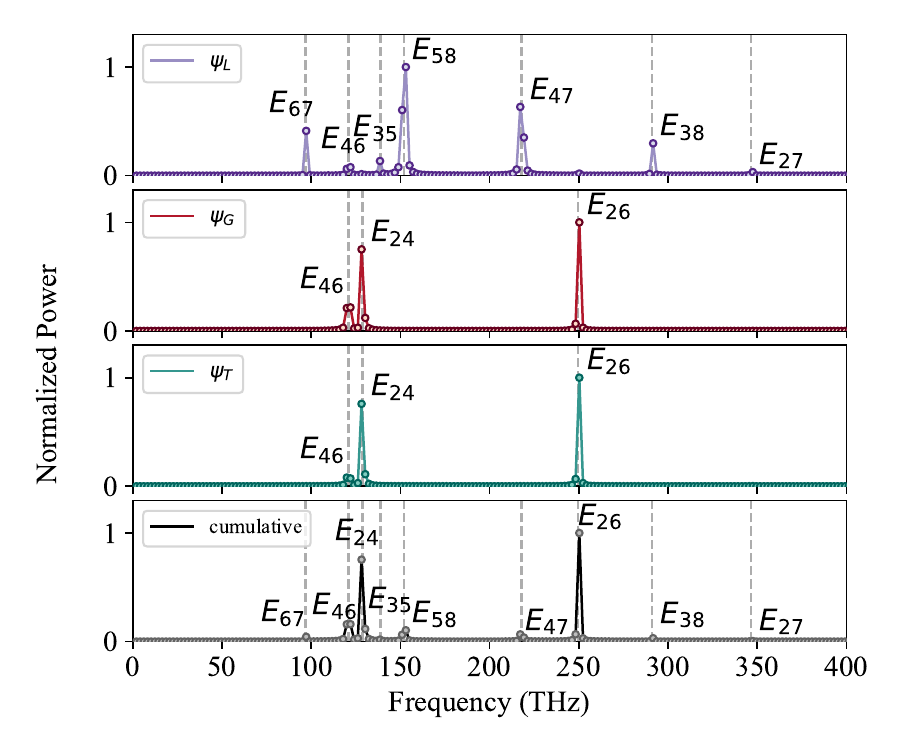}
    \caption{The Fourier spectra showing all possible vibrational frequencies (as in equation \ref{FTequation}) for the proton dynamics in the hydroxide (left) and protonated Zundel (right) water cluster corresponding to the different initial wavepackets given in Table \ref{InitWPs} as obtained from the mapping algorithm. The dashed lines in gray are computed from the exact diagonalization of the Hamiltonian corresponding to $N=3$.}
    \label{FFT_h3o2_proton_map}
\end{figure*}

\begin{figure*}[htbp]
    \includegraphics[width=\textwidth]{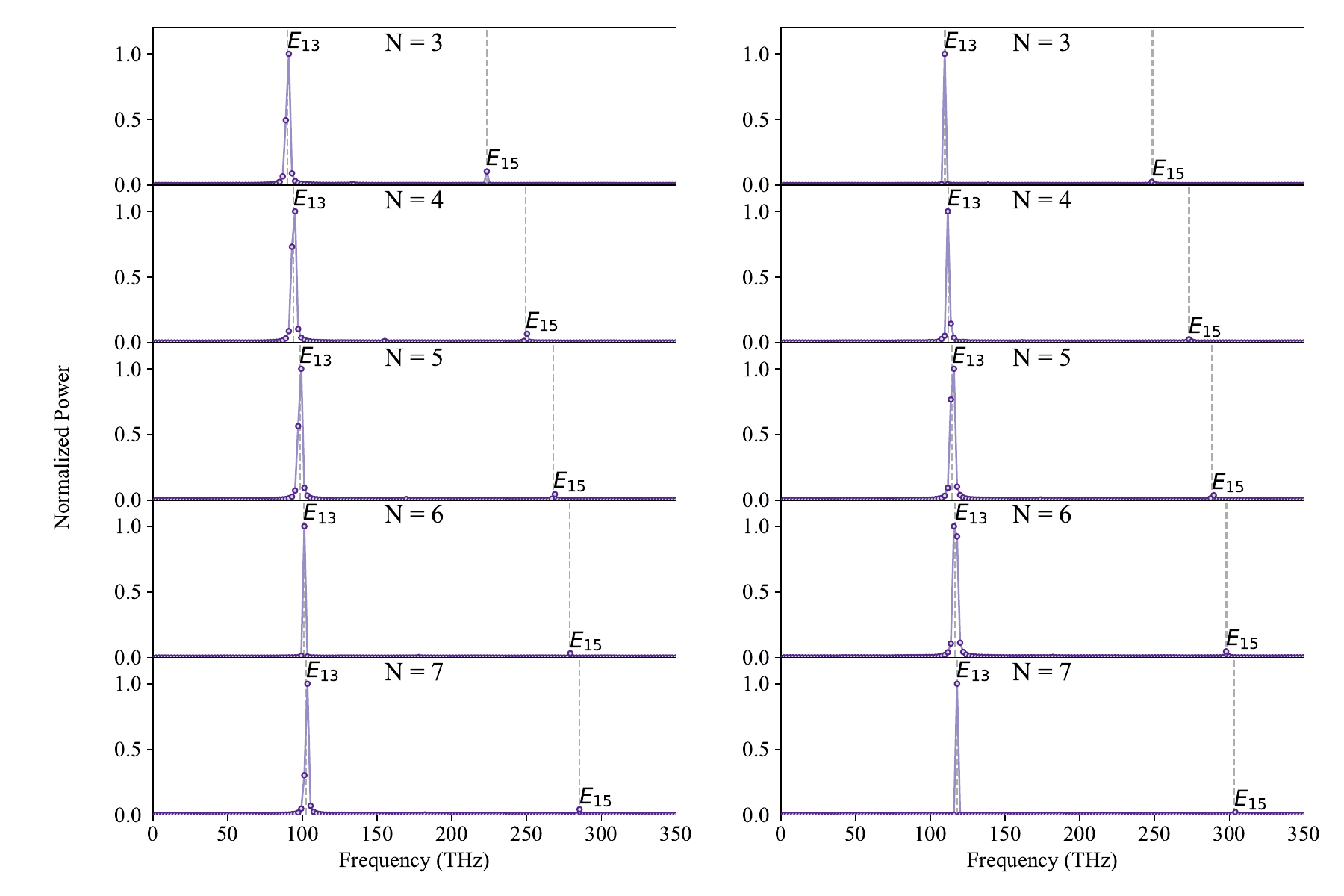}
    \caption{The Fourier spectra showing all possible vibrational frequencies (as in equation \ref{FTequation}) for the proton dynamics in the hydroxide (left) and Zundel (right) water cluster corresponding to the initial wavepacket $\psi_{G}(x;0)$. The peaks correspond to the Fourier spectra for the time evolution computed using the QSD method on IBM's QASM simulator. The gray dotted lines are the frequency differences computed using exact diagonalization.}
    \label{FFT_zundel_proton_QSD}
\end{figure*}

Note that the term on the first line of Eq. (\ref{Density-timecorrelation-FT-2}), that is
\begin{align}  
    {\cal I} (\omega;x) & = \int_{-\infty}^{+\infty} dt\,  e^{{\imath \omega t}} \;
    \sum_{i,j}c_{i}(0)c_{j}^{*}(0) 
    e^{\imath (E_{i}-E_{j}) t/{\hbar}}
    {\phi_{i}(x)}
    {\phi_{j}(x)} \nonumber \\
    & = \sum_{i,j} \delta(\omega-(E_{i}-E_{j})) c_{i}(0)c_{j}^{*}(0) {\phi_{i}(x)}
    {\phi_{j}(x)},
    \label{Density-timecorrelation-FT-3} 
\end{align}
is identical to Eq. \ref{FTequation} which makes the first term in Eq. \ref{Density-timecorrelation-FT-2},
\begin{align}  
{\cal P} (\omega) = \int dx \; {\left\vert {\cal I} (\omega; x) \right\vert}^2
    \label{Eq:integ-Density-timecorrelation-FT-3} 
\end{align}
While standard (classical) approaches may compute the results from Eq. (\ref{Density-timecorrelation-FT-2}), a quantum computer can obtain information at a finer resolution and compute the individual terms inside the integral as shown in Eq. (\ref{Density-timecorrelation-FT-3}) separately. Figures \ref{FFT_h3o2_proton_map} and \ref{FFT_zundel_proton_QSD} show the Fourier transforms of the

%% file: qsphere_1qubit.tex
    \begin{blochsphere}[radius=1.5 cm,tilt=15,rotation=-20,opacity=0.1]
        \node at (0,0) {};
        \draw [fill][red!100] (0,1.5) circle (1.8pt);
        \draw [fill][blue!100] (0,-1.5) circle (1.8pt);
        \labelLatLon{11}{90}{0};
        \labelLatLon{00}{-90}{90};
        \node[above=1mm,font=\fontsize{8}{10}\selectfont] at (11) {{$\left|1\right>$ }};
        \node[below=1mm,font=\fontsize{8}{10}\selectfont] at (00) {{$\left|0\right>$}};
    \end{blochsphere}

%% file: qsphere_2qubits.tex
\begin{blochsphere}[radius=1.5 cm,tilt=15,rotation=-20,opacity=0.1][scale=0.5]
    \node at (0,0) {};
    \drawLatitudeCircle[style={solid}]{0};
    \draw [fill][blue!100] (0,1.5) circle (1.8pt);
    \draw [fill][blue!100] (0,-1.5) circle (1.8pt);
    \labelLatLon{11}{90}{0};
    \labelLatLon{00}{-90}{90};
    \node[above=1mm,font=\fontsize{8}{10}\selectfont] at (11) {{$\left|11\right>$ }};
    \node[below=1mm,font=\fontsize{8}{10}\selectfont] at (00) {{$\left|00\right>$}};
    \draw [fill][red!100] (-1.5,0) circle (1.8pt);
    \draw [fill][red!100] (1.5,0) circle (1.8pt);
    \labelLatLon{01}{0}{180};
    \node[left=1mm,font=\fontsize{8}{10}\selectfont] at (01) {{$\left|01\right>$ }};
    \labelLatLon{10}{0}{0};
    \node[right=1mm,font=\fontsize{8}{10}\selectfont] at (10) {{$\left|10\right>$ }};
\end{blochsphere}

%% file: qsphere_3qubits.tex
\begin{blochsphere}[radius=1.5 cm,tilt=15,rotation=-20,opacity=0.1]
    
    \drawLatitudeCircle[style={solid}]{30};
    \drawLatitudeCircle[style={solid}]{-30};
    \draw [fill][red!100] (0,1.5) circle (1.8pt);
    \draw [fill][blue!100] (0,-1.5) circle (1.8pt);
    \draw [fill][blue!100] (0,0.4) circle (1.8pt);
    \draw [fill][red!100] (0,-0.4) circle (1.8pt);
    \draw [fill][blue!100] (-0.85,.975) circle (1.8pt);
    \draw [fill][blue!100] (0.85,.975) circle (1.8pt);
    \draw [fill][red!100] (-0.85,-.975) circle (1.8pt);
    \draw [fill][red!100] (0.85,-.975) circle (1.8pt);



    \labelLatLon{111}{90}{0};
    \labelLatLon{000}{-90}{90};
    \node[above=1mm,font=\fontsize{8}{10}\selectfont] at (111) {{$\left|111\right>$ }};
    \node[below=1mm,font=\fontsize{8}{10}\selectfont] at (000) {{$\left|000\right>$}};

    \labelLatLon{011}{30}{70};
    \node[below=1mm,font=\fontsize{8}{10}\selectfont] at (011) {{$\left|101\right>$ }};
    \labelLatLon{101}{50}{-30};
    \node[right=1mm,font=\fontsize{8}{10}\selectfont] at (101) {{$\left|011\right>$ }};
    \labelLatLon{110}{50}{-180};
    \node[left=1mm,font=\fontsize{8}{10}\selectfont] at (110) {{$\left|110\right>$ }};

    \labelLatLon{001}{-50}{330};
    \node[right=2mm,font=\fontsize{8}{10}\selectfont] at (001) {{$\left|001\right>$ }};
    \labelLatLon{010}{-45}{-110};
    \node[above=-1.5mm,font=\fontsize{8}{10}\selectfont] at (010) {{$\left|010\right>$ }};
    \labelLatLon{100}{-50}{180};
    \node[left=2mm,font=\fontsize{8}{10}\selectfont] at (100) {{$\left|100\right>$ }};

\end{blochsphere}

%% file: qsphere_4qubits.tex
\begin{blochsphere}[radius=1.5 cm,tilt=15,rotation=-20,opacity=0.1]
    \node at (0,0) {};
    \drawLatitudeCircle[style={solid}]{0};
    \drawLatitudeCircle[style={solid}]{45};
    \drawLatitudeCircle[style={solid}]{-45};
    \draw [fill][blue!100] (0,1.5) circle (1.8pt);
    \draw [fill][blue!100] (0,-1.5) circle (1.8pt);
    
    \draw [fill][red!100] (-0.95,0.90) circle (1.8pt);
    \draw [fill][red!100] (0.95,0.90) circle (1.8pt);
    \draw [fill][red!100] (-0.55,1.25) circle (1.8pt);
    \draw [fill][red!100] (0.55,1.25) circle (1.8pt);
    \draw [fill][red!100] (-0.95,-1.15) circle (1.8pt);
    \draw [fill][red!100] (0.95,-1.15) circle (1.8pt);
    \draw [fill][red!100] (-0.55,-0.80) circle (1.8pt);
    \draw [fill][red!100] (0.55,-0.80) circle (1.8pt);
    \draw [fill][blue!100] (0.75,0.35) circle (1.8pt);
    \draw [fill][blue!100] (0.75,-0.35) circle (1.8pt);
    \draw [fill][blue!100] (-0.75,0.35) circle (1.8pt);
    \draw [fill][blue!100] (-0.75,-0.35) circle (1.8pt);
    \draw [fill][blue!100] (-1.5,0) circle (1.8pt);
    \draw [fill][blue!100] (1.5,0) circle (1.8pt);
    \labelLatLon{1111}{90}{0};
    \labelLatLon{0000}{-90}{90};
    \node[above=1mm,font=\fontsize{4}{10}\selectfont] at (1111) {{$\left|1111\right>$ }};
    \node[below=1mm,font=\fontsize{4}{10}\selectfont] at (0000) {{$\left|0000\right>$}};
    
    \labelLatLon{1011}{28}{-55};
    \node[below,font=\fontsize{4}{10}\selectfont] at (1011) {{$\left|1011\right>$ }};
    \labelLatLon{0111}{50}{-45};
    \node[below,font=\fontsize{4}{10}\selectfont] at (0111) {{$\left|0111\right>$ }};
   \labelLatLon{1101}{28}{165};
    \node[right,font=\fontsize{4}{10}\selectfont] at (1101) {{$\left|1101\right>$ }};
    \labelLatLon{1110}{50}{-155};
    \node[left,font=\fontsize{4}{10}\selectfont] at (1110) {{$\left|1110\right>$ }};    

    \labelLatLon{0001}{-50}{5};
    \node[below,font=\fontsize{4}{10}\selectfont] at (0001) {{$\left|0001\right>$ }};
    \labelLatLon{0010}{-36}{0};
    \node[above,font=\fontsize{4}{10}\selectfont] at (0010) {{$\left|0010\right>$ }};
    \labelLatLon{0100}{-40}{155};
    \node[above,font=\fontsize{4}{10}\selectfont] at (0100) {{$\left|0100\right>$ }};
    \labelLatLon{1000}{-50}{155};
    \node[below,font=\fontsize{4}{10}\selectfont] at (1000) {{$\left|1000\right>$ }};

    \labelLatLon{1100}{5}{215};
    \node[left,font=\fontsize{4}{10}\selectfont] at (1100) {{$\left|1100\right>$}};
    \labelLatLon{1010}{-5}{180};
    \node[left=1mm,font=\fontsize{4}{10}\selectfont] at (1010) {{$\left|1010\right>$}};
    \labelLatLon{1001}{-30}{210};
    \node[left,font=\fontsize{4}{10}\selectfont] at (1001) {{$\left|1001\right>$}};
    \labelLatLon{0110}{30}{30};
    \node[right,font=\fontsize{4}{10}\selectfont] at (0110) {{$\left|0110\right>$}};
    \labelLatLon{0101}{5}{0};
    \node[right=1mm,font=\fontsize{4}{10}\selectfont] at (0101) {{$\left|0101\right>$}};
    \labelLatLon{0011}{-5}{15};
    \node[below,font=\fontsize{4}{10}\selectfont] at (0011) {{$\left|0011\right>$}};
\end{blochsphere}

%% file: 2qubit-Ising.tex
\begin{tikzpicture}
[scale=0.8,auto=center,every node/.style={circle,scale=0.5,fill=red!20}] 
    \node [fill=gray!30](a1) at (1,-0.01) {$\ket{00}$};  
    \node [fill=gray!30](a2) at (3,-0.01) {$\ket{11}$};
    \node [fill=gray!30](a3) at (5,-0.01) {$\ket{01}$};
    \node [fill=gray!30](a4) at (7,-0.01) {$\ket{10}$};
    \node (a1) at (1,-1) {$\ket{0}$};  
    \node (a2) at (3,-1) {$\ket{3}$};
    \node (a3) at (5,-1) {$\ket{1}$};
    \node (a4) at (7,-1) {$\ket{2}$};
    \node (a1) at (1,1) {$\ket{\downarrow \downarrow}$};  
    \node (a2) at (3,1) {$\ket{\uparrow \uparrow}$};
    \node (a3) at (5,1) {$\ket{\downarrow \uparrow}$};
    \node (a4) at (7,1) {$\ket{\uparrow \downarrow}$};
    \node (a5) at (2,2) [fill=blue!30] {$J_{12}^{x}-J_{12}^{y}$};
    \node (a6) at (6,2) [fill=blue!30] {$J_{12}^{x}+J_{12}^{y}$};
    \node (a13) at (4,2) [fill=green!30] {$B_{2}^{x}+iB_{2}^{y}$};
    \node (a15) at (3,3) [fill=green!30] {$B_{1}^{x}-iB_{1}^{y}$};
    \node (a16) at (5,3) [fill=green!30] {$B_{1}^{x}+iB_{1}^{y}$};
    \node (a17) at (4,4) [fill=green!30] {$B_{2}^{x}-iB_{2}^{y}$};
    \draw (a1) -- (a5);
    \draw (a5) -- (a15);
    \draw (a15) -- (a17);
    \draw (a2) -- (a5);
    \draw (a2) -- (a13);
    \draw (a3) -- (a13);
    \draw (a3) -- (a6);
    \draw (a6) -- (a4);
    \draw (a6) -- (a16);
    \draw (a13) -- (a15);
    \draw (a13) -- (a16);
    \draw (a16) -- (a17);
\end{tikzpicture}

%% file: 3qubit-Ising.tex
\begin{tikzpicture}
[scale=0.8,auto=center,every node/.style={circle,scale=0.5,fill=red!20}]
     \node (a1) at (1,1) {$\ket{\downarrow \downarrow \downarrow}$};  
    \node (a2) at (3,1) {$\ket{\downarrow \uparrow \uparrow}$};    
    \node (a3) at (5,1) {$\ket{\uparrow \downarrow \uparrow}$};
    \node (a4) at (7,1) {$\ket{\uparrow \uparrow \downarrow}$};
    \node (i1) at (1,-1) {$\ket{0}$};  
    \node (i2) at (3,-1) {$\ket{3}$};    
    \node (i3) at (5,-1) {$\ket{5}$};
    \node (i4) at (7,-1) {$\ket{6}$};
    \node (bi1) [fill=gray!30] at (1,0) {$\ket{000}$};  
    \node (bi2) [fill=gray!30] at (3,0) {$\ket{011}$};    
    \node (bi3) [fill=gray!30] at (5,0) {$\ket{101}$};
    \node (bi4) [fill=gray!30] at (7,0) {$\ket{110}$};
    \node  (a5) [fill=blue!30] at (2,2) {$J_{12}^{x}-J_{12}^{y}$};
    \node  (a6) [fill=blue!30] at (6,2) {$J_{12}^{x}+J_{12}^{y}$};
    \node  (a13) [fill=blue!20] at (4,2) {$J_{23}^{x}+J_{23}^{y}$};
    \node (a15) [fill=blue!20] at (3,3) {$J_{13}^{x}-J_{13}^{y}$};
    \node (a16) [fill=blue!20] at (5,3) {$J_{13}^{x}+J_{13}^{y}$};
    \node (a17) [fill=blue!20] at (4,4) {$J_{23}^{x}-J_{23}^{y}$};
    \draw (a1) -- (a5);
    \draw (a5) -- (a15);
    \draw (a15) -- (a17);
    \draw (a2) -- (a5);
    \draw (a2) -- (a13);
    \draw (a3) -- (a13);
    \draw (a3) -- (a6);
    \draw (a6) -- (a4);
    \draw (a6) -- (a16);
    \draw (a13) -- (a15);
    \draw (a13) -- (a16);
    \draw (a16) -- (a17);
    \node (a7) at (9,1) {$\ket{\downarrow \downarrow \uparrow}$};  
    \node (a8) at (11,1) {$\ket{\downarrow \uparrow \downarrow}$};   
    \node (a9) at (13,1) {$\ket{\uparrow \downarrow \downarrow}$};
    \node (a10) at (15,1) {$\ket{\uparrow \uparrow \uparrow}$};
    \node (i1) at (9,-1) {$\ket{1}$};  
    \node (i2) at (11,-1) {$\ket{2}$};    
    \node (i3) at (13,-1) {$\ket{4}$};
    \node (i4) at (15,-1) {$\ket{7}$};
    \node (bi1) [fill=gray!30] at (9,0) {$\ket{001}$};  
    \node (bi2) [fill=gray!30] at (11,0) {$\ket{010}$};    
    \node (bi3) [fill=gray!30] at (13,0) {$\ket{100}$};
    \node (bi4) [fill=gray!30] at (15,0) {$\ket{111}$};
    \node (a11) [fill=blue!30] at (10,2) {$J_{12}^{x}+J_{12}^{y}$};
    \node (a12) [fill=blue!30] at (14,2) {$J_{12}^{x}-J_{12}^{y}$};
    \node (a14) [fill=blue!20] at (12,2) {$J_{23}^{x}+J_{23}^{y}$};
    \node (a18) [fill=blue!20] at (12,4) {$J_{23}^{x}-J_{23}^{y}$};
    \node (a19) [fill=blue!20] at (11,3) {$J_{13}^{x}+J_{13}^{y}$};
    \node (a20) [fill=blue!20] at (13,3) {$J_{13}^{x}-J_{13}^{y}$};
    \draw (a7) -- (a11);
    \draw (a8) -- (a11);
    \draw (a9) -- (a12);
    \draw (a10) -- (a12);
    \draw (a8) -- (a14);
    \draw (a9) -- (a14);
    \draw (a14) -- (a19);
    \draw (a14) -- (a20);
    \draw (a11) -- (a19);
    \draw (a12) -- (a20);
    \draw (a20) -- (a18);
    \draw (a19) -- (a18);
    \node (a21) [fill=green!20] at (8,2) {    };
    \node (a22) [fill=green!20] at (7,3) {$B_{3}^{x}+iB_{3}^{y}$};
    \node (a23) [fill=green!20] at (9,3) {$B_{3}^{x}+iB_{3}^{y}$};
    \node (a24) [fill=green!30] at (6,4) {$B_{2}^{x}+iB_{2}^{y}$};
    \node (a25) [fill=green!20] at (8,4) {   };
    \node (a26) [fill=green!30] at (10,4) {$B_{2}^{x}+iB_{2}^{y}$};
    \node (a27) [fill=green!30] at (5,5) {$B_{1}^{x}-iB_{1}^{y}$};
    \node (a28) [fill=green!30] at (7,5) {$B_{1}^{x}+iB_{1}^{y}$};
    \node (a29) [fill=green!30] at (9,5) {$B_{1}^{x}+iB_{1}^{y}$};
    \node (a30) [fill=green!30] at (11,5) {$B_{1}^{x}-iB_{1}^{y}$};
    \node (a31) [fill=green!30] at (6,6) {$B_{2}^{x}-iB_{2}^{y}$};
    \node (a32) [fill=green!20] at (8,6) {   };
    \node (a33) [fill=green!30] at (10,6) {$B_{2}^{x}-iB_{2}^{y}$};
    \node (a34) [fill=green!20] at (8,8) {    };
    \node (a35) [fill=green!20] at (7,7) {$B_{3}^{x}-iB_{3}^{y}$};
    \node (a36) [fill=green!20] at (9,7) {$B_{3}^{x}-iB_{3}^{y}$};
    \draw (a17) -- (a27);
    \draw (a16) -- (a24);
    \draw (a6) -- (a22);
    \draw (a4) -- (a21);
    \draw (a31) -- (a27);
    \draw (a28) -- (a24);
    \draw (a25) -- (a22);
    \draw (a23) -- (a21);
    \draw (a31) -- (a35);
    \draw (a28) -- (a32);
    \draw (a25) -- (a29);
    \draw (a23) -- (a26);
    \draw (a34) -- (a35);
    \draw (a36) -- (a32);
    \draw (a33) -- (a29);
    \draw (a30) -- (a26);
    \draw (a7) -- (a21);
    \draw (a11) -- (a23);
    \draw (a19) -- (a26);
    \draw (a18) -- (a30);
    \draw (a22) -- (a21);
    \draw (a25) -- (a23);
    \draw (a29) -- (a26);
    \draw (a33) -- (a30);
    \draw (a22) -- (a24);
    \draw (a25) -- (a28);
    \draw (a29) -- (a32);
    \draw (a33) -- (a36);
    \draw (a34) -- (a36);
    \draw (a27) -- (a24);
    \draw (a31) -- (a28);
    \draw (a35) -- (a32);
\end{tikzpicture}

%% file: 4qubit-Ising.tex
\begin{tikzpicture}
[scale=0.58,auto=center,every node/.style={circle,scale=0.44,fill=red!20}]
     \node (a0) at (1,1) {$\ket{\downarrow \downarrow \downarrow \downarrow}$};  
    \node (a3) at (3,1) {$\ket{\downarrow \downarrow \uparrow \uparrow}$};    
    \node (a5) at (5,1) {$\ket{\downarrow \uparrow \downarrow \uparrow}$};
    \node (a6) at (7,1) {$\ket{\downarrow \uparrow \uparrow \downarrow}$};      \node (a9) at (9,1) {$\ket{\uparrow \downarrow \downarrow \uparrow}$};  
    \node (a10) at (11,1) {$\ket{\uparrow \downarrow \uparrow \downarrow}$};    
    \node (a12) at (13,1) {$\ket{\uparrow \uparrow \downarrow \downarrow}$};
    \node (a15) at (15,1) {$\ket{\uparrow \uparrow \uparrow \uparrow}$};    \node (i0) at (1,-1) {$\ket{0}$};  
    \node (i3) at (3,-1) {$\ket{3}$};    
    \node (i5) at (5,-1) {$\ket{5}$};
    \node (i6) at (7,-1) {$\ket{6}$};
    \node (i9) at (9,-1) {$\ket{9}$};  
    \node (i10) at (11,-1) {$\ket{10}$};    
    \node (i12) at (13,-1) {$\ket{12}$};
    \node (i15) at (15,-1) {$\ket{15}$};
    \node (bi0) [fill=gray!30] at (1,0) {$\ket{0000}$};  
    \node (bi3) [fill=gray!30] at (3,0) {$\ket{0011}$};    
    \node (bi5) [fill=gray!30] at (5,0) {$\ket{0101}$};
    \node (bi6) [fill=gray!30] at (7,0) {$\ket{0110}$};
    \node (bi9) [fill=gray!30] at (9,0) {$\ket{1001}$};  
    \node (bi10) [fill=gray!30] at (11,0) {$\ket{1010}$};    
    \node (bi12) [fill=gray!30] at (13,0) {$\ket{1100}$};
    \node (bi15) [fill=gray!30] at (15,0) {$\ket{1111}$};
    \node  (h_0-3) [fill=blue!30] at (2,2) {$J_{12}^{x}-J_{12}^{y}$};
    \node  (h_5-6) [fill=blue!30] at (6,2) {$J_{12}^{x}+J_{12}^{y}$};
    \node  (h_3-5) [fill=blue!20] at (4,2) {$J_{23}^{x}+J_{23}^{y}$};
    \node  (h_6-9) [fill=blue!10] at (8,2) {};
    \node  (h_9-10) [fill=blue!30] at (10,2) {$J_{12}^{x}+J_{12}^{y}$};
    \node  (h_10-12) [fill=blue!20] at (12,2) {$J_{23}^{x}+J_{23}^{y}$};    
    \node  (h_12-15) [fill=blue!30] at (14,2) {$J_{12}^{x}-J_{12}^{y}$};

    \node  (h_0-5) [fill=blue!20] at (3,3) {$J_{13}^{x}-J_{13}^{y}$};
    \node  (h_3-6) [fill=blue!20] at (5,3) {$J_{13}^{x}+J_{13}^{y}$};
    \node  (h_5-9) [fill=blue!10] at (7,3) {$J_{34}^{x}+J_{34}^{y}$};
    \node  (h_6-10) [fill=blue!10] at (9,3) {$J_{34}^{x}+J_{34}^{y}$};
    \node  (h_9-12) [fill=blue!20] at (11,3) {$J_{13}^{x}+J_{13}^{y}$};
    \node  (h_10-15) [fill=blue!20] at (13,3) {$J_{13}^{x}-J_{13}^{y}$};
    \node  (h_0-6) [fill=blue!20] at (4,4) {$J_{23}^{x}-J_{23}^{y}$};
    \node  (h_3-9) [fill=blue!10] at (6,4) {$J_{24}^{x}+J_{24}^{y}$};
    \node  (h_5-10) [fill=blue!10] at (8,4) {};
    \node  (h_6-12) [fill=blue!10] at (10,4) {$J_{24}^{x}+J_{24}^{y}$};
    \node  (h_9-15) [fill=blue!20] at (12,4) {$J_{23}^{x}-J_{23}^{y}$};
    \node  (h_0-9) [fill=blue!10] at (5,5) {$J_{14}^{x}-J_{14}^{y}$};
    \node  (h_3-10) [fill=blue!10] at (7,5) {$J_{14}^{x}+J_{14}^{y}$};
    \node  (h_5-12) [fill=blue!10] at (9,5) {$J_{14}^{x}+J_{14}^{y}$};
    \node  (h_6-15) [fill=blue!10] at (11,5) {$J_{14}^{x}-J_{14}^{y}$};
    \node  (h_0-10) [fill=blue!10] at (6,6) {$J_{24}^{x}-J_{24}^{y}$};
    \node  (h_3-12) [fill=blue!10] at (8,6) {};
    \node  (h_5-15) [fill=blue!10] at (10,6) {$J_{24}^{x}-J_{24}^{y}$};
    \node (h_0-12) [fill=blue!10] at (7,7) {$J_{34}^{x}-J_{34}^{y}$};
    \node (h_3-15) [fill=blue!10] at (9,7) {$J_{34}^{x}-J_{34}^{y}$};
    \node (h_0-15) [fill=blue!10] at (8,8) {};

    \draw (a0) -- (h_0-3);
    \draw (a3) -- (h_0-3);
    \draw (a3) -- (h_3-5);
    \draw (a5) -- (h_3-5);
    \draw (a5) -- (h_5-6);
    \draw (a6) -- (h_5-6);
    \draw (a6) -- (h_6-9);
    \draw (a9) -- (h_6-9);
    \draw (a9) -- (h_9-10);
    \draw (a10) -- (h_9-10);
    \draw (a10) -- (h_10-12);
    \draw (a12) -- (h_10-12);
    \draw (a12) -- (h_12-15);
    \draw (a15) -- (h_12-15);
    \draw (h_0-3) -- (h_0-5);
    \draw (h_3-5) -- (h_0-5);
    \draw (h_3-5) -- (h_3-6);
    \draw (h_5-6) -- (h_3-6);
    \draw (h_6-9) -- (h_5-9);
    \draw (h_5-6) -- (h_5-9);
    \draw (h_6-9) -- (h_6-10);
    \draw (h_9-10) -- (h_6-10);
    \draw (h_9-10) -- (h_9-12);
    \draw (h_10-12) -- (h_9-12);
    \draw (h_10-12) -- (h_10-15);
    \draw (h_12-15) -- (h_10-15);
    \draw (h_0-5) -- (h_0-6);
    \draw (h_3-6) -- (h_0-6);
    \draw (h_3-6) -- (h_3-9);
    \draw (h_5-9) -- (h_3-9);
    \draw (h_5-9) -- (h_5-10);
    \draw (h_6-10) -- (h_5-10);
    \draw (h_6-10) -- (h_6-12);
    \draw (h_9-12) -- (h_6-12);
    \draw (h_9-12) -- (h_9-15);
    \draw (h_10-15) -- (h_9-15);
    \draw (h_0-6) -- (h_0-9);
    \draw (h_3-9) -- (h_0-9);
    \draw (h_3-9) -- (h_3-10);
    \draw (h_5-10) -- (h_3-10);
    \draw (h_5-10) -- (h_5-12);
    \draw (h_6-12) -- (h_5-12);
    \draw (h_6-12) -- (h_6-15);
    \draw (h_9-15) -- (h_6-15);
    \draw (h_0-9) -- (h_0-10);
    \draw (h_3-10) -- (h_0-10);
    \draw (h_3-10) -- (h_3-12);
    \draw (h_5-12) -- (h_3-12);
    \draw (h_5-12) -- (h_5-15);
    \draw (h_6-15) -- (h_5-15);
    \draw (h_0-10) -- (h_0-12);
    \draw (h_3-12) -- (h_0-12);
    \draw (h_3-12) -- (h_3-15);
    \draw (h_5-15) -- (h_3-15);
    \draw (h_0-12) -- (h_0-15);
    \draw (h_3-15) -- (h_0-15);
    \node (a1) at (17,1) {$\ket{\downarrow \downarrow \downarrow \uparrow}$};  
    \node (a2) at (19,1) {$\ket{\downarrow \downarrow \uparrow \downarrow}$};   
    \node (a4) at (21,1) {$\ket{\downarrow \uparrow \downarrow \downarrow}$};
    \node (a7) at (23,1) {$\ket{\downarrow \uparrow \uparrow \uparrow}$};
    \node (a8) at (25,1) {$\ket{\uparrow \downarrow \downarrow \downarrow }$};  
    \node (a11) at (27,1) {$\ket{\uparrow \downarrow \uparrow \uparrow}$};   
    \node (a13) at (29,1) {$\ket{\uparrow \uparrow \downarrow \uparrow}$};
    \node (a14) at (31,1) {$\ket{\uparrow \uparrow \uparrow \downarrow}$};
    \node (i1) at (17,-1) {$\ket{1}$};  
    \node (i2) at (19,-1) {$\ket{2}$};    
    \node (i4) at (21,-1) {$\ket{4}$};
    \node (i7) at (23,-1) {$\ket{7}$};
    \node (i8) at (25,-1) {$\ket{8}$};  
    \node (i11) at (27,-1) {$\ket{11}$};    
    \node (i13) at (29,-1) {$\ket{13}$};
    \node (i14) at (31,-1) {$\ket{14}$};
    \node (bi1) [fill=gray!30] at (17,0) {$\ket{0001}$};  
    \node (bi2) [fill=gray!30] at (19,0) {$\ket{0010}$};    
    \node (bi4) [fill=gray!30] at (21,0) {$\ket{0100}$};
    \node (bi7) [fill=gray!30] at (23,0) {$\ket{0111}$};
    \node (bi8) [fill=gray!30] at (25,0) {$\ket{1000}$};  
    \node (bi11) [fill=gray!30] at (27,0) {$\ket{1011}$};    
    \node (bi13) [fill=gray!30] at (29,0) {$\ket{1101}$};
    \node (bi14) [fill=gray!30] at (31,0) {$\ket{1110}$};
    \node  (h_1-2) [fill=blue!30] at (18,2) {$J_{12}^{x}+J_{12}^{y}$};
    \node  (h_2-4) [fill=blue!20] at (20,2) {$J_{23}^{x}+J_{23}^{y}$};
    \node  (h_4-7) [fill=blue!30] at (22,2) {$J_{12}^{x}-J_{12}^{y}$};
    \node  (h_7-8) [fill=blue!10] at (24,2) {};
    \node  (h_8-11) [fill=blue!30] at (26,2) {$J_{12}^{x}-J_{12}^{y}$};
    \node  (h_11-13) [fill=blue!20] at (28,2) {$J_{23}^{x}+J_{23}^{y}$};
    \node  (h_13-14) [fill=blue!30] at (30,2) {$J_{12}^{x}+J_{12}^{y}$};
    \draw (a1) -- (h_1-2);
    \draw (a2) -- (h_1-2);
    \draw (a2) -- (h_2-4);
    \draw (a4) -- (h_2-4);
    \draw (a4) -- (h_4-7);
    \draw (a7) -- (h_4-7);
    \draw (a7) -- (h_7-8);
    \draw (a8) -- (h_7-8);
    \draw (a8) -- (h_8-11);
    \draw (a11) -- (h_8-11);
    \draw (a11) -- (h_11-13);
    \draw (a13) -- (h_11-13);
    \draw (a13) -- (h_13-14);
    \draw (a14) -- (h_13-14);
    \node  (h_1-4) [fill=blue!20] at (19,3) {$J_{13}^{x}+J_{13}^{y}$};
    \node  (h_2-7) [fill=blue!20] at (21,3) {$J_{13}^{x}-J_{13}^{y}$};
    \node  (h_4-8) [fill=blue!10] at (23,3) {$J_{34}^{x}+J_{34}^{y}$};
    \node  (h_7-11) [fill=blue!10] at (25,3) {$J_{34}^{x}+J_{34}^{y}$};
    \node  (h_8-13) [fill=blue!20] at (27,3) {$J_{13}^{x}-J_{13}^{y}$};
    \node  (h_11-14) [fill=blue!20] at (29,3) {$J_{13}^{x}+J_{13}^{y}$};
    \node  (h_1-7) [fill=blue!20] at (20,4) {$J_{23}^{x}-J_{23}^{y}$};
    \node  (h_2-8) [fill=blue!10] at (22,4) {$J_{24}^{x}+J_{24}^{y}$};
    \node  (h_4-11) [fill=blue!10] at (24,4) {};
    \node  (h_7-13) [fill=blue!10] at (26,4) {$J_{24}^{x}+J_{24}^{y}$};
    \node  (h_8-14) [fill=blue!20] at (28,4) {$J_{23}^{x}-J_{23}^{y}$};
    \node  (h_1-8) [fill=blue!10] at (21,5) {$J_{14}^{x}+J_{14}^{y}$};
    \node  (h_2-11) [fill=blue!10] at (23,5) {$J_{14}^{x}-J_{14}^{y}$};
    \node  (h_4-13) [fill=blue!10] at (25,5) {$J_{14}^{x}-J_{14}^{y}$};
    \node  (h_7-14) [fill=blue!10] at (27,5) {$J_{14}^{x}+J_{14}^{y}$};
    \node  (h_1-11) [fill=blue!10] at (22,6) {$J_{24}^{x}-J_{24}^{y}$};
    \node  (h_2-13) [fill=blue!10] at (24,6) {};
    \node  (h_4-14) [fill=blue!10] at (26,6) {$J_{24}^{x}-J_{24}^{y}$};
    \node (h_1-13) [fill=blue!10] at (23,7) {$J_{34}^{x}-J_{34}^{y}$};
    \node (h_2-14) [fill=blue!10] at (25,7) {$J_{34}^{x}-J_{34}^{y}$};
    \node (h_1-14) [fill=blue!10] at (24,8) {};
    \draw (h_1-2) -- (h_1-4);
    \draw (h_2-4) -- (h_1-4);
    \draw (h_2-4) -- (h_2-7);
    \draw (h_4-7) -- (h_2-7);
    \draw (h_4-7) -- (h_4-8);
    \draw (h_7-8) -- (h_4-8);
    \draw (h_7-8) -- (h_7-11);
    \draw (h_8-11) -- (h_7-11);
    \draw (h_8-11) -- (h_8-13);
    \draw (h_11-13) -- (h_8-13);
    \draw (h_11-13) -- (h_11-14);
    \draw (h_13-14) -- (h_11-14);
    
    \draw (h_1-4) -- (h_1-7);
    \draw (h_2-7) -- (h_1-7);
    \draw (h_2-7) -- (h_2-8);
    \draw (h_4-8) -- (h_2-8);
    \draw (h_4-8) -- (h_4-11);
    \draw (h_7-11) -- (h_4-11);
    \draw (h_7-11) -- (h_7-13);
    \draw (h_8-13) -- (h_7-13);
    \draw (h_8-13) -- (h_8-14);
    \draw (h_11-14) -- (h_8-14);
    
    \draw (h_1-7) -- (h_1-8);
    \draw (h_2-8) -- (h_1-8);
    \draw (h_2-8) -- (h_2-11);
    \draw (h_4-11) -- (h_2-11);
    \draw (h_4-11) -- (h_4-13);
    \draw (h_7-13) -- (h_4-13);
    \draw (h_7-13) -- (h_7-14);
    \draw (h_8-14) -- (h_7-14);
    
    \draw (h_1-8) -- (h_1-11);
    \draw (h_2-11) -- (h_1-11);
    \draw (h_2-11) -- (h_2-13);
    \draw (h_4-13) -- (h_2-13);
    \draw (h_4-13) -- (h_4-14);
    \draw (h_7-14) -- (h_4-14);
    
    \draw (h_1-11) -- (h_1-13);
    \draw (h_2-13) -- (h_1-13);
    \draw (h_2-13) -- (h_2-14);
    \draw (h_4-14) -- (h_2-14);
    
    \draw (h_1-13) -- (h_1-14);
    \draw (h_2-14) -- (h_1-14);

    \node  (h_15-1) [fill=green!20] at (16,2) {};
    \node  (h_12-1) [fill=green!20] at (15,3) {};
    \node  (h_15-2) [fill=green!20] at (17,3) {};
    \node  (h_10-1) [fill=green!20] at (14,4) {};
    \node  (h_12-2) [fill=green!20] at (16,4) {};
    \node  (h_15-4) [fill=green!20] at (18,4) {};
    \node  (h_9-1) [fill=green!10] at (13,5) {$B_{4}^{x}+iB_{4}^{y}$};
    \node  (h_10-2) [fill=green!10] at (15,5) {$B_{4}^{x}+iB_{4}^{y}$};
    \node  (h_12-4) [fill=green!10] at (17,5) {$B_{4}^{x}+iB_{4}^{y}$};
    \node  (h_15-7) [fill=green!10] at (19,5) {$B_{4}^{x}+iB_{4}^{y}$};
    \node  (h_6-1) [fill=green!20] at (12,6) {};
    \node  (h_9-2) [fill=green!20] at (14,6) {};
    \node  (h_10-4) [fill=green!20] at (16,6) {};
    \node  (h_12-7) [fill=green!20] at (18,6) {};
    \node  (h_15-8) [fill=green!20] at (20,6) {};
    \node  (h_5-1) [fill=green!20] at (11,7) {$B_{3}^{x}+iB_{3}^{y}$};
    \node  (h_6-2) [fill=green!20] at (13,7) {$B_{3}^{x}+iB_{3}^{y}$};
    \node  (h_9-4) [fill=green!20] at (15,7) {};
    \node  (h_10-7) [fill=green!20] at (17,7) {};
    \node  (h_12-8) [fill=green!20] at (19,7) {$B_{3}^{x}+iB_{3}^{y}$};
    \node  (h_15-11) [fill=green!20] at (21,7) {$B_{3}^{x}+iB_{3}^{y}$};

    \node (h_3-1) [fill=green!30] at (10,8) {$B_{2}^{x}+iB_{2}^{y}$};
    \node (h_5-2) [fill=green!20] at (12,8) {};
    \node (h_6-4) [fill=green!30] at (14,8) {$B_{2}^{x}+iB_{2}^{y}$};
    \node (h_9-7) [fill=green!20] at (16,8) {};
    \node (h_10-8) [fill=green!30] at (18,8) {$B_{2}^{x}+iB_{2}^{y}$};
    \node (h_12-11) [fill=green!20] at (20,8) {};
    \node (h_15-13) [fill=green!30] at (22,8) {$B_{2}^{x}+iB_{2}^{y}$};

    \node (h_0-1) [fill=green!30] at (9,9) {$B_{1}^{x}-iB_{1}^{y}$};
    \node (h_3-2) [fill=green!30] at (11,9) {$B_{1}^{x}+iB_{1}^{y}$};
    \node (h_5-4) [fill=green!30] at (13,9) {$B_{1}^{x}+iB_{1}^{y}$};
    \node (h_6-7) [fill=green!30] at (15,9) {$B_{1}^{x}-iB_{1}^{y}$};
    \node (h_9-8) [fill=green!30] at (17,9) {$B_{1}^{x}+iB_{1}^{y}$};
    \node (h_10-11) [fill=green!30] at (19,9) {$B_{1}^{x}-iB_{1}^{y}$};
    \node (h_12-13) [fill=green!30] at (21,9) {$B_{1}^{x}-iB_{1}^{y}$};
    \node (h_15-14) [fill=green!30] at (23,9) {$B_{1}^{x}+iB_{1}^{y}$};

    \node (h_0-2) [fill=green!30] at (10,10) {$B_{2}^{x}-iB_{2}^{y}$};
    \node (h_3-4) [fill=green!20] at (12,10) {};
    \node (h_5-7) [fill=green!30] at (14,10) {$B_{2}^{x}-iB_{2}^{y}$};
    \node (h_6-8) [fill=green!20] at (16,10) {};
    \node (h_9-11) [fill=green!30] at (18,10) {$B_{2}^{x}-iB_{2}^{y}$};
    \node (h_10-13) [fill=green!20] at (20,10) {};
    \node (h_12-14) [fill=green!30] at (22,10) {$B_{2}^{x}-iB_{2}^{y}$};

    \node (h_0-4) [fill=green!20] at (11,11) {$B_{3}^{x}-iB_{3}^{y}$};
    \node (h_3-7) [fill=green!20] at (13,11) {$B_{3}^{x}-iB_{3}^{y}$};
    \node (h_5-8) [fill=green!20] at (15,11) {};
    \node (h_6-11) [fill=green!20] at (17,11) {};
    \node (h_9-13) [fill=green!20] at (19,11) {$B_{3}^{x}-iB_{3}^{y}$};
    \node (h_10-14) [fill=green!20] at (21,11) {$B_{3}^{x}-iB_{3}^{y}$};

    \node (h_0-7) [fill=green!20] at (12,12) {};
    \node (h_3-8) [fill=green!10] at (14,12) {};
    \node (h_5-11) [fill=green!10] at (16,12) {};
    \node (h_6-13) [fill=green!10] at (18,12) {};
    \node (h_9-14) [fill=green!10] at (20,12) {};

    \node (h_0-8) [fill=green!10] at (13,13) {$B_{4}^{x}-iB_{4}^{y}$};
    \node (h_3-11) [fill=green!10] at (15,13) {$B_{4}^{x}-iB_{4}^{y}$};
    \node (h_5-13) [fill=green!10] at (17,13) {$B_{4}^{x}-iB_{4}^{y}$};
    \node (h_6-14) [fill=green!10] at (19,13) {$B_{4}^{x}-iB_{4}^{y}$};

    \node (h_0-11) [fill=green!10] at (14,14) {};
    \node (h_3-13) [fill=green!10] at (16,14) {};
    \node (h_5-14) [fill=green!10] at (18,14) {};

    \node (h_0-13) [fill=green!10] at (15,15) {};
    \node (h_3-14) [fill=green!10] at (17,15) {};

    \node (h_0-14) [fill=green!10] at (16,16) {};

    \draw (a15) -- (h_15-1);
    \draw (a1) -- (h_15-1);
    \draw (h_12-15) -- (h_12-1);
    \draw (h_15-1) -- (h_12-1);
    \draw (h_15-1) -- (h_15-2);
    \draw (h_1-2) -- (h_15-2);
    \draw (h_10-15) -- (h_10-1);
    \draw (h_12-1) -- (h_10-1);
    \draw (h_15-2) -- (h_12-2);
    \draw (h_12-1) -- (h_12-2);
    \draw (h_15-2) -- (h_15-4);
    \draw (h_1-4) -- (h_15-4);
    \draw (h_9-15) -- (h_9-1);
    \draw (h_10-1) -- (h_9-1);
    \draw (h_12-2) -- (h_10-2);
    \draw (h_10-1) -- (h_10-2);
    \draw (h_12-2) -- (h_12-4);
    \draw (h_15-4) -- (h_12-4);
    \draw (h_15-4) -- (h_15-7);
    \draw (h_1-7) -- (h_15-7);
    \draw (h_6-15) -- (h_6-1);
    \draw (h_9-1) -- (h_6-1);
    \draw (h_10-2) -- (h_9-2);
    \draw (h_9-1) -- (h_9-2);
    \draw (h_10-2) -- (h_10-4);
    \draw (h_12-4) -- (h_10-4);
    \draw (h_12-4) -- (h_12-7);
    \draw (h_15-7) -- (h_12-7);
    \draw (h_1-8) -- (h_15-8);
    \draw (h_15-7) -- (h_15-8);
    \draw (h_5-15) -- (h_5-1);
    \draw (h_6-1) -- (h_5-1);
    \draw (h_9-2) -- (h_6-2);
    \draw (h_6-1) -- (h_6-2);
    \draw (h_9-2) -- (h_9-4);
    \draw (h_10-4) -- (h_9-4);
    \draw (h_12-7) -- (h_10-7);
    \draw (h_10-4) -- (h_10-7);
    \draw (h_12-7) -- (h_12-8);
    \draw (h_15-8) -- (h_12-8);
    \draw (h_1-11) -- (h_15-11);
    \draw (h_15-8) -- (h_15-11);
    \draw (h_3-15) -- (h_3-1);
    \draw (h_5-1) -- (h_3-1);
    \draw (h_6-2) -- (h_5-2);
    \draw (h_5-1) -- (h_5-2);
    \draw (h_6-2) -- (h_6-4);
    \draw (h_9-4) -- (h_6-4);
    \draw (h_10-7) -- (h_9-7);
    \draw (h_9-4) -- (h_9-7);
    \draw (h_10-7) -- (h_10-8);
    \draw (h_12-8) -- (h_10-8);
    \draw (h_15-11) -- (h_12-11);
    \draw (h_12-8) -- (h_12-11);
    \draw (h_15-11) -- (h_15-13);
    \draw (h_1-13) -- (h_15-13);
    \draw (h_0-15) -- (h_0-1);
    \draw (h_3-1) -- (h_0-1);
    \draw (h_5-2) -- (h_3-2);
    \draw (h_3-1) -- (h_3-2);
    \draw (h_5-2) -- (h_5-4);
    \draw (h_6-4) -- (h_5-4);
    \draw (h_9-7) -- (h_6-7);
    \draw (h_6-4) -- (h_6-7);
    \draw (h_9-7) -- (h_9-8);
    \draw (h_10-8) -- (h_9-8);
    \draw (h_12-11) -- (h_10-11);
    \draw (h_10-8) -- (h_10-11);
    \draw (h_12-11) -- (h_12-13);
    \draw (h_15-13) -- (h_12-13);
    \draw (h_1-14) -- (h_15-14);
    \draw (h_15-13) -- (h_15-14);

    \draw (h_0-1) -- (h_0-2);
    \draw (h_3-2) -- (h_0-2);
    \draw (h_3-2) -- (h_3-4);
    \draw (h_5-4) -- (h_3-4);
    \draw (h_5-4) -- (h_5-7);
    \draw (h_6-7) -- (h_5-7);
    \draw (h_6-7) -- (h_6-8);
    \draw (h_9-8) -- (h_6-8);
    \draw (h_9-8) -- (h_9-11);
    \draw (h_10-11) -- (h_9-11);
    \draw (h_10-11) -- (h_10-13);
    \draw (h_12-13) -- (h_10-13);
    \draw (h_12-13) -- (h_12-14);
    \draw (h_15-14) -- (h_12-14);
    
    \draw (h_0-2) -- (h_0-4);
    \draw (h_3-4) -- (h_0-4);
    \draw (h_3-4) -- (h_3-7);
    \draw (h_5-7) -- (h_3-7);
    \draw (h_5-7) -- (h_5-8);
    \draw (h_6-8) -- (h_5-8);
    \draw (h_6-8) -- (h_6-11);
    \draw (h_9-11) -- (h_6-11);
    \draw (h_9-11) -- (h_9-13);
    \draw (h_10-13) -- (h_9-13);
    \draw (h_10-13) -- (h_10-14);
    \draw (h_12-14) -- (h_10-14);

    \draw (h_0-4) -- (h_0-7);
    \draw (h_3-7) -- (h_0-7);
    \draw (h_3-7) -- (h_3-8);
    \draw (h_5-8) -- (h_3-8);
    \draw (h_5-8) -- (h_5-11);
    \draw (h_6-11) -- (h_5-11);
    \draw (h_6-11) -- (h_6-13);
    \draw (h_9-13) -- (h_6-13);
    \draw (h_9-13) -- (h_9-14);
    \draw (h_10-14) -- (h_9-14);

    \draw (h_0-7) -- (h_0-8);
    \draw (h_3-8) -- (h_0-8);
    \draw (h_3-8) -- (h_3-11);
    \draw (h_5-11) -- (h_3-11);
    \draw (h_5-11) -- (h_5-13);
    \draw (h_6-13) -- (h_5-13);
    \draw (h_6-13) -- (h_6-14);
    \draw (h_9-14) -- (h_6-14);

    \draw (h_0-8) -- (h_0-11);
    \draw (h_3-11) -- (h_0-11);
    \draw (h_3-11) -- (h_3-13);
    \draw (h_5-13) -- (h_3-13);
    \draw (h_5-13) -- (h_5-14);
    \draw (h_6-14) -- (h_5-14);

    \draw (h_0-11) -- (h_0-13);
    \draw (h_3-13) -- (h_0-13);
    \draw (h_3-13) -- (h_3-14);
    \draw (h_5-14) -- (h_3-14);

     \draw (h_0-13) -- (h_0-14);
    \draw (h_3-14) -- (h_0-14);
\end{tikzpicture}

%% file: qsdfull_tree_3qubits.tex
\begin{tikzpicture}
[
    level 1/.style = {red, sibling distance = 5.5cm},
    level 2/.style = {blue, sibling distance = 2.8cm},
    level 3/.style = {red, sibling distance = 1.5cm},
    level 4/.style = {blue, sibling distance = 0.63cm},
    level 5/.style = {black, sibling distance = 0.63cm},
    edge from parent fork down
]
 
\node(parent){U}
    child {node [circle,fill=red!10!, minimum size = 1pt] {}
    child {node [circle,fill=blue!10!, minimum size = 1pt] {} 
    child {node [circle,fill=red!10!, minimum size = 1pt] {}
    child {node [circle,fill=blue!10!, minimum size = 1pt] {}
    child {node (-15) [circle,fill=gray!40!, minimum size = 4pt] [label=below:ZYZ] {-15}}}
    child {node (-14) [circle,fill=blue!20!, minimum size = 4.5pt] [label=below:M-Rz] {-14}}
    child {node [circle,fill=blue!10!, minimum size = 1pt] {}
    child {node (-13) [circle,fill=gray!40!, minimum size = 4pt] [label=below:ZYZ] {-13}}}}
    child {node (-12) [circle,fill=red!20!, minimum size = 4.5pt] [label=below:M-Ry] {-12}}
    child {node [circle,fill=red!10!, minimum size = 1pt] {}
    child {node [circle,fill=blue!10!, minimum size = 1pt] {}
    child {node (-11) [circle,fill=gray!40!, minimum size = 4pt] [label=below:ZYZ] {-11}}}
    child {node (-10) [circle,fill=blue!20!, minimum size = 4.5pt] [label=below:M-Rz] {-10}}
    child {node [circle,fill=blue!10!, minimum size = 1pt] {}
    child {node (-9) [circle,fill=gray!40!, minimum size = 4pt] [label=below:ZYZ] {-9}}}}
    edge from parent [dashed]}
    child {node (-8) [circle,fill=blue!20!, minimum size = 4.5pt] [label=below:M-Rz] {-8}}
    child {node [circle,fill=blue!10!, minimum size = 1pt] {}
    child {node [circle,fill=red!10!, minimum size = 1pt] {}
    child {node [circle,fill=blue!10!, minimum size = 1pt] {}
    child {node (-7) [circle,fill=gray!40!, minimum size = 4pt] [label=below:ZYZ] {-7}}}
    child {node (-6) [circle,fill=blue!20!, minimum size = 4.5pt] [label=below:M-Rz] {-6}}
    child {node [circle,fill=blue!10!, minimum size = 1pt] {}
    child {node (-5) [circle,fill=gray!40!, minimum size = 4pt] [label=below:ZYZ] {-5}}}}
    child {node (-4) [circle,fill=red!20!, minimum size = 4.5pt] [label=below:M-Ry] {-4}}
    child {node [circle,fill=red!10!, minimum size = 1pt] {}
    child {node [circle,fill=blue!10!, minimum size = 1pt] {}
    child {node (-3) [circle,fill=gray!40!, minimum size = 4pt] [label=below:ZYZ] {-3}}}
    child {node (-2) [circle,fill=blue!20!, minimum size = 4.5pt] [label=below:M-Rz] {-2}}
    child {node [circle,fill=blue!10!, minimum size = 1pt] {}
    child {node (-1) [circle,fill=gray!40!, minimum size = 4pt] [label=below:ZYZ] {-1}}}}}
    edge from parent [dashed]} 
    child {node (0) [circle,fill=red!20!, minimum size = 4.5pt] [label=below:M-Ry] {0}}
    child {node [circle,fill=red!10!, minimum size = 1pt] {}
    child {node [circle,fill=blue!10!, minimum size = 1pt] {}
    child {node [circle,fill=red!10!, minimum size = 1pt] {}
    child {node [circle,fill=blue!10!, minimum size = 1pt] {}
    child {node (1) [circle,fill=gray!40!, minimum size = 4pt] [label=below:ZYZ] {1}}}
    child {node (2) [circle,fill=blue!20!, minimum size = 4.5pt] [label=below:M-Rz] {2}}
    child {node [circle,fill=blue!10!, minimum size = 1pt] {}
    child {node (3) [circle,fill=gray!40!, minimum size = 4pt] [label=below:ZYZ] {3}}}}
    child {node (4) [circle,fill=red!20!, minimum size = 4.5pt] [label=below:M-Ry] {4}}
    child {node [circle,fill=red!10!, minimum size = 1pt] {}
    child {node [circle,fill=blue!10!, minimum size = 1pt] {}
    child {node (5) [circle,fill=gray!40!, minimum size = 4pt] [label=below:ZYZ] {5}}}
    child {node (6) [circle,fill=blue!20!, minimum size = 4.5pt] [label=below:M-Rz] {6}}
    child {node [circle,fill=blue!10!, minimum size = 1pt] {}
    child {node (7) [circle,fill=gray!40!, minimum size = 4pt] [label=below:ZYZ] {7}}}}}
    child {node (8) [circle,fill=blue!20!, minimum size = 4.5pt] [label=below:M-Rz] {8}}
    child {node [circle,fill=blue!10!, minimum size = 1pt] {}
    child {node [circle,fill=red!10!, minimum size = 1pt] {}
    child {node [circle,fill=blue!10!, minimum size = 1pt] {}
    child {node (9) [circle,fill=gray!40!, minimum size = 4pt] [label=below:ZYZ] {9}}}
    child {node (10) [circle,fill=blue!20!, minimum size = 4.5pt] [label=below:M-Rz] {10}}
    child {node [circle,fill=blue!10!, minimum size = 1pt] {}
    child {node (11) [circle,fill=gray!40!, minimum size = 4pt] [label=below:ZYZ] {11}}}}
    child {node (12) [circle,fill=red!20!, minimum size = 4.5pt] [label=below:M-Ry] {12}}
    child {node [circle,fill=red!10!, minimum size = 1pt] {}
    child {node [circle,fill=blue!10!, minimum size = 1pt] {}
    child {node (13) [circle,fill=gray!40!, minimum size = 4pt] [label=below:ZYZ] {13}}}
    child {node (14) [circle,fill=blue!20!, minimum size = 4.5pt] [label=below:M-Rz] {14}}
    child {node [circle,fill=blue!10!, minimum size = 1pt] {}
    child {node (15) [circle,fill=gray!40!, minimum size = 4pt] [label=below:ZYZ] {15}}}}}
    edge from parent [dashed]};
\end{tikzpicture}

%% file: 3qubits_circuit_blackbox.tex
\resizebox{\textwidth}{!}{
\begin{quantikz}
\lstick{$q[0]$}&\qw &\qw &\qw &\qw &\qw & \qw & \qw &\gate[wires=3]{8} &\qw &\qw &\qw &\qw &\qw &\qw &\qw &\gate[wires=3]{0}&\qw &\qw &\qw &\qw &\qw &\qw &\qw &\gate[wires=3]{-8} & \qw & \qw& \qw & \qw& \qw & \qw & \qw & \qw\\
\lstick{$q[1]$}& \qw &\gate[wires=2]{14} &\qw &\gate[wires=2]{12} &\qw &\gate[wires=2]{10} &\qw & \qw & \qw &\gate[wires=2]{6} &\qw &\gate[wires=2]{4}&\qw &\gate[wires=2]{2}&\qw &\qw &\qw &\gate[wires=2]{-2} &\qw &\gate[wires=2]{-4} &\qw &\gate[wires=2]{-6} &\qw & \qw & \qw & \gate[wires=2]{-10} & \qw & \gate[wires=2]{-12} & \qw & \gate[wires=2]{-14} & \qw & \qw\\
\lstick{$q[2]$}& \gate[wires=1]{15} & \qw & \gate[wires=1]{13} & \qw & \gate[wires=1]{11} & \qw & \gate[wires=1]{9} & \qw &\gate[wires=1]{7}&\qw &\gate[wires=1]{5}&\qw &\gate[wires=1]{3}&\qw &\gate[wires=1]{1}&\qw &\gate[wires=1]{-1}&\qw &\gate[wires=1]{-3}&\qw &\gate[wires=1]{-5}&\qw &\gate[wires=1]{-7} & \qw & \gate[wires=1]{-9} & \qw & \gate[wires=1]{-11} & \qw & \gate[wires=1]{-13} & \qw & \gate[wires=1]{-15} & \qw
\end{quantikz}
}

%% file: refs.tex
\providecommand{\latin}[1]{#1}
\providecommand*\mcitethebibliography{\thebibliography}
\csname @ifundefined\endcsname{endmcitethebibliography}  {\let\endmcitethebibliography\endthebibliography}{}